\documentclass[a4paper,11pt]{article}
\pdfoutput=1

\usepackage{jheppub}
\usepackage[T1]{fontenc}
\usepackage{tensor}
\usepackage{subcaption}
\hypersetup{bookmarksnumbered}
\usepackage{bbold}
\usepackage{float}
\usepackage{mathtools}
\usepackage[table]{xcolor}
\usepackage{longtable}

\DeclareFontFamily{OT1}{pzc}{}
\DeclareFontShape{OT1}{pzc}{m}{it}{<-> s * [1.10] pzcmi7t}{}
\DeclareMathAlphabet{\mathpzc}{OT1}{pzc}{m}{it}

\usepackage[vcentermath]{youngtab}

\def\tyoung#1{\,{\scriptsize \young(#1)}\,}
\def\myoung#1{\,{\young(#1)}\,}
\def\eone{e_1}
\def\etwo{e_2}

\def\efour{e_4}
\def\n{n}
\def\np{n'}

\newcommand{\be}{\begin{equation}}
\newcommand{\ee}{\end{equation}}
\newcommand{\bea}{\begin{eqnarray}}
\newcommand{\eea}{\end{eqnarray}}
\newcommand{\ba}{\begin{equation}\begin{aligned}}
\newcommand{\ea}{\end{aligned}\end{equation}}

\newcommand{\MM}{{\mathcal{M}}}

\newcommand{\R}{\mathcal R}
\newcommand{\T}{\mathcal T}

\newcommand{\im}{\text{Im}}
\newcommand{\J}{\mathcal J}
\newcommand{\Ach}{A_{\text{ch}}}
\newcommand{\Bch}{B_{\text{ch}}}
\newcommand{\gOne}{\mathpzc{g}_1}

\newcommand{\bZ}{\mathbb Z}

\def\legP{\mathcal{P}}
\def\legQ{\mathcal{Q}}
\def\avg#1{\left<#1\right>}
\def\tr#1{\text{Tr}\left(#1\right)}

\title{\boldmath Bootstrapping Pions at Large $N$  \\ {\Large Part II: Background Gauge Fields and the Chiral Anomaly}}

\author[a,b]{Jan Albert}
\author[a]{and Leonardo Rastelli}

\affiliation[a]{C. N. Yang Institute for Theoretical Physics, Stony Brook University,\newline Stony Brook, NY 11794-3840, U.S.A.}

\affiliation[b]{Simons Center for Geometry and Physics, Stony Brook University,\newline Stony Brook, NY 11794-3636, U.S.A.}

\preprint{YITP-SB-2023-15}

\abstract{We continue the program~\cite{Albert:2022oes} of carving out the space of large~$N$ confining gauge theories
by modern S-matrix bootstrap methods, 
 with the ultimate goal of cornering large~$N$ QCD.  In this paper, we focus on the effective field theory of massless pions coupled to background electromagnetic fields. We derive the full set of positivity constraints encoded in the  system of 2 $\to$ 2 scattering amplitudes of pions and photons. This system probes 
  a larger set of intermediate meson states,
and is thus sensitive to intricate large $N$ selection rules,
especially when supplemented with expectations from Regge theory.
  It also has access to the
 coefficient of the chiral anomaly.
We find novel numerical bounds on several ratios of Wilson coefficients, in units of the rho mass.
By matching the chiral anomaly with the microscopic theory, we also derive bounds that contain an explicit $N$ dependence. 
}

\begin{document} 
\maketitle

\section{Introduction and summary}

The physical picture for the large $N$
limit of QCD\footnote{By QCD we mean the four-dimensional $SU(N)$ gauge theory with a fixed number $N_f$ of fundamental Dirac fermions (quarks). For simplicity, we further assume that the quarks are massless, though it would be straightforward to lift this assumption.} has been well-established since the 1970s~\cite{tHooft:1973alw, Witten:1979kh}.
For strictly infinite   $N$, 
QCD is a theory of infinitely many stable color-singlet glueballs and mesons.
(Baryons are heavy at large $N$).
Their interactions, 
suppressed by inverse powers of $N$, are captured to leading order by meromorphic S-matrices, whose poles correspond to single-hadron exchanges.
Regge theory
predicts that hadrons should organize themselves into
Regge trajectories, whose intercepts
control large energy, 
small angle scattering.
This picture gives a surprisingly good caricature of the real $(N=3)$ world, where mesons really do lie in (almost linear) Regge trajectories, Regge theory fits hadron scattering very well, 
and  large $N$ selection rules (such as the OZI rule \cite{OKUBO1963165,Zweig:1964jf,Iizuka}) are approximately obeyed.
The same physical picture applies to a broad class of large $N$ confining gauge theories. 

This is the perfect scenario that calls for an application of the bootstrap paradigm. There is a space of putative large $N$ theories, parametrized by an infinite set of data (the spectrum of hadrons, encoded in their masses and other quantum numbers,
 and their on-shell cubic couplings) subject to an infinite set of consistency conditions, such as crossing and unitarity of all $2 \to 2$ meromorphic S-matrices. 
This general framework is really quite parallel to that of the very successful conformal bootstrap~\cite{Rattazzi:2008pe, Poland:2018epd, Poland:2022qrs}, and analogous
 semidefinite programming methods can be used to carve out theory space. The ultimate hope is that with sufficient physical input (such as suitable spectral assumptions) one may be able to corner large $N$ QCD and other theories of physical interest, much like the 3D Ising model is recognized to sit at a kink of the exclusion boundary~\cite{El-Showk:2012cjh}  and is in fact
 constrained to lie in a tiny island in parameter space
 once more elaborate constraints are imposed~\cite{Kos:2014bka}.

Broadly understood, these ideas have  a venerable history, indeed we are just declining a version of the old S-matrix bootstrap program for the strong interactions.
More concretely, it has long been appreciated that causality and unitarity of the S-matrix constrain theory space,
leading for example to interesting inequalities for the higher-derivative coefficients of the chiral Lagrangian, see~e.g.~\cite{Martin1969,Pham:1985cr,Ananthanarayan:1994hf,Pennington:1994kc,Comellas:1995hq,Dita:1998mh}. 
But a truly systematic analysis of large $N$ theory space was only started in our first paper~\cite{Albert:2022oes}. Our work 
builds on the general philosophy developed during the
 modern
 S-matrix bootstrap  renaissance
 (see~e.g.~\cite{Paulos:2016fap, Paulos:2016but, Paulos:2017fhb, Homrich:2019cbt, Cordova:2018uop, Guerrieri:2018uew, Bercini:2019vme, Cordova:2019lot, Guerrieri:2020bto, Hebbar:2020ukp, Guerrieri:2020kcs, He:2021eqn, Kruczenski:2022lot})
 and especially on the recent 
 developments~\cite{Bellazzini:2020cot, Tolley:2020gtv, Caron-Huot:2020cmc, Arkani-Hamed:2020blm}
 that have systematized the set of ``positivity constraints'' that follow from consistency of $2\to 2$ scattering.\footnote{
See e.g.~\cite{Sinha:2020win,Li:2021lpe,Caron-Huot:2021rmr,Bern:2021ppb,Chiang:2021ziz,Caron-Huot:2021enk,Henriksson:2021ymi,Davighi:2021osh,Du:2021byy,Chowdhury:2021ynh,Caron-Huot:2022ugt,deRham:2022hpx, Henriksson:2022oeu,Chiang:2022ltp,Caron-Huot:GravPW,Bern:2022yes, CarrilloGonzalez:2022fwg, Creminelli:2022onn, deRham:2022sdl, EliasMiro:2022xaa, Fernandez:2022kzi, deRham:2022gfe, Li:2022aby, Hong:2023zgm, Bellazzini:2023nqj} for recent developments on positivity constraints.
}

\subsection*{Bootstrapping the mesons with positivity bounds}

Our immediate focus is on the meson sector, which is a consistent subsector  at large $N$. There are several reasons to start with the mesons. First,
a basic piece of spectral information comes from
spontaneous chiral symmetry breaking, which implies the existence of massless Goldstone bosons (the pions) in the adjoint representation of the $U(N_f)$ flavor group.
Second, meson scattering is constrained by large $N$ selection rules, encoding the two related facts
that only $q \bar q$ states are stable resonances at large $N$ and 
that only diagrams with a disk topology contribute. For $2\to 2$ pion scattering, this reduces to the statement that the flavor-ordered amplitude has poles just in two kinematic channels,\footnote{These are the $s$ and $u$ channels in our conventions; $t$-channel poles are absent.} but for  more general external states 
the selection rules are quite elaborate and capture a robust physical feature of large $N$ gauge theories.
Last but not least, there is a wealth of real-world data 
that can guide our intuition, and against which compare our results.\footnote{Ideally, we would rely instead on large $N$ lattice results (and in the chiral limit!), but as far as we are aware they are only available for a limited number of observables, see e.g.~\cite{Lucini:2012gg,DeGrand:2016pur,Hernandez:2019qed,Perez:2020vbn,Baeza-Ballesteros:2022azb}.
 }
Such comparison is perhaps not too far-fetched, as for many purposes 
$N=3$ appears to be quite close to $N=\infty$.

The most straightforward parametrization of theory space  would be in terms of the complete spectrum of ``single-trace'' mesons (i.e.~$q \bar q$ states) and their cubic couplings\footnote{This is a slight oversimplification. Cubic couplings would suffice if  $2 \to 2$ scattering amplitudes limited to zero in the Regge limit of $|s| \to \infty$ with fixed $u <0$. The Regge behavior is slightly worse (in general, amplitudes only decay faster than $O(s)$), and some higher-point couplings are needed. See section \ref{sec:ReggeBehavior} and appendix \ref{app:Regge} for a detailed discussion of the expected Regge behavior in meson scattering.} (which are of order $1/\sqrt{N}$). 
These data could be encoded in a Lagrangian with infinitely many fields, one for each meson. 
The full bootstrap problem involves imposing all the constraints that follow from $2 \to 2$ scattering of arbitrary external states.
In practice,
 we can only study a finite number of scattering 
processes at a time, and it makes sense to start with those of the lightest states.
This suggests an equivalent and  perhaps more useful way to organize the bootstrap problem, using
 the language of effective field theory (EFT).  We introduce a UV cut-off scale $M$, and divide the mesons into light states
with masses smaller than $M$, and heavy states with masses larger than $M$. If
 we had complete knowledge of the  large $N$ theory, the EFT of the light states would arise by integrating
out the heavy states at tree level. 
Instead, we decide to be agnostic about the heavy data and constrain the low-energy EFT by imposing its compatibility with consistent
scattering of the light states. In this approach, a point in theory space is parametrized by the infinite set of Wilson coefficients of the EFT.

We emphasize that for us $N$ is always the largest number.  The EFT must be treated at {\it tree level}, because all Wilson coefficients are $O(1/N)$ and EFT loops are obviously subleading.\footnote{This is in contrast with the usual framework of chiral perturbation theory, where calculations to a given order in $E/M$ ($E$ being the typical energy scale of a physical process and $M$ the UV cut-off) comprise both (a finite number of) higher-derivative tree-level interactions and  loops involving the lower-derivative couplings.} 
By the same token, the spectral density for the heavy data (entering the partial wave decomposition of  $2 \to 2$ scattering of the light states) is $O(1/N)$. In this limit, what survives of 
 unitarity  is (semidefinite) {\it positivity} of the spectral density. Further imposing crossing
leads to an infinite set of ``null constraints'' for the heavy data \cite{Tolley:2020gtv, Caron-Huot:2020cmc}.
Semidefinite programming methods can then be applied to {\it rigorously carve out} the space of Wilson coefficients,
in close conceptual and technical analogy with the conformal bootstrap. More precisely,
given that all conditions are inherently homogeneous, this method constrains 
{\it ratios} of Wilson coefficients, rendered dimensionless by appropriate powers of~$M$.

In \cite{Albert:2022oes}, we carried out this program in the simplest setup, where the only light states are the massless pions and the cut-off 
$M$ is identified with the mass of the first state that appears in pion-pion
scattering, which in large $N$ QCD is the rho vector meson. The EFT is just the celebrated chiral Lagrangian for the massless pions. We found novel bounds carving out a convex region in the space of the two four-derivative couplings (normalized by the pion decay constant and in units of $M$). 
Real-world data appear to be compatible with the theoretically allowed region, but their error bars are too large to meaningfully place QCD on our plot.
The allowed region in this parameter space  (see figure \ref{fig:BchIntro}, ignoring the color coding for now) has an interesting geometry of sharp corners (which we were able to understand in terms of simple spurious solutions to crossing) and a tantalizing prominent kink, where the exclusion boundary has a qualitatively change of shape, from straight to curved. 
Could the kink correspond to large $N$ QCD?

Our best current hypothesis is that the kink may also be explained by two  unphysical amplitudes
that  exchange dominance there.
But while there appears to be a satisfactory analytic explanation~\cite{Albert:2022oes, Fernandez:2022kzi} for
the straight segment (in terms of the UV completion of a single tree-level rho exchange)\footnote{In \cite{Albert:2022oes}, we identified a portion of the straight segment in terms of the single rho exchange, but  it appeared that the numerical kink lied outside that ruled in portion.  The authors of \cite{Fernandez:2022kzi} have argued that this is due to slow numerical convergence --  with infinite numerical precision the kink will sit at the endpoint of the ruled-in straight segment.}, 
understanding the curved segment has proved more elusive. In fact, in this paper we are going to find a hint that even that straight segment might hide
some richer physics as soon as we go beyond the four-pion sector.

Fortunately, there is much more that can be done to corner large $N$ QCD.
We can enlarge the setup and input further physical features specific to large $N$ QCD.
A natural next step is to include the rho meson among the light states of the EFT;
the cut-off
$M$ is now interpreted as the mass of the next stable resonance, which in large $N$ QCD is the $f_2$ meson. 
This is the subject of upcoming work~\cite{rhos}, which studies the full set of constraints encoded in the mixed system of  $2 \to 2$ scattering amplitudes of $\pi$s and $\rho$s.

\subsection*{Including background gauge fields}
In this paper, we proceed in another direction. 
The continuous global symmetry of large~$N$ QCD is  $U(N_f)_L\times U(N_f)_R$, which is spontaneously broken to the diagonal subgroup $U(N_f)_V$. 
While still keeping only the pions among the light physical states, we can enlarge the set of observables by including 
matrix elements for the Noether currents of  $U(N_f)_L\times U(N_f)_R$. Both for simplicity and because of its immediate physical significance, we focus here on matrix elements of the electromagnetic current $J^\mu_{\rm em}$, which corresponds to a diagonal $U(1)_Q$ subgroup\footnote{For $N_f =3$, we can take $U(1)_Q$ as the diagonal subgroup of $U(3)_V$ that corresponds to
the usual electric charge assignments of the $u$, $d$ and $s$ quarks, so that $J^\mu_{\rm em}$ is the literal electromagnetic current. 
But there is no additional computational  cost in keeping $U(1)_Q$ general.} of the linearly realized vector symmetry $U(N_f)_V$. 
The expedient way to proceed is to introduce a background gauge field $A_\mu$ for $J^\mu_{\rm em}$ and   study scattering processes involving both pions and ``photons'' (the quanta of $A_\mu$). These photons are not dynamical. They enter only as external states and could in principle be taken  off-shell.  
We keep them on-shell only because this leads to significant  kinematic simplifications.

The task is clear. We are instructed to consider the general EFT that describes pions in a background electromagnetic field. The space of its Wilson coefficients is carved out by the  positivity constraints that follow from  the full mixed system of 
$2 \to 2$  scattering amplitudes of pions and photons.
This is of course not a new problem, and in fact some aspects of it were already studied by  Gourdin and Martin in the 1960s \cite{Martin:1959,Martin:1960}.
We wish to revisit it with the full machinery of the modern S-matrix bootstrap, and explore systematically all its positivity bounds.

There are compelling physical motivations to embark in this technically challenging problem.
First, as the pion/photon system probes a set of intermediate mesons with more general quantum numbers than those appearing in pure pion scattering,
large $N$ selection rules
become much more interesting and  
amount to a non-trivial physical input, especially 
 when supplemented with expectations from Regge theory. 
Second, this system  
 has  access to the value the chiral anomaly, which must be matched
 between the microscopic UV theory and the low-energy EFT.\footnote{
 The idea of inputing anomalies into the S-matrix bootstrap has appeared in
 \cite{Karateev:2022jdb}, which studies  the $a$ Weyl-anomaly coefficient
 in the non-perturbative unitarity framework.
 }

 Famously, in the EFT of pions the anomaly is accounted for by the Wess-Zumino-Witten (WZW) term, which however only contributes to $n$-point pion scattering for odd $n \geq 5$. It is a very interesting open problem to develop the S-matrix bootstrap for higher-point amplitudes, but for the time being we are limited to four-point amplitudes ($2 \to 2$ scattering). Incorporating the background electromagnetic field is a neat shortcut, as now the
 anomaly coefficient shows up in the $4\gamma$ and  $3 \pi 1 \gamma$
 amplitudes. We can then  input a quantitative  feature of large $N$ QCD into our bootstrap problem.
Finally,  we can even hope to make contact with the rich real-world phenomenology of $\pi \gamma$ scattering, see e.g.~\cite{Bijnens:1987dc,PhysRevD.42.1350,Donoghue:1993kw,Volkov:1997is,Moinester:2019sew}.\footnote{The approximation of non-dynamical photons is of course perfectly justified in the real world, where the electromagnetic coupling is  small.} 
 
If we have dwelt on the motivations, it's because the task at hand 
looks daunting. 
We need to deal with a big system of mixed correlators, both  with an intricate flavor and photon-polarization dependence. 
The general strategy is to first parametrize the amplitudes and then work out their partial wave expansion so that we can write down dispersion relations linking the UV to the IR.
For aesthetic {\it and} pragmatic reasons, we  work fully covariantly,
both in the parametrization of the amplitudes and in their high-energy expansion in partial waves. For the parametrization, we first take care of flavor by the usual Chan-Paton method. Following~\cite{Chowdhury:2019kaq,Chowdhury:2020ddc}, we then deal with 
Lorentz kinematics 
by expanding the amplitudes in a covariant basis 
of tensor structures, built out of external polarizations and external momenta. When the dust settles, the entire system is encoded in nine ``reduced'' scalar amplitudes, with transparent transformation properties under crossing.

For the the partial wave expansion, we adopt the beautiful formalism of \cite{Caron-Huot:2022ugt}, which we generalize to deal with mixed correlators and flavor dependence. The idea is to construct the partial waves by gluing the tensors that represent the three-point vertices with the initial and final states in an $s$-channel exchange. 
In this approach, it is immediate to keep track of the spin, parity and charge conjugation quantum numbers of the intermediate states, making it effortless to impose the large $N$ selection rules. Positivity is also completely transparent. We hope that the technical lessons learnt here will be valuable in other S-matrix bootstrap studies.

With the full kinematics in place, we can follow a by now familiar blueprint~\cite{Caron-Huot:2020cmc}. We
write suitably subtracted dispersion relations,
and derive from them sum rules for the low-energy couplings as well as null constraints (encoding crossing) for the heavy spectral densities. To get optimal results, we wish to perform the minimal amount of subtractions needed to  drop the contour at infinity in the complex $s$ plane. For meson scattering, the  behavior at large $s$ (and fixed $u$) is at worst that of the rho Regge trajectory, whose intercept is approximately 0.5 < 1, so one subtraction always suffices. It turns out that in some case we can do better: we have found  linear combinations of amplitudes where the rho trajectory drops out. Regge theory predicts that their large-$s$ behavior is controlled  
by the {\it next} trajectory, which in QCD is that of the pion, whose  exactly zero intercept (in the chiral limit) guarantees the validity of {\it unsubtracted} dispersion relations. These ``improved Regge channels'' allow us to derive sum rules for otherwise inaccessible low-lying EFT couplings, as well as new towers of null constraints. Finally,
a novel set of null constraints 
follows by imposing that the pion is a Goldstone boson.
The sigma model structure of the EFT relates otherwise arbitrary low-energy coefficients; equating the corresponding sum rules yields new ``Goldstone constraints'' amongst the  spectral densities.

\subsection*{Results}

We end this introduction with a brief overview of our numerical results.
In the $4\gamma$ system (which forms a closed subsector) we obtain  bounds on normalized EFT coefficients with up to six derivatives.
Exclusion plots for two-dimensional slices are displayed in figures \ref{fig:a31a23}, \ref{fig:a32a23} and \ref{fig:a33a23}. Their polygonal shapes call for an analytic understanding. While we are able to ``rule in''
a few corners and segments with simple amplitudes,  a more comprehensive analysis remains an interesting future direction.
Comparing with previous work on the $S$-matrix bootstrap for $4\gamma$ scattering \cite{Haring:2022sdp,Henriksson:2021ymi,Henriksson:2022oeu}, we find that our bounds do not improve on the previous results (in particular the bounds of \cite{Henriksson:2022oeu} in the forward limit),
despite the fact that our assumptions are somewhat different (e.g.~we impose a stronger Regge behavior). This suggests that in both cases the bounds are saturated by the same amplitudes, which might or might not be simple solutions to crossing.

\begin{figure}[ht]
\centering
\includegraphics[scale=1]{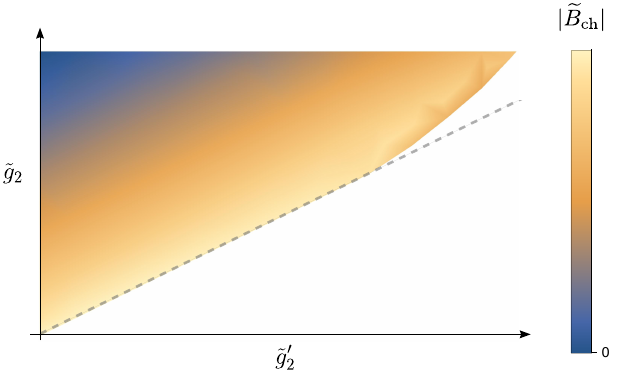}
\caption{Our old exclusion 
plot \cite{Albert:2022oes} for the normalized
four-derivative couplings $\tilde g_2$ and $\tilde g'_2$ in the chiral Lagrangian, now color-coded to 
represent  the bound on the
 normalized anomaly  coefficient $|\widetilde B_{\text{ch}}|$. The bound is maximized along the straight oblique segment
  ($\tilde g_2 = \frac{1}{4}\tilde g_2'$ up to the kink) and decreases away from it. }
\label{fig:BchIntro}
\end{figure}

Another closed subsector is given by the $\pi\gamma \to \pi\gamma$ channel,
whose spectral density is positive on its own. Figure \ref{fig:d21d22} shows the corresponding bounds in a the space of normalized four-derivative couplings. A novel feature of this plot 
is that the diagonal boundary arises from large $N$ selection rules -- without imposing them we would find a larger rectangular allowed region. 
In particular, a naively ruled-in amplitude gets excluded by this new constraint, because its UV completion would necessitate higher-spin resonances
that are forbidden by the selection rules.

More interesting are the 
bounds that are sensitive to the  structure of the full mixed correlator system.  An insightful game to play is to 
combine them with our old bounds \cite{Albert:2022oes} in the $4 \pi$ sector, extending the exclusion region in a third direction. For example, figure \ref{fig:c22-c21} shows a 3D contour plot where the
the color-coding represents the upper bound on a certain combination of four-derivative Wilson coefficients for the $2\pi 2\gamma$ amplitude.
Interestingly, these new bounds are still ruled in (at least in part) by the same simple ``theories'' which saturated the $4\pi$ scattering bounds, suggesting no degeneracies of the extremal theories.

Finally, we arrive at 
the  most exciting bound: that on the  chiral anomaly coefficient.\footnote{
The anomaly coefficient controls both the two-derivative term
in the $4 \gamma$ amplitude, and the lowest contact term in $3\pi\gamma$. For  reasons that we explain 
in sections \ref{sec:4photonsPlots} and \ref{sec:lowderivative} the former decouples and we cannot bound it. Fortunately we  {\it are} able to constrain the anomaly coefficient that appears in the $3 \pi 1\gamma$ amplitude.} If we treat it as any of the other EFT coefficients, we get numerical upper bounds on homogeneous ratios involving the anomaly,~i.e. on the anomaly coefficient divided by suitable positive combinations of other Wilson coefficients. If we remember that the anomaly must be matched with the UV theory, we can input its value in large $N$ QCD and obtain 
``inhomogenous'' bounds with an explicit $N$ dependence, such as 
(\ref{eq:nonhomogeneousbound}).
  It is also rewarding to see how the upper bound on the anomaly coefficient changes as we move in the space of the other EFT couplings, e.g.~as  a function of the two $4 \pi$ four-derivative coefficients. 
Figure~\ref{fig:BchIntro} (a simplified version of
figure~\ref{fig:Bch3D}) shows  the corresponding 3D contour plot. The anomaly is maximized on the straight segment of the exclusion boundary. In the $4\pi$ system, one can show~\cite{Fernandez:2022kzi} that the unique spectral density that corresponds to the straight segment is the (UV completion of) a single tree-level rho exchange. Curiously, 
integrating out a single rho in $\pi \pi \to \pi \gamma$ 
yields an anomaly coefficient strictly smaller than  our numerical bound, suggesting the need for additional light states (which decouple in the $4 \pi$ system).

It is worth  pointing out that we have encountered two obstructions, whose resolution would give even stronger results. First, a few
 EFT couplings that occur at lowest  derivative order have non-positive definite sum rules. As we shall explain, this means that they decouple from the semidefinite problem and cannot be bounded. This is a pity, because they are among the most physically relevant. An example is a four-derivative coupling that controls $\pi \gamma$ scattering and has been measured in the real world.\footnote{To the best of our knowledge, there are no phenomenological estimates for the six-derivative couplings in $2\pi 2 \gamma$ that we {\it are} able to bound. This however may also be a reflection of the different way chiral perturbation theory is matched to experimental data at finite $N$.}
It is conceivable that 
taking the photons off-shell or
considering 
a larger set of background gauge fields might yield positive sum rules for these couplings. 
Second, technical numerical reasons (presumably, sign oscillations with spin) prevent our bounds to reflect some of the most interesting null constraints 
derived in section \ref{sec:Dispersion},
namely
those obtained by taking into account ``improved Regge channels'' and those that follow by identifying the pion with a Goldstone boson.

\bigskip
\noindent
The remainder of the paper is organized as follows. In section~\ref{sec:background} we review standard material about the  (large $N$) chiral Lagrangian in the presence of background gauge fields, 
with an emphasis on the matching of 't Hooft anomalies. 
In section~\ref{sec:Parametrizations} we find a covariant parametrization for the full mixed system of $2 \to 2$ scattering amplitudes in terms of a set of reduced scalar amplitudes.  In section~\ref{sec:PartialWaves} we work out covariantly the high-energy partial wave decomposition,  dealing in one fell swoop with the Lorentz, flavor, and multi-correlator structures
by gluing three-point vertices in 
an $s$-channel exchange. This section is complemented with an ancillary file providing a pedagogical explicit construction of the partial waves.
In section~\ref{sec:Dispersion} we write down the dispersion relations that relate the low-energy EFT to the heavy spectral densities, deriving sum rules for the Wilson coefficients as well as the complete set of null constraints that follow from crossing symmetry. We carefully discuss the Regge behavior.
In section~\ref{sec:Positivity} we implement the positivity bounds by numerical semidefinite 
methods (using SDBP), obtaining several exclusion plots for ratios of Wilson coefficients in units of the cut-off.  
By matching the chiral anomaly, we also derive 
bounds that contain an explicit $N$ dependence. 
We give a preliminary discussion of how some features of our plots may be explained analytically in terms of  simple ``ruled-in'' amplitudes.
In section~\ref{sec:Discussion} we provide a summary of the new physical constraints and we discuss  the obstructions that we have found in implementing them. We end in section~\ref{sec:Conclusions} with some  concluding remarks.
To make the paper self-contained, we offer in appendix~\ref{app:Regge} an informal review of some basic results of Regge theory.

\section{The chiral Lagrangian with background gauge fields}
\label{sec:background}

Consider large $N$ 
 QCD in the chiral limit: the four-dimensional $SU(N)$ gauge theory with $N_f$  {\it massless} Dirac fermions in the fundamental representation, where $N \to \infty$  keeping
$N_f$ and  $\Lambda_{\rm QCD}$ fixed 
\cite{tHooft:1973alw}. 
It is universally believed (and confirmed by lattice studies for increasing values of $N$) that QCD remains confining at large $N$, and undergoes spontaneous chiral symmetry breaking\footnote{Under reasonable general assumptions, one can in fact prove~\cite{Coleman:1980mx} that chiral symmetry breaking to the diagonal subgroup must occur in the large $N$ theory.} according to the pattern
\begin{equation} \label{pattern}
    U(N_f)_L\times U(N_f)_R \longrightarrow {\rm diag} (U(N_f)_L\times U(N_f)_R) \equiv  U(N_f)_V\, ,
\end{equation}
where  $U(N_f)_L$ and $U(N_f)_R$ act respectively on the left-handed and right-handed components of the fermions.
Note that the global symmetry group of the large $N$ theory includes 
the axial $U(1)_A$, which is explicitly broken at finite $N$ by the ABJ anomaly.
Spontaneous symmetry breaking implies
 the existence of 
 $N_f^2$ massless Goldstone bosons $\pi^a$ parametrizing the coset space $U(N_f)$. They  are the only degrees of freedom that survive at very low energies. We will keep referring to them  as ``pions'' for general $N_f$.

\subsection{The chiral Lagrangian}
The physics of QCD  is described at low energies by an effective field theory (EFT) for the pions, the celebrated chiral Lagrangian (see e.g.\ \cite{Gasser:1983yg,Gasser:1984gg,Kaiser:2000gs}). The identification of the pions
with the Goldstone bosons of the spontaneously broken chiral symmetry implies it must take the form of a non-linear sigma model with $U(N_f)$ target space,
\begin{align}
    \label{eq:Lch}
	S_\text{Ch}= \int d^4 x \bigg[-\frac{f_\pi^2}{4}&\tr{\partial_\mu U \partial^\mu U^\dagger}
	+\kappa_1 \tr{\partial_\mu U \partial^\mu U^\dagger \partial_\nu U \partial^\nu U^\dagger}\nonumber\\
	+\kappa_2 &\tr{\partial_\mu U \partial_\nu U^\dagger \partial^\mu U \partial^\nu U^\dagger} + \cdots\bigg]\,,
\end{align}
where   $U(x)=\exp \left[i\frac{2}{f_\pi} T_a\pi^a(x)\right]\in U(N_f)$, with $T_a$ a $\mathfrak u(N_f)$ generator.\footnote{\label{foot:u(Nf)gens}We take the generators $T_a\in \mathfrak u(N_f)$ in the defining representation as the usual $N_f^2-1$ generators of $\mathfrak{su}(N_f)$ together with the normalized identity $T_0\equiv \frac{1}{\sqrt{2N_f}}\mathbb 1$ so that they are all normalized by
\begin{equation}
    \tr{T_aT_b}=\frac{1}{2}\delta_{ab}\,.
\end{equation}
They then satisfy
\begin{equation}
    T_a T_b = \frac{1}{2}\sum_{c=0}^{N_f^2-1}(if_{abc} + d_{abc})T_c\,, \qquad \text{where} \quad 
    \begin{cases}
        d_{abc} \equiv 2\tr{T_a\{T_b,T_c\}}\,, \\
    f_{abc} \equiv \frac{2}{i}\tr{T_a[T_b,T_c]}\,.
    \end{cases}
\end{equation}}
The chiral Lagrangian comprises all  terms compatible with the symmetries of the theory, arranged in a derivative expansion.  Importantly, at large $N$ only single-trace terms survive. They correspond to a single quark loop, resumming Feynman diagrams with a disk topology.\footnote{In the language of string theory, where the open string coupling 
$g_o \sim 1/\sqrt{N}$, the leading contribution corresponds to tree-level open string scattering amplitudes, computed by a disk worldsheet with  string vertex operators inserted on its boundary.}
Each additional quark loop (additional trace) would be suppressed by a power of $1/N$.

At finite $N$, the $U(1)_A$ symmetry is broken by the ABJ anomaly \cite{tHooft:1976rip}, and the spontaneous symmetry breaking pattern becomes $SU(N_f)_L\times SU(N_f)_R\times U(1)_V \to SU(N_f)_V\times U(1)_V$. In this case, there are only $N_f^2-1$ Goldstone bosons spanning the coset manifold $SU(N_f)$. When $N\to \infty$, an additional meson ---the $\eta'$, associated to the generator $T_0\sim \mathbb{1}_{N_f}$--- becomes massless, completing the $U(N_f)$ sigma model of \eqref{eq:Lch} \cite{Witten:1979vv}. Thus, in the strict large $N$ limit the $\eta'$ is on equal footing to the pions, and they can be treated democratically. One may add corrections to \eqref{eq:Lch} that account for the explicit breaking of the $U(N_f)_L\times U(N_f)_R$ symmetry. These come in two types: first, \textit{quark mass} corrections giving a mass to the pions, and second, $1/N$ corrections disentangling the $\eta'$ from the pions, accounting for the breaking of the $U(1)_A$ by the anomaly. While these corrections are interesting in their own right (e.g.\ they encode a rich interplay between the $\eta'$, the quark masses and the theta angle \cite{Witten:1980sp,DiVecchia:1980yfw}), here we will not be concerned by them. We will work only at leading order in $1/N$ and in the strict chiral limit.

The Wilson coefficients $f_\pi^2, \kappa_1, \kappa_2, \ldots$ in \eqref{eq:Lch} are not fixed by symmetry and depend on the microscopic details of the UV theory.
Computing them from the QCD Lagrangian is of course a very hard 
problem 
---though some progress has been made on the lattice~\cite{Lucini:2012gg,DeGrand:2016pur,Hernandez:2019qed,Perez:2020vbn,Baeza-Ballesteros:2022azb}.
We will proceed instead by constraining their allowed values by requiring self-consistency of certain scattering amplitudes,
in the spirit of the S-matrix bootstrap.
This general approach dates back to the 60s, and some  constraints have long been known, see e.g.~\cite{Martin1969,Pham:1985cr,Ananthanarayan:1994hf,Pennington:1994kc,Comellas:1995hq,Dita:1998mh}.
A truly systematic 
analysis was initiated only in our first paper~\cite{Albert:2022oes}, leveraging new
ideas and technical tools developed in recent years.

The chiral Lagrangian \eqref{eq:Lch} describes the low-energy scattering of Goldstone bosons, i.e.\ matrix elements of asymptotic pion states. It admits a  natural extension that allows to compute matrix elements of the Noether currents $J_L^\mu, J_R^\mu$ associated to the symmetry $U(N_f)_L\times U(N_f)_R$, which acts as 
\begin{equation}
    U(x)\mapsto L^\dagger U(x) R \,, \qquad L\in U(N_f)_L,\, R\in U(N_f)_R\,.
\end{equation}
The idea is to introduce \textit{background gauge fields} $A^L_\mu, A^R_\mu$ in \eqref{eq:Lch} coupling to the conserved currents, so that their matrix elements can be obtained via functional derivatives $\delta/\delta A^{L,R}_\mu$. We introduce the covariant derivative
\begin{equation}
    D_\mu U \equiv \partial_\mu U + iA^L_\mu U - iUA^R_\mu\,,
\end{equation}
and we simply promote all the derivatives in \eqref{eq:Lch} to covariant derivatives. In addition, we must introduce new terms proportional to the field-strength tensors,
\begin{align}
    L_{\mu\nu} &\,\equiv \partial_\mu A_\nu^L - \partial_\nu A_\mu^L + i[A_\mu^L,A_\nu^L]\,,\\
    R_{\mu\nu} &\,\equiv \partial_\mu A_\nu^R - \partial_\nu A_\mu^R + i[A_\mu^R,A_\nu^R]\,,\nonumber
\end{align}
with new unfixed coefficients.
At lowest derivative order  we thus have (see e.g.\ \cite{Donoghue:1992dd})
\begin{align}
    \label{eq:lrgaugeLch}
	S_\text{Ch}[A_L,A_R]= \int d^4 x \bigg[&-\frac{f_\pi^2}{4}\tr{D_\mu U D^\mu U^\dagger}
	+\kappa_1 \tr{D_\mu U D^\mu U^\dagger D_\nu U D^\nu U^\dagger}\nonumber\\
	&+\kappa_2 \tr{D_\mu U D_\nu U^\dagger D^\mu U D^\nu U^\dagger} \nonumber\\
    &+i\kappa_3 \tr{L_{\mu\nu}D^\mu U D^\nu U^\dagger + R_{\mu\nu}D^\mu U^\dagger D^\nu U} \nonumber \\
    &+\kappa_4 \tr{L_{\mu\nu}UR^{\mu\nu}U^\dagger} + \kappa_5 \tr{L_{\mu\nu}L^{\mu\nu} + R_{\mu\nu}R^{\mu\nu}}+\cdots\bigg]\,.
\end{align}
While the terms containing field strengths decouple from pure pion scattering, they contribute to  correlation functions with currents 
and encode information of the full underlying theory, albeit in a very intricate way. In fact, we will see that these couplings capture information from sectors of the spectrum that completely decouple from pion scattering.

Alternatively to the left and right currents, we may introduce the axial and vector currents
\begin{equation}
    J_A^\mu = J_L^\mu - J_R^\mu \,, \qquad J_V^\mu = J_L^\mu + J_R^\mu\,,
\end{equation}
and their corresponding gauge fields
\begin{equation}
    A^A_\mu = \frac{1}{2}(A^L_\mu - A^R_\mu)\,, \qquad A^V_\mu = \frac{1}{2}(A^L_\mu + A^R_\mu)\,.
\end{equation}
A particularly interesting choice of background gauge fields is $A^L_\mu = A^R_\mu = eQ A_\mu$, with $Q$ a \textit{charge matrix} valued in the Cartan subalgebra of $\mathfrak u(N_f)$. This is associated with the ``electromagnetic current'' $J_{\text{em}}^\mu$ for a $U(1)_Q$ subgroup of the vector symmetry $U(N_f)_V$. In real-world QCD with $N_f=3$, one takes
\begin{equation}\label{eq:charneM}
    Q=\begin{pmatrix}
        2/3 & & \\
        & -1/3 & \\
        & & -1/3
    \end{pmatrix}\,,
\end{equation}
assigning the electromagnetic charges of the $u$, $d$ and $s$ quarks. Then, correlation functions involving $J_{\text{em}}^\mu$ correspond to scattering processes with off-shell probe photons. More generally, we will take $Q$ to be any diagonal $N_f\times N_f$ matrix, and by a slight abuse of notation, we will continue to call ``photon'' the $U(1)_Q$ gauge boson $A_\mu$.

With this choice of background gauge fields, \eqref{eq:lrgaugeLch} becomes
\begin{align}
    \label{eq:gaugeLch}
	S_\text{Ch}[A]= \int d^4 x \bigg[&-\frac{f_\pi^2}{4}\tr{D_\mu U D^\mu U^\dagger}
	+\kappa_1 \tr{D_\mu U D^\mu U^\dagger D_\nu U D^\nu U^\dagger}\nonumber\\
	&+\kappa_2 \tr{D_\mu U D_\nu U^\dagger D^\mu U D^\nu U^\dagger} \nonumber\\
    &+ie\kappa_3 F_{\mu\nu}\tr{Q\left[D^\mu U, D^\nu U^\dagger\right]} \nonumber \\
    &+e^2 \kappa_4 F_{\mu\nu}F^{\mu\nu} \tr{QUQU^\dagger} + e^2\kappa_5' F_{\mu\nu}F^{\mu\nu}+\cdots\bigg]\,,
\end{align}
where now the covariant derivative is given by
\begin{equation}
    D_\mu U\equiv \partial_\mu U + ie A_\mu \left[Q,U\right]\,.
\end{equation}
To make physical sense of \eqref{eq:gaugeLch}, we can  ``weakly gauge'' the $U(1)_Q$. That is, we make $A_\mu$ dynamical but keep the gauge coupling $e$ parametrically small. From this point of view, \eqref{eq:gaugeLch} is simply an effective action describing the low-energy interactions of pions and $U(1)_Q$ photons. Importantly, though, we are instructed to work to lowest order in $e$, such that the photons only appear as external states in any scattering process. Moreover, since they are avatars of the conserved current $J_{\text{em}}^\mu$, these photons can be taken off-shell.

\subsection{The Wess-Zumino-Witten term}\label{sec:WZWterm}
It is well known that \eqref{eq:gaugeLch} does not describe all the processes observed in nature involving pions and photons. Apart from the continuous $U(N_f)_L\times U(N_f)_R$ chiral symmetry, \eqref{eq:gaugeLch} enjoys some discrete symmetries, which should reflect those of QCD. First, it is invariant under charge conjugation symmetry
\begin{equation}
    C:\qquad U\mapsto U^T,\; A_\mu \mapsto -A_\mu\,.
\end{equation}
Note that this exchanges $\pi^+$ and $\pi^-$, as one would expect. Second, it is separately invariant under
\begin{align}
    \text{Naive parity } P_0:& \qquad t\leftrightarrow t, \vec{x}\leftrightarrow -\vec{x}\,,\\
    \text{Pion number } (-1)^{N_\pi}:& \qquad U\leftrightarrow U^\dagger\,.
\end{align}
But QCD is only invariant under the true parity operator $P=(-1)^{N_\pi}P_0$, which takes into account that pions are pseudo-scalars. Indeed, there are low-energy processes observed in nature, such as $K^+ + K^- \to \pi^+ + \pi^- + \pi^0$, which break both $(-1)^{N_\pi}$ and $P_0$ but preserve $P$. These processes are not captured by \eqref{eq:gaugeLch}.

Relaxing the symmetry assumption to $P$ alone rather than $P_0$ and $(-1)^{N_\pi}$ separately is what prompted Witten to introduce a new term in the chiral Lagrangian; the famous Wess-Zumino-Witten (WZW) term \cite{Wess:1971yu,Witten:1983tw},\footnote{The normalization is chosen with respect to the generator of $\pi_5(U(N_f))\simeq \mathbb Z$ (for $N_f\geq 3$). Note that it is the same generator as for $\pi_5(SU(N_f))\simeq \mathbb Z$ because $U(N_f)$ as a manifold is diffeomorphic to $S^1\times SU(N_f)$ and $\pi_5(S^1)=0$. See \cite{Lee:2020ojw} for a discussion of homology vs homotopy for the normalization of this term, as well as for a modern perspective on more general WZW terms.}
\begin{equation}\label{eq:WZW}
    S_{\text{WZW}} = -\frac{ik}{240\pi^2} \int_{D_5} d^5y\, \varepsilon^{\mu\nu\rho\sigma\tau}\tr{U^\dagger \frac{\partial U}{\partial y^\mu} U^\dagger \frac{\partial U}{\partial y^\nu} U^\dagger \frac{\partial U}{\partial y^\rho} U^\dagger \frac{\partial U}{\partial y^\sigma} U^\dagger \frac{\partial U}{\partial y^\tau}}\,.
\end{equation}
This term is written as an integral over a 5-dimensional manifold $D_5$ whose boundary is our four-dimensional space-time $\partial D_5=M_4$, but it is a topological term that truly depends only on the boundary data. The coefficient $k$ is a priori unfixed like the couplings $\kappa_i$ of \eqref{eq:gaugeLch}, but it is required to be quantized $k\in \bZ$ so that \eqref{eq:WZW} is independent of the extension manifold $D_5$. We will shortly review how this coupling can actually be fixed by matching the anomalies in the UV \cite{Witten:1983tw}.

Before getting to that, though, let us point out that this term plays no role in $2\to 2$ pion scattering. Expanding it out in pion fields and integrating by parts yields
\begin{equation}
    S_{\text{WZW}} = \frac{2k}{15\pi^2f_\pi^5} \tr{T_a T_b T_c T_d T_e}
    \int_{M_4} d^4x\, \varepsilon^{\mu\nu\rho\sigma}\pi^a\partial_\mu\pi^b\partial_\nu\pi^c\partial_\rho\pi^d\partial_\sigma\pi^e + O(\pi^7)\,,
\end{equation}
which starts off with a five-point pion vertex that indeed describes the process $K^+ + K^- \to \pi^+ + \pi^- + \pi^0$. For this reason, the coefficient of the WZW term cannot be directly accessed with the bootstrap methods currently available. One would need to develop the S-matrix bootstrap for five-point scattering amplitudes. We will now see that this coefficient can actually be accessed indirectly with $2\to 2$ scattering processes involving currents.

Coupling the WZW term to background gauge fields is more involved than for \eqref{eq:gaugeLch}. The general result can be found in \cite{Witten:1983tw}, here we just quote the result for the $U(1)_Q$ background discussed above,\footnote{For $N_f=2$, there is no five-dimensional WZW term because $U(2)$ is a four-dimensional Lie group and so one cannot construct the suitable five-form. (There is still a notion of a $\mathbb Z_2$ WZW term responsible for the statistics of Skyrmions \cite{Witten:1983tx,Lee:2020ojw}, but that will not be relevant for our purposes.) Nevertheless, the parity-odd couplings to the gauge fields do exist, and like for general $N_f$, they are responsible for matching the anomalies \cite{Kaiser:2000ck}, as we review below.}
\begin{align}\label{eq:gaugeWZW}
    S_{\text{WZW}}[A] = \,& S_\text{WZW}
	+ \frac{e\,k}{48\pi^2}\int d^4 x\,  A_\mu\,\varepsilon^{\mu\nu\rho\sigma} \tr{\{Q,U^\dagger\}\partial_\nu U \partial_\rho U^\dagger\partial_\sigma U}\\
	& + \frac{ie^2k}{48\pi^2}\int d^4 x\, \varepsilon^{\mu\nu\rho\sigma} F_{\mu\nu}A_\rho
	\tr{\{Q^2,U^\dagger\}\partial_\sigma U
	-\tfrac{1}{2}( QUQ\partial_\sigma U^\dagger - QU^\dagger Q\partial_\sigma U) }\,.\nonumber
\end{align}
The term in the first line is roughly a term $\sim A_\mu J^\mu$ coupling the gauge field to a current and the remaining terms are required by gauge invariance. Expanding in pion fields, we find
\begin{align}\label{eq:gaugeWZWpion}
    S_{\text{WZW}}[A] =S_\text{WZW} + &\, \int d^4 x \Bigg(\frac{e^2 k}{16\pi^2 f_\pi} \tr{T_a Q^2} \varepsilon^{\mu\nu\rho\sigma}F_{\mu\nu}F_{\rho\sigma} \pi^a  \\
    & +i\frac{e\,k}{3\pi^2 f_\pi^3} \tr{Q T_a T_b T_c} \varepsilon^{\mu\nu\rho\sigma}A_\mu \partial_\nu \pi^a \partial_\rho \pi^b \partial_\sigma \pi^c  + \cdots\Bigg)\,. \nonumber
\end{align}
The first of these terms describes the decay  $\pi\to \gamma\gamma$, whereas the second one describes a $3\pi\gamma$ contact interaction. Both of these processes are observed in nature. Since these interactions contribute to processes like $\gamma + \gamma \to \gamma +\gamma$ and $\pi + \pi \to \pi + \gamma $, their coefficients (which match that of the WZW term) will be accessible in a mixed system of $2\to 2$ scattering of pions and photons, which is tractable with standard bootstrap techniques.

\subsection{Anomalies of QCD and their matching}\label{sec:anomalyMatching}
An important tool when looking for the low-energy description of a known UV theory are 't Hooft anomalies, for they are robust under RG flow. Matching these anomalies thus fixes some of the coefficients of the EFT that were a priori unknown. In the case of QCD, the anomalies are matched by the WZW term, as first noticed by Witten \cite{Witten:1983tw}, and as we now briefly review. For this reason, the coefficient of the WZW term is singularly interesting among the low-energy couplings.

The perturbative anomalies of four-dimensional QCD (in flat space) are captured by the anomaly polynomial\footnote{See e.g.\ \cite{Bilal:2008qx,Bertlmann} for an introduction to the anomaly polynomial and descent equations.}
\begin{equation}\label{eq:Apoly}
    I^{\text{QCD}}_6 = \left[\text{tr}\, e^{\frac{1}{2\pi}\left(g_{\text{YM}}\, \mathsf G + \mathsf F_L + \mathsf F_R\right)}\right]_{\text{6-form part}} = \frac{1}{3!}\text{tr}\left[\frac{1}{\left(2\pi\right)^3}\left(g_{\text{YM}}\,\mathsf G + \mathsf F_L + \mathsf F_R\right)^3\right]\,,
\end{equation}
where $\mathsf G, \mathsf F_L, \mathsf F_R$ are the respective field strengths of the (continuous) symmetry groups of QCD; $SU(N)\times U(N_f)_L\times U(N_f)_R$. Here we are using differential form notation for the antihermitean fields $\mathsf A, \mathsf F$,
\begin{equation}
    \mathsf A \equiv -i A_\mu^a T_a\, dx^\mu\,, \qquad 
    \mathsf F \equiv \text d \mathsf A + \mathsf A^2 = -\frac{i}{2} F_{\mu\nu}^a T_a\, dx^\mu \wedge dx^\nu\,,
\end{equation}
and the trace is taken both over the color and flavor representations of the fermions of QCD. Before we proceed, it is important to emphasize the distinction between the color and flavor parts of the symmetry group. On the one hand, the $SU(N)$ piece is a \textit{dynamical gauge group}, and we are therefore instructed to sum over its gauge configurations in the path integral. On the other hand, $U(N_f)_L\times U(N_f)_R$ is a global symmetry and the associated gauge fields $A_L,A_R$ are left as \textit{background gauge fields}.

In the chiral limit QCD has $N_f$ left-handed 
massless Weyl fermions $\psi_L$ 
and $N_f$ right-handed massless Weyl fermions 
$\overline \psi_R$ transforming in the following representations:
\begin{center}
\begin{tabular}{c|c|c|c}
& $SU(N)$ & $U(N_f)_L$ & $U(N_f)_R$\\
\hline
$\psi_L$ & $\Box$ & $\Box$ & \textbf{1}\\
\hline
$\psi_R$ & $\overline\Box$ & \textbf{1} & $\overline \Box$
\end{tabular}
\end{center}
Performing the sum in \eqref{eq:Apoly}, we find only the following non-vanishing terms,
\begin{equation}\label{eq:anomalies}
    I^{\text{QCD}}_6 = \frac{1 }{(2\pi)^3}\frac{g_{\text{YM}}^2}{2}\text{tr}_{f} (\mathsf F_L- \mathsf F_R)\ \text{tr}_{c} (\mathsf G^2) + \frac{1}{(2\pi)^3}\frac{N}{3!}\text{tr}_{f} (\mathsf F_L^3 - \mathsf F_R^3)\,,
\end{equation}
where we used the fact that $\text{tr}_{\overline \Box} (M) = - \text{tr}_{\Box}(M)$ for an antihermitean matrix $M$ to bring all the traces to the fundamental rep of color $SU(N)$ and flavor $U(N_f)$, which we denote by $\text{tr}_c, \text{tr}_f$ respectively.

The first term of \eqref{eq:anomalies}, which involves the dynamical gauge fields $G_{\mu\nu}$, corresponds to the ABJ triangle anomaly of the axial current with two gluons. At finite $N$, this anomaly breaks the $U(1)_A$ symmetry and it gives a mass to the $\eta'$ meson. However, it is proportional to $\sim g_{\text{YM}}^2 \equiv \frac{\lambda}{N}$, and so it vanishes at large $N$ \cite{Witten:1979vv}. The other term in \eqref{eq:anomalies}, in contrast, involves only background gauge fields. It thus captures 't Hooft anomalies of QCD, and it survives in the large $N$ limit.

Using the descent equations
\begin{equation}
    \text d I_5 = I^{\text{QCD}}_6 \,, \qquad \text d I_4^{(\lambda_L,\lambda_R)} = \delta_{(\lambda_L,\lambda_R)} I_5\,,
\end{equation}
where $\delta_{(\lambda_L,\lambda_R)}$ indicates a gauge transformation with parameters $\lambda_L, \lambda_R$,\footnote{Explicitly, the gauge fields transform by $\delta \mathsf A_L = \text d \lambda_L + \left[\mathsf A_L, \lambda_L\right]$ and $\delta \mathsf A_R = \text d \lambda_R + \left[\mathsf A_R, \lambda_R\right]$, where $\lambda_{L,R}(x)$ are antihermitean. In components, the gauge variations read $\delta {A_L^a}_\mu(x) = \left(D_\mu \,i\lambda_{L}(x)\right)^a$, $\delta {A_R^a}_\mu(x) = \left(D_\mu \,i\lambda_{R}(x)\right)^a$.} we determine
\begin{equation}\label{eq:AnomPhase}
    I_4^{(\lambda_L,\lambda_R)} = \frac{1}{(2\pi)^3}\frac{N}{3!}\Big(\text{tr}_{f} \left[\lambda_L \,\text d(\mathsf A_L \text d\mathsf A_L + \tfrac{1}{2} \mathsf A_L^3)\right] - \text{tr}_{f} \left[\lambda_R \,\text d(\mathsf A_R \text d\mathsf A_R + \tfrac{1}{2} \mathsf A_R^3)\right]\Big)\,.
\end{equation}
We conclude that under the gauge transformations parameterized by $\lambda_L,\lambda_R$ the partition function of the theory coupled to the background gauge fields $A_L,A_R$ picks up a phase
\begin{equation}
    Z_{\text{QCD}}\left[A_L',A_R'\right] = e^{-2\pi \int_{M_4} I_4^{(\lambda_L,\lambda_R)}} Z_{\text{QCD}}\left[A_L,A_R\right]\,.
\end{equation}
This is the hallmark of a 't Hooft anomaly. Now, recalling that the background gauge fields couple to currents by a term $\sim A_\mu^a J^{\mu}_a$ we will relate this anomalous phase to the failure of the currents to be conserved.

Defining the effective action $Z_{\text{QCD}}[A_L,A_R] \equiv e^{i W[A_L,A_R]}$, we may obtain the expectation value of the left and right currents by the variations
\begin{equation}\label{eq:ConsistentJ}
    \frac{\delta W}{\delta {A_L^a}_\mu (x)} = \left< {J_L^\mu}_a(x)\right>\,, \qquad \frac{\delta W}{\delta {A_R^a}_\mu (x)} = \left< {J_R^\mu}_a(x)\right>\,.
\end{equation}
The gauge variation of the effective action can then be related to the covariant derivative of the currents by
\begin{equation}
    \delta_{\lambda_L} W = \int d^4 x\,  \delta {A_L}_\mu^a(x) \frac{\delta W}{\delta {A_L^a}_\mu (x)} = -\int d^4 x\, i\lambda_L^a(x) \left(D_\mu \left< {J_L^\mu}(x)\right>\right)_a\,,
\end{equation}
and similarly for $\delta_{\lambda_R}$. Comparing to the anomalous phase \eqref{eq:AnomPhase}, i.e.\
\begin{equation}
    \delta_{\lambda_L} W = \frac{i}{(2\pi)^2}\frac{N}{3!}\int_{M_4}\text{tr}_{f} \left[\lambda_L \,\text d(\mathsf A_L \text d\mathsf A_L + \tfrac{1}{2} \mathsf A_L^3)\right]\,,
\end{equation}
we conclude that the anomaly breaks the covariant conservation of the currents by
\begin{align}
    \left(D_\mu \left< J_L^\mu(x) \right>\right)_a =&\, \frac{N}{24\pi^2} \varepsilon^{\mu\nu\rho\sigma} \,\text{Tr} \big[T_a\, \partial_\mu \left({A_L}_\nu \partial_\rho {A_L}_\sigma - \tfrac{i}{2}{A_L}_\nu {A_L}_\rho {A_L}_\sigma\right)\big]\,, \\
    \left(D_\mu \left< J_R^\mu(x) \right>\right)_a =&\, \frac{-N}{24\pi^2} \varepsilon^{\mu\nu\rho\sigma} \,\text{Tr} \big[T_a\, \partial_\mu \left({A_R}_\nu \partial_\rho {A_R}_\sigma - \tfrac{i}{2}{A_R}_\nu {A_R}_\rho {A_R}_\sigma\right)\big]\,.
\end{align}
In the literature these are known as \textit{consistent anomalies}, referring to the fact that they satisfy the Wess-Zumino consistency conditions \cite{Bertlmann}. However, the currents $J_L,J_R$ defined by \eqref{eq:ConsistentJ} do not transform covariantly. They pick up an anomalous term under gauge transformations. They can be made covariant by including an improvement term depending only on the background gauge fields \cite{Bardeen:1984pm},
\begin{equation}
    {J_L}_a^\mu \to {J_L}_a^\mu + P_a^\mu\left[A_L\right]\,, \qquad {J_R}_a^\mu \to {J_R}_a^\mu - P_a^\mu\left[A_R\right]\,,
\end{equation}
where
\begin{equation}
    P_a^\mu \left[A\right] \equiv \frac{N}{48\pi^2}\varepsilon^{\mu\nu\rho\sigma} \, \text{Tr} \big[T_a\, \left({A}_\nu F_{\rho\sigma} + F_{\rho\sigma}A_\nu + \tfrac{i}{2}{A}_\nu {A}_\rho {A}_\sigma\right)\big]\,.
\end{equation}
Taking the covariant derivative of the improved currents then yields the so-called \textit{covariant anomalies}. Restricting to the $U(1)_Q$ ``photon'' background $A^L_\mu = A^R_\mu = eQ A_\mu$, we find for the (covariant) vector and axial currents
\begin{subequations}
\begin{align}
    \left(D_\mu \left< J_V^\mu(x) \right>\right)_a =&\, 0\,,\\
    \left(D_\mu \left< J_A^\mu(x) \right>\right)_a =&\, \frac{e^2 N}{16\pi^2} \tr{T_a Q^2} \varepsilon^{\mu\nu\rho\sigma} F_{\mu\nu}F_{\rho\sigma}\,, \label{eq:UVanomaly}
\end{align}
\end{subequations}
and thus we recover the familiar \textit{chiral anomaly}. This matches the diagrammatic computation of the anomaly from an $AVV$ triangle diagram \cite{Bardeen:1969md}. 

Being a 't Hooft anomaly, \eqref{eq:UVanomaly} must be matched in the IR. To perform the matching we first note that under RG, the axial current in the UV is simply mapped to the one in the IR; which we may compute by taking variations of the chiral Lagrangian \eqref{eq:lrgaugeLch} with respect to $A_L, A_R$,\footnote{Note that in the presence of a background $A_\mu$ the current must involve the covariant derivative so that it transforms covariantly under gauge transformations.}
\begin{equation}
    (J_A^a)_\mu = \left(J_L^a - J_R^a\right)_\mu
    = -i\, \frac{f_\pi^2}{2}\tr{T^a (U D_\mu U^\dagger + D_\mu U^\dagger U)} = -f_\pi (D_\mu \pi)^a + O(\pi^3)\,.
\end{equation}
We see in this way that the axial current---which is a current for a spontaneously broken symmetry---generates the corresponding Goldstone boson $\pi^a$. The anomalous conservation equation \eqref{eq:UVanomaly} then implies a non-trivial overlap between a pion state and two photons, which would equivalently arise from a Lagrangian with an interaction term
\begin{equation}\label{eq:Lanom}
    \mathcal L = -\frac{1}{2}(D_\mu \pi)^a (D^\mu \pi)_a 
    + \frac{e^2N}{16\pi^2f_\pi}\tr{T_aQ^2} \pi^a\varepsilon^{\mu\nu\rho\sigma}F_{\mu\nu}F_{\rho\sigma} + \cdots\,.
\end{equation}
Indeed, one can check that the equations of motion for such a theory reproduce \eqref{eq:UVanomaly}.

Now we recall that such a term appeared in the expansion in pion fields of the WZW term coupled to gauge fields \eqref{eq:gaugeWZWpion}. Comparing it with \eqref{eq:Lanom} we conclude that to match the anomaly, the coupling of the WZW term must be
\begin{equation}
    k=N\,.
\end{equation}
In a more involved analysis one can show that with this choice of coefficient, the WZW term matches all the remaining perturbative 't Hooft anomalies of QCD \cite{Witten:1983tw}. This is a powerful constraint. It fixes one  of the directions in the space of EFT couplings, which were all a priori unknown.

\section{Covariant kinematics and reduced amplitudes}\label{sec:Parametrizations}
The Wilson coefficients of the chiral Lagrangian coupled to background gauge fields are captured by scattering amplitudes of pions and probe photons. In order to enforce unitarity, we 
must consider the \textit{full} mixed system of $2\to 2$ scattering amplitudes involving pions and photons. There are four independent processes: $\pi\pi\to\pi\pi$, $\gamma\gamma\to\gamma\gamma$, $\pi\pi\to\gamma\gamma$ and $\pi\pi\to\pi\gamma$, and crossed versions thereof. The process $\pi\gamma \to \gamma\gamma$ is disallowed by charge conjugation symmetry, as we will see below. In this section we discuss the parametrization of the amplitudes for each of these processes in the large $N$ limit. There are two aspects to take care of: the flavor dependence of the external pions $\pi^a$ ($a=1,...,N_f^2$) and the dependence on the polarizations of the photons $\gamma^\lambda$ ($\lambda=\pm$). For the former, we note that in the large $N$ limit the only diagrams that contribute to these amplitudes have the topology of a disk, with a single quark loop running through the external legs. We can therefore write these amplitudes in terms of basic ``disk'' amplitudes, much like in tree-level open-string scattering, and introduce the flavor dependence via Chan-Paton factors.

For the latter, i.e.\ the dependence on the helicity of the external photons, there are two possible approaches. One is to pick a frame and work out which external polarizations for the amplitudes are independent, exploiting the symmetries of the problem. This is the approach followed by \cite{Hebbar:2020ukp,Henriksson:2021ymi,Henriksson:2022oeu,Haring:2022sdp}, among others. The other approach is to work covariantly keeping contractions of the polarization vectors $\epsilon_i^{\lambda_i}$ and the momenta $p_i$ unevaluated, as advocated in \cite{Chowdhury:2019kaq,Chowdhury:2020ddc}.\footnote{In $D=4$ there is yet another approach which is to use the spinor-helicity formalism, see e.g.\ \cite{Caron-Huot:2022ugt}.} The strategy consists in classifying all the possible independent tensor structures built out from the $\epsilon_i^{\lambda_i}$ and $p_i$ consistent with the symmetries, and then expanding the amplitudes in this basis. Among the advantages of this approach are that it is fully covariant, it makes crossing symmetry very apparent, and it is easy to generalize to higher dimensions (as exploited in \cite{Caron-Huot:GravPW}). Although we will stick to $D=4$ dimensions, this is the approach that will be more convenient for our purposes.

\subsection{Four pions}\label{sec:param4Pi}
Four-pion scattering at large $N$ was already discussed in \cite{Albert:2022oes}. The parametrization that will prove useful for us is
\begin{align} \label{eq:T4pi}
{\mathcal T}_{abcd}  = \,& 4\left[\tr{T_aT_bT_cT_d}+\tr{T_aT_dT_cT_b}\right] M_{4\pi}(s, t)\nonumber \\ 
+\,& 4\left[\tr{T_aT_cT_dT_b}+\tr{T_aT_bT_dT_c}\right] M_{4\pi}(s, u)\nonumber \\
+\,& 4\left[\tr{T_aT_dT_bT_c}+\tr{T_aT_cT_bT_d}\right] M_{4\pi}(t, u)\,,
\end{align}
where $M_{4\pi}(s,u)$ is the basic ``disk'' amplitude and $T_a$ are the $\mathfrak u(N_f)$ generators associated to the external pions, defined in footnote \ref{foot:u(Nf)gens}. Note that these combinations of traces are compatible with charge conjugation symmetry, which acts as $T_a\mapsto T_a^\intercal$. Here and throughout, our conventions for the Mandelstam invariants are the following:
\be
s = - (p_1 + p_2)^2 \, , \quad t = - (p_2 + p_3)^2\, , \quad u = - (p_1 + p_3)^2 \,,
\ee
and we use \textit{all incoming} conventions, i.e.\ $p_1+p_2+p_3+p_4=0$.
By the large $N$ assumption, $M_{4\pi}(s,u)$ is a meromorphic function in $s$ and $u$ with poles corresponding to physical exchanges of higher mesons. Since it is a flavor-ordered amplitude, $M_{4\pi}(s,u)$ is $s\leftrightarrow u$ symmetric but not fully $s\leftrightarrow t \leftrightarrow u$ symmetric. Moreover, it has no $t$-channel poles \cite{Albert:2022oes}, as a result of the \textit{large-N selection rules}, which forbid the exchange of glueballs (the so-called Zweig's or OZI rule \cite{OKUBO1963165,Zweig:1964jf,Iizuka}) and exotic mesons.

\subsection{Four photons}\label{sec:4photons}
The $2\to 2$ scattering of dynamical photons has been studied from a bootstrap point of view in the recent years both at weak coupling with positivity methods \cite{Henriksson:2021ymi,Henriksson:2022oeu} and non-perturbatively using full unitarity \cite{Haring:2022sdp}.
Here we will treat the photons as  probes. So, effectively, we will be studying the correlation function of four conserved currents $\left< J_{\text{em}}^\mu J_{\text{em}}^\nu J_{\text{em}}^\rho J_{\text{em}}^\sigma\right>$. In this setup we are allowed to take the photons off-shell and study the internal momentum dependence of this correlation function, but for simplicity we will restrict the photons to be on shell. Off-shell photons remain an exciting future direction.

\begin{figure}[htb]
\centering
\includegraphics[scale=0.4]{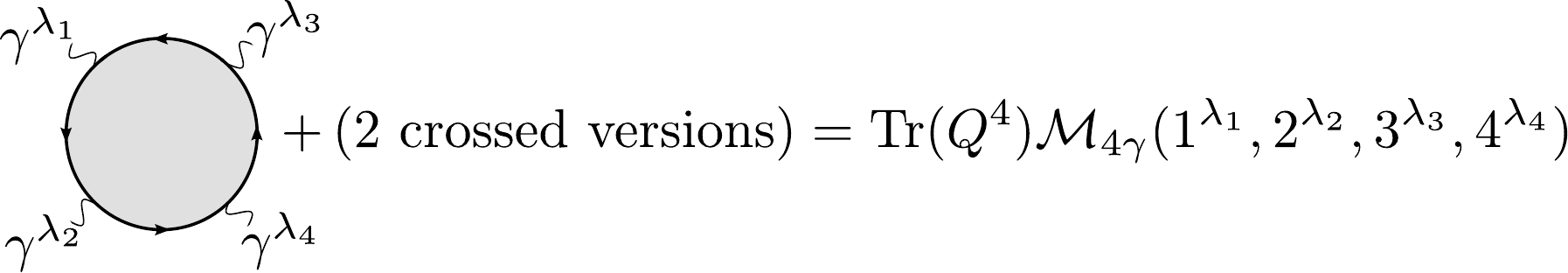}
\caption{Schematic representation of the basic disk amplitude for large $N$ four-photon scattering.}
\label{fig:4g}
\end{figure}

At large $N$, the diagrams contributing to this amplitude have again the topology of a disk with four photons inserted along the boundary (see figure \ref{fig:4g}). While pions are associated to a $U(N_f)$ generator $T_a$, photon insertions come with a factor of the charge matrix $Q$. Since all photons carry the same matrix $Q$, the flavor structure of the four-photon amplitude is just a factor $\sim \tr{Q^4}$ for any ordering of the external legs. The sum of all the crossed versions of the disk then combines into a fully crossing-symmetric amplitude,
\begin{equation} \label{eq:T4gamma}
    {\mathcal T}^{\lambda_1\lambda_2\lambda_3\lambda_4}  = 8\tr{Q^4} \MM_{4\gamma}(1^{\lambda_1},2^{\lambda_2},3^{\lambda_3},4^{\lambda_4})\,,
\end{equation}
invariant under the exchange of any two photons.

Since we eventually want to deal with functions of the Mandelstam invariants alone, we next must take care of the dependence on the polarizations. We do this by identifying all the possible covariant tensors $E_k^{4\gamma}$ involving the polarization vectors $\epsilon_i$ that are compatible with the symmetries of the process and then expanding the amplitude in this basis as \cite{Chowdhury:2019kaq}
\begin{equation}\label{eq:covExpans}
    \MM_{4\gamma}=\sum_k M_{4\gamma}^{(k)}(s,u)E_k^{4\gamma}\,.
\end{equation}
The idea is that the polarization structures $E_k^{4\gamma}$ contain all the dependence on the polarizations so that the coefficients $M_{4\gamma}^{(k)}(s,u)$ are functions only of the Mandelstam invariants. We refer to these functions as \textit{reduced amplitudes} throughout the text and we denote them by the letter $M$, while we reserve $\MM$ for the amplitudes that still depend on the polarizations.

What are the structures $E_k^{4\gamma}$ that we need for four-photon scattering? To answer this we need to examine the symmetries of $\MM_{4\gamma}(1^{\lambda_1},2^{\lambda_2},3^{\lambda_3},4^{\lambda_4})$:
\begin{itemize}
    \item \textbf{Little group:} To give the amplitude $\MM_{4\gamma}$ the correct little group scaling \cite{Elvang:2013cua} (and the correct dependence on the polarizations), the tensors $E_k^{4\gamma}$ must be homogeneous of degree one in each $\epsilon_i$.
    
    \item \textbf{Crossing:} $\MM_{4\gamma}$ is invariant under all permutations of the external photons, forming the group $\text S_4$. This group has a normal subgroup $\bZ_2\times\bZ_2$ generated by the \textit{double exchanges} $(1^{\lambda_1},2^{\lambda_2})\leftrightarrow (3^{\lambda_3},4^{\lambda_4})$ and $(1^{\lambda_1},3^{\lambda_3})\leftrightarrow (2^{\lambda_2},4^{\lambda_4})$. Since these transformations leave the Mandelstam invariants $s,t,u$ (and hence the reduced amplitudes $M_{4\gamma}^{(k)}(s,u)$) unchanged, the basis tensors $E_k^{4\gamma}$ themselves must also be invariant under them. We will impose the remaining symmetry $\text S_4/(\bZ_2\times\bZ_2)\cong \text S_3$, which exchanges $s\leftrightarrow t \leftrightarrow u$, momentarily.
    
    \item \textbf{Gauge invariance:} At the level of the amplitude, gauge invariance is imposed by the Ward identity, which requires that $\MM_{4\gamma}$ vanishes when we replace $\epsilon_i\to p_i$ for any external photon. This property is directly inherited by the basis tensors $E_k^{4\gamma}$.
    
    \item \textbf{Parity:} Since the external states carry no intrinsic parity, the polarization structures $E_k^{4\gamma}$ should be \textit{parity even}, i.e.\ they should not involve the epsilon symbol $\varepsilon^{\mu\nu\rho\sigma}$.
\end{itemize}

\noindent
A basis of covariant tensors satisfing all these properties was identified in \cite{Chowdhury:2019kaq} (see also \cite{Caron-Huot:GravPW}). In general there are seven independent tensors with the desired properties,
\begin{gather}\label{eq:4gPolStr}
    E_{1,s}^{4\gamma} = \frac{1}{4s^2} H_{12}H_{34}\, \quad E_{1,u}^{4\gamma} = \frac{1}{4u^2} H_{13}H_{24}\,, \quad E_{1,t}^{4\gamma} = \frac{1}{4t^2}H_{14}H_{23}\,,\nonumber\\
    E_2^{4\gamma} = \frac{1}{stu}\left(V_1 H_{234} + V_2 H_{341} + V_3 H_{412} + V_4 H_{123}\right)\,,\nonumber\\
    E_{3,s}^{4\gamma} = H_{1324} -\frac{1}{4}\left(H_{12}H_{34} + H_{13}H_{24} +H_{14}H_{23}\right)\,,\nonumber\\
    E_{3,u}^{4\gamma} = H_{1234} -\frac{1}{4}\left(H_{12}H_{34} + H_{13}H_{24} +H_{14}H_{23}\right)\,,\nonumber\\
    E_{3,t}^{4\gamma} = H_{1243} -\frac{1}{4}\left(H_{12}H_{34} + H_{13}H_{24} +H_{14}H_{23}\right)\,,
\end{gather}
where
\begin{alignat}{2}
    H_{12}=F^\mu_{1\nu}F^\nu_{2\mu}\,, \quad
    H_{123}=F^\mu_{1\nu}F^\nu_{2\rho}F^\rho_{3\mu}\,, \quad
    H_{1234}=F^\mu_{1\nu}F^\nu_{2\rho}F^\rho_{3\sigma}F^\sigma_{4\mu}\,, \quad
    V_1 = {p_4}_\mu F^{\mu\nu}_1 {p_2}_\nu\,,
\end{alignat}
and $F^{\mu\nu}_i\equiv p_i^\mu \epsilon_i^\nu - p_i^\nu \epsilon_i^\mu$ ($H$'s and $V$'s with other indices are defined by permutations of the above). However, in $D=4$ not all of these structures are independent; they obey the relations
\begin{equation}\label{eq:4Drels}
    \frac{1}{s^2} E_{3,s}^{4\gamma} = \frac{1}{u^2} E_{3,u}^{4\gamma} = \frac{1}{t^2} E_{3,t}^{4\gamma}\equiv E_3^{4\gamma}\,.
\end{equation}
This can be checked for example by evaluating the $E_k^{4\gamma}$ tensors in the center of mass frame.

Alternatively, one can use the relation with the ``bare module'' of \cite{Chowdhury:2019kaq} to verify \eqref{eq:4Drels}. This is a different basis of polarization structures constructed directly from the physical data encoded in the polarization vectors $\epsilon_i$. We first decompose each $\epsilon_i$ into a part parallel and a part transverse to the $3$-plane spanned by the momenta $p_i$; $\epsilon_i = \epsilon^{\parallel}_i + \epsilon_i^\bot$. The condition $\epsilon_i \cdot p_i = 0$ then removes a dimension from $\epsilon^{\parallel}_i$, and we can use the parametrization
\begin{alignat}{2}
    \epsilon_1^\parallel &= \alpha_1 \sqrt{\frac{su}{t}}\left(\frac{p_2}{s} - \frac{p_3}{u}\right) + a_1 p_1\,, \qquad
    \epsilon_2^\parallel &&= \alpha_2 \sqrt{\frac{su}{t}}\left(\frac{p_1}{s} - \frac{p_4}{u}\right) + a_2 p_2\,, \\
    \epsilon_3^\parallel &= \alpha_3 \sqrt{\frac{su}{t}}\left(\frac{p_4}{s} - \frac{p_1}{u}\right) + a_3 p_3\,, \qquad
    \epsilon_4^\parallel &&= \alpha_4 \sqrt{\frac{su}{t}}\left(\frac{p_3}{s} - \frac{p_2}{u}\right) + a_4 p_4\,. \nonumber
\end{alignat}
Gauge invariance requires that all the $a_i$ drop from any physical polarization structures. The bare module is constructed out of the $\alpha_i$ and $\epsilon_i^\bot$, which in $D=4$ are one-dimensional. These correspond to the two physical degrees of freedom of the photon. See \cite{Chowdhury:2019kaq} for further details on the construction of the bare module.

Either way, the relations \eqref{eq:4Drels} reduce the number of independent structures to five, and so in $D=4$ we can expand the four-photon amplitude in polarization structures as
\begin{align}\label{eq:4g_PolParam}
    \MM_{4\gamma}(1^{\lambda_1},2^{\lambda_2},3^{\lambda_3},4^{\lambda_4}) =&\, M_{4\gamma}^{(1)}(t,u)E_{1,s}^{4\gamma} + M_{4\gamma}^{(1)}(s,t)E_{1,u}^{4\gamma} + M_{4\gamma}^{(1)}(s,u)E_{1,t}^{4\gamma}\nonumber\\
    +&\, M_{4\gamma}^{(2)}(s,u) E_{2}^{4\gamma} + M_{4\gamma}^{(3)}(s,u) E_{3}^{4\gamma}\,.
\end{align}
The remaining $\text S_3$ crossing symmetry forces the coefficients in front of the $E_{1,i}^{4\gamma}$ tensors to be crossed versions of the same function $M_{4\gamma}^{(1)}$. It further implies that $M_{4\gamma}^{(2)}(s,u)$ and $M_{4\gamma}^{(3)}(s,u)$ are fully crossing symmetric while $M_{4\gamma}^{(1)}(s,u)$ is only $s\leftrightarrow u$ symmetric. As in the four-pion case, the large $N$ planarity implies that these reduced amplitudes are meromorphic functions with poles corresponding to physical exchanges of heavy mesons.\footnote{\label{foot:Poly}In general one should beware of spurious poles introduced when ``stripping off'' the polarization tensors. This happens when the $E_k^{4\gamma}$ carry away too many powers of the momenta, and so these spurious poles can be removed by suitable rescalings of the $E_k^{4\gamma}$ by negative powers of the Mandelstam invariants. In practice, the basis can be assessed a posteriori by checking that the partial waves (to be discussed below) are polynomial in $u$.}

\subsection{Two pions and two photons}
\begin{figure}[htb]
\centering
\includegraphics[scale=0.4]{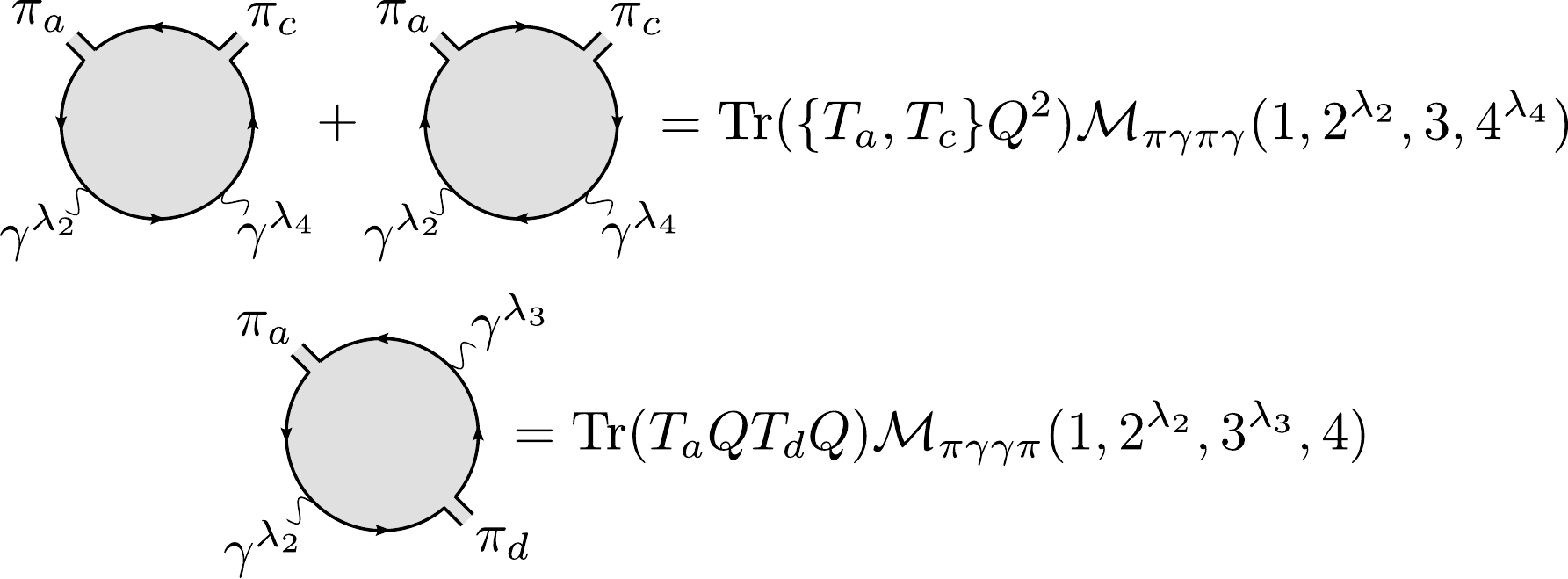}
\caption{Schematic representation of the basic disk amplitudes for the scattering of two pions and two photons. There are two inequivalent orderings of the external legs.}
\label{fig:2p2g}
\end{figure}

Let us now turn to the mixed amplitude involving two pions and two photons. The large $N$ structure of this amplitude is largely the same as before, namely it is made of disk amplitudes proportional to traces of the Chan-Paton factors. But there is one significant difference; there are now \textit{two} different basic amplitudes corresponding to two inequivalent orderings of the external legs, see figure \ref{fig:2p2g}. The full amplitude reads
\begin{align} \label{eq:T2p2g}
    {\cal T}_{ad}^{\lambda_2\lambda_3}  =&\, 2 \tr{\{T_a,T_d\}Q^2} \MM_{\pi\gamma\pi\gamma}(1,2^{\lambda_2},4,3^{\lambda_3}) \nonumber\\
    &+ 4\tr{T_aQT_dQ}\MM_{\pi\gamma\gamma\pi}(1,2^{\lambda_2},3^{\lambda_3},4) \,,
\end{align}
where $\MM_{\pi\gamma\pi\gamma}$ and $\MM_{\pi\gamma\gamma\pi}$ are two independent amplitudes that are separately invariant under the exchange of the two pions or of the two photons. Note that these trace combinations are again invariant under $T_a\to T_a^\intercal$, as required by charge conjugation symmetry.

The next step is to deal with the polarization indices of these amplitudes. In the spirit of \cite{Chowdhury:2019kaq} (and as reviewed above) we do this by classifying all the possible polarization structures and expanding the amplitudes like in \eqref{eq:covExp}. Since both $2\pi2\gamma$ amplitudes share the same symmetries, we only need to do this once. In the current case, we are looking for independent tensors that:
\begin{itemize}
    \item are homogeneous of degree one in $\epsilon_2$ and $\epsilon_3$,
    \item are invariant under the double exchange $(1,2^{\lambda_2})\leftrightarrow (4,3^{\lambda_3})$,
    \item are gauge invariant (i.e. they satisfy the Ward identity),
    \item and are parity even (i.e.\ they do not involve the epsilon symbol).
\end{itemize}
With these conditions, there exist only two independent tensor structures:
\begin{subequations}\label{eq:basistnsrs}
\begin{align}
    E_{1,t}^{2\pi2\gamma}\,&= \epsilon_2\cdot \epsilon_3 - \frac{2}{s}\left(p_1\cdot \epsilon_2\right)\left(p_4\cdot \epsilon_3\right) - \frac{2}{u}\left(p_4\cdot \epsilon_2\right)\left(p_1\cdot \epsilon_3\right)\,,\nonumber\\
    E_{2,t}^{2\pi2\gamma}\,&= \epsilon_2\cdot \epsilon_3 + \frac{2}{t}\left(p_3\cdot \epsilon_2\right)\left(p_2\cdot \epsilon_3\right)\,.
    \end{align}
\end{subequations}
So the two independent disk amplitudes of \eqref{eq:T2p2g} can be decomposed as
\begin{subequations}\label{eq:covExp}
\begin{align}
    \MM_{\pi\gamma\pi\gamma}(1,2^{\lambda_2},4,3^{\lambda_3})&\,= M_{\pi\gamma\pi\gamma}^{(1)}(s,u)E_{1,t}^{2\pi2\gamma} + M_{\pi\gamma\pi\gamma}^{(2)}(s,u)E_{2,t}^{2\pi2\gamma}\,, \\
    \MM_{\pi\gamma\gamma\pi}(1,2^{\lambda_2},3^{\lambda_3},4)&\,= M_{\pi\gamma\gamma\pi}^{(1)}(s,u)E_{1,t}^{2\pi2\gamma} + M_{\pi\gamma\gamma\pi}^{(2)}(s,u)E_{2,t}^{2\pi2\gamma}\,.
\end{align}
\end{subequations}
Since both $E_{1,t}^{2\pi2\gamma}$ and $E_{2,t}^{2\pi2\gamma}$ are $2^{\lambda_2}\leftrightarrow 3^{\lambda_3}$ symmetric, the remaining crossing symmetry demands that all $M_{\pi\gamma...}^{(i)}(s,u)$ be $s\leftrightarrow u$ crossing-symmetric. They are meromorphic functions whose poles correspond once more to the physical mesons in the spectrum and, importantly, they have \textit{no photon pole}. This is because we treat the photon as a probe (or, equivalently, as a very weakly coupled dynamical field) and so it only appears as an external state.

\subsection{Three pions and one photon}
\begin{figure}[htb]
\centering
\includegraphics[scale=0.4]{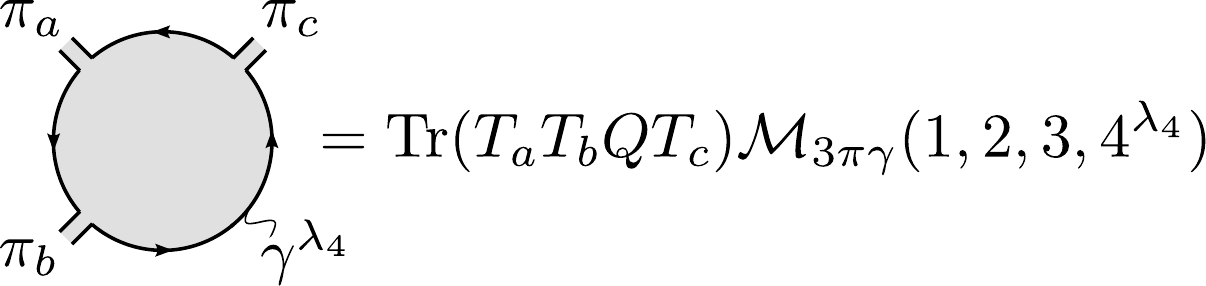}
\caption{Disk amplitude for the scattering of three pions and one photon.}
\label{fig:3p1g}
\end{figure}

Last, but not least, we consider the scattering of two pions into a pion and a photon. This process is important as far as the chiral anomaly is concerned because the Wess-Zumino term induces a $3\pi\gamma$ low-energy coupling proportional to it (recall \eqref{eq:gaugeWZWpion}). At large $N$ the full amplitude can be written in terms of the single disk amplitude of figure \ref{fig:3p1g},
\begin{align} \label{eq:T3pi1g}
    \mathcal T_{abc}^{\lambda_4}  = &\,\left[\tr{T_bT_aQT_c} - \tr{T_bT_cQT_a}\right] \MM_{3\pi\gamma}(2,1,3,4^{\lambda_4}) \nonumber\\
   + &\,\left[\tr{T_aT_bQT_c} - \tr{T_aT_cQT_b}\right] \MM_{3\pi\gamma}(1,2,3,4^{\lambda_4}) \nonumber\\
   + &\,\left[\tr{T_cT_bQT_a} - \tr{T_cT_aQT_b}\right] \MM_{3\pi\gamma}(3,2,1,4^{\lambda_4}) \,.
\end{align}
The minus signs between the traces are required by charge conjugation symmetry, recalling that the photon is odd under $C$. Then Bose symmetry of the exchange of any of the three pions implies that the disk amplitude must be $\textit{crossing anti-symmetric}$, i.e.\
\begin{equation}
    \MM_{3\pi\gamma}(1,2,3,4^{\lambda_4})=- \MM_{3\pi\gamma}(1,3,2,4^{\lambda_4})\,.
\end{equation}
This is a consequence of the external states having overall negative intrinsic charge conjugation.

To deal with the polarization index of $\MM_{3\pi\gamma}(1,2,3,4^{\lambda_4})$, we expand it in covariant tensors involving the polarization vector $\epsilon_4$. The tensors should be
\begin{itemize}
    \item homogeneous of degree one in $\epsilon_4$,
    \item gauge invariant or, equivalently, vanish when we replace $\epsilon_4\to p_4$ (Ward identity),
    \item and parity odd, for the external states have overall negative intrinsic parity.\footnote{Indeed, being a pseudoscalar, the pion transforms as $P:\pi^a \mapsto -\pi^a$, but the photon field remains unchanged, $P: A_\mu \mapsto A_\mu$.}
\end{itemize}
There is only one such tensor;
\begin{equation}
    E_1^{3\pi\gamma}\equiv \varepsilon^{\mu\nu\rho\sigma}{p_1}_\mu
{p_2}_{\nu} {p_3}_\rho {\epsilon_4}_\sigma\,,
\end{equation}
so the disk amplitude can be written as
\begin{equation}
    \MM_{3\pi\gamma}(1,2,3,4^{\lambda_4}) = M_{3\pi\gamma}(s,u)E_1^{3\pi\gamma}\,,
\end{equation}
where $M_{3\pi\gamma}(s,u)$ is crossing symmetric in $s\leftrightarrow u$ but not fully crossing symmetric.

This completes the parametrizations of all the amplitudes involved in the pion-photon mixed scattering system. The remaining process, $\pi\gamma \to \gamma\gamma$, is disallowed by charge conjugation symmetry. Indeed, the only possible trace factor for these external states is $\tr{T_a Q^3}$, which is even under $C:T_a\mapsto T_a^\intercal$, but charge conjugation symmetry requires an odd structure to balance the intrinsic charge conjugation of the external states. Note that this is not intrinsic to large $N$. We reach the same conclusion even when allowing for multi-trace operators like $\tr{T_a Q}\tr{Q^2}$ or $\tr{T_a}\tr{Q^3}$.

\section{Covariant partial waves and large \texorpdfstring{\boldmath $N$}{N} selection rules}
\label{sec:PartialWaves}
Our goal will be to constrain low-energy physics by exploiting unitarity at high energies. In order to express the consequences of unitarity, though, we first need a basis of partial waves on which we can expand the amplitudes described above. This is rather non-trivial because we are dealing with a system of mixed correlators with spin and flavor dependence. Various subsets of these complications have been studied before e.g\ in \cite{Hebbar:2020ukp,Henriksson:2021ymi,Henriksson:2022oeu,Caron-Huot:2022ugt,Haring:2022sdp,Karateev:2022jdb}. The method that we will use to construct the partial waves for this system is based on the formalism of \cite{Caron-Huot:GravPW}, which was originally targeted for spinning states in higher dimensions. The beauty of their approach is that it keeps everything covariant,
making it easy to track the quantum numbers of the intermediate states. We generalize this  method to deal with mixed correlators and flavor dependence.

The idea is to expand every amplitude in a sum over all the possible $s$-channel exchanges. In the current case, this means summing over representations $\mathcal R$ of $U(N_f)$ and representations of the massive little group $SO(3)$, characterized by the spin $J$. That is,
\begin{equation}\label{eq:PWexp}
	\mathcal T = \sum_{\R,J}n_J^{\R}\sum_{ij}\left(a^{\R}_J(s)\right)_{ji}\left(\pi^{\R}_J\right)^{ij}\,,
\end{equation}
where the (generalized) partial waves $\pi^{\R}_J$ take care of the polarization and flavor dependence of $\mathcal T$, and can in general be matrices.
 The partial wave coefficients $a^{\R}_J(s)$ are then also matrices, and unitarity is expressed as a matrix constraint,
\begin{equation}
	\rho_J^{\R}(s)=\text{Im}\,a^{\R}_J(s) \succeq 0\,.
\end{equation}
This states that the spectral density must be a positive-semidefinite matrix. This constraint usually goes under the name of \textit{positivity}, since full unitarity involves additional non-linear constraints (see e.g.\ \cite{Hebbar:2020ukp}). However, at large $N$ the spectral density is suppressed as $\rho_J^{\R}\sim \frac{1}{N}$ and all non-linear constraints trivialize. The constant $n_J^{\R}$ in \eqref{eq:PWexp} simply fixes the normalization of the partial waves.

To construct the partial waves $\pi^{\R}_J$ we recall that the contribution to the amplitude of an exchange in a specific representation factorizes into the contraction of two invariant tensors of that representation. We can therefore construct the partial waves by gluing tensors $v^{i,\alpha,A}$, representing three-point vertices for the initial and final states \cite{Caron-Huot:GravPW}, in the following way
\begin{equation}\label{eq:gluing}
	\left(\pi_J^{\R}\right)^{ij}=\overline{v^{i,\alpha,A}}(3,4)\, g_{\alpha,\beta}g_{A,B}\, v^{j,\beta,B}(1,2)\,.
\end{equation}
The notation here is the following: $\alpha,\beta$ are spin-$J$ little group indices while $A,B$ are indices of the representation $\R$; $g_{\alpha,\beta}$ and $g_{A,B}$ are respectively $SO(3)$ and $U(N_f)$-invariant metrics and the bar means complex conjugation. The remaining labels $i,j$ count the different vertices there can be for a given spin $J$ and representation $\R$. This multiplicity has two sources. First, we can have different external states of the mixed system; $\pi\pi$, $\pi\gamma$ or $\gamma\gamma$. Second, the vertices involving photons have further multiplicities due to their polarizations. This is what promotes the partial waves to matrices.\footnote{As a side comment, we mention that the invariant tensors $v^{i,\alpha,A}$ correspond to intertwiners for the representations of $SO(3)\times U(N_f)$, i.e.\ $2\to 1$ morphisms from the external representations to the internal one which commute with the group action. Similarly, partial waves are a basis on which to decompose $2\to 2$ intertwiners. This suggests that a construction of the $S$-matrix bootstrap in the language of category theory should be possible, in analogy to the one for the conformal bootstrap \cite{AlbertChoi}.}

In the presence of global discrete symmetries (such as parity $P$ and charge conjugation $C$) the spectral density $\rho_J^{\R}$ further decomposes into a direct sum of blocks according to the quantum numbers of the states being exchanged. Positivity is satisfied in each of these blocks independently. In this way, the partial wave decomposition gives us control over what is being exchanged. This is important because at large $N$ only single-meson states should be exchanged, which should in turn be purely $q\bar q$ bound states. This results in some \textit{large $N$ selection rules} that constrain which representations can be exchanged. We will impose these constraints simply by removing the corresponding (blocks in the) spectral densities of \eqref{eq:PWexp}.

Below we explain in detail how to compute the partial waves for the problem at hand. We first classify all the possible three-point vertices $v^{i,\alpha,A}$ and then we glue them to construct the entries of the partial wave matrices. Since the little group and flavor symmetries couple in a trivial way ---a consequence of the Coleman-Mandula theorem--- the three point vertices (and their partial waves) factorize into a flavor and a spin part, which we can study separately. Once the partial waves are constructed, one only has to expand the amplitudes $\mathcal T$ for the different processes using the partial waves and compare to the parametrizations in section \ref{sec:Parametrizations} to extract the corresponding partial waves for the reduced amplitudes $M_i(s,u)$.

\subsection{Flavor} \label{sec:flavor}
\begin{figure}[htb]
\centering
\includegraphics[scale=0.4]{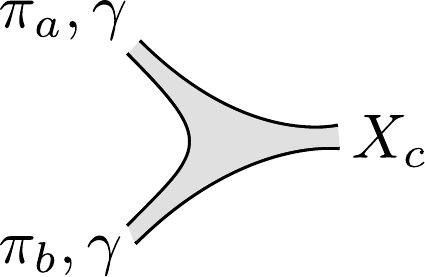}
\caption{At large $N$ also three-point couplings must be planar with a quark loop running along the boundary, making the coupling proportional to a single trace.}
\label{fig:3pt}
\end{figure}

At large $N$ we can only have physical exchanges of single-meson states. Since these are strictly $q_\alpha\bar q^\beta$ pairs, they all carry \textit{adjoint} representations of $U(N_f)$ or, equivalently, they are valued in the $\mathfrak u(N_f)$ Lie algebra.
Therefore, in the sum over $U(N_f)$ representations of \eqref{eq:PWexp}, only $\R=\text{adj}$ will contribute with a non-vanishing spectral density. What are the possible three-point vertices for this representation? We can get them through traces of ``Chan-Paton factors'', as depicted in figure \ref{fig:3pt}. Concretely, we associate a generator $T_a\in \mathfrak u(N_f)$ to every leg corresponding to a meson and the charge matrix $Q$, valued in the Cartan subalgebra, to every photon leg. Then we take the trace. Note that photons can only appear as external legs since we are treating them as probes.

The possible flavor structures for the three-point vertices are
\begin{subequations}\label{eq:flPW}
\begin{itemize}
    \item \textbf{Two pions:}
    \begin{equation}\label{eq:2pifl}
        \pi_a \pi_b\to X_c :\begin{cases}
        d_{abc} = 2\tr{T_a\{T_b,T_c\}} & C=+\,,\\
        f_{abc} = \frac{2}{i}\tr{T_a[T_b,T_c]} & C=-\,,\\
        \end{cases}
    \end{equation}
    
    \item \textbf{Pion-photon:}
    \begin{equation}
        \pi_a \gamma\to X_c :\begin{cases}
        f_{aQc} = \frac{2}{i}\tr{T_a[Q,T_c]} & C=+\,,\\
        d_{aQc} = 2\tr{T_a\{Q,T_c\}} & C=-\,,
        \end{cases}
    \end{equation}
    
    \item \textbf{Two photons:}
    \begin{equation}
        \gamma \gamma\to X_c :\begin{cases}
        d_{QQc} = 2\tr{Q^2T_c} & C=+\,.
        \end{cases}
    \end{equation}
\end{itemize}
\end{subequations}
Here $C$ stands for the charge conjugation eigenvalue of the intermediate meson $X$. To determine it we use that $C:\, T_a\mapsto T_a^\intercal$ and recall that photons carry negative intrinsic charge conjugation. The transformation of the vertex itself times the intrinsic charge conjugation of the three legs should trivialize in a theory with charge conjugation symmetry. The overall normalization of the vertices will not be important.

These vertices will come along with the little group structures that we discuss below, but since they factorize, we can already see the flavor part of the partial waves by gluing these vertices. To do so, we simply consider the flavor vertices relevant for the processes $1,2\to X_c$ and $3,4\to X_{c'}$ and we glue them using the Killing metric $\delta_{c c'}$.\footnote{Note that, since we are using conventions where the generators $T_a$ are hermitian, both $d_{abc}$ and $f_{abc}$ are real and we need not worry about taking the complex conjugate of the $3,4\to X_{c'}$ flavor vertex.} To preserve charge conjugation symmetry, we only allow for gluings between vertices with the same internal $C$. To give an example, the flavor partial waves for the four-pion amplitude are $d\indices{_{ab}^e}d_{cde}$ and $f\indices{_{ab}^e}f_{cde}$.\footnote{\label{foot:completeness}To compare with the parametrizations of section \ref{sec:Parametrizations}, it is useful to note that the generators $T_a$ satisfy the following completeness relation. For any elements $A,B \in \mathfrak u(N_f)$, we have
\begin{equation}
    \sum_{c}\tr{AT_c}\tr{BT_c}=\frac{1}{2}\tr{AB}\,.
\end{equation}
The proof is straightforward. We first expand $A$ as $A=\sum_c a^c T_c$ and similarly for $B$. Then we use the orthogonality relation for the generators,
\begin{equation}
    \sum_{c}\tr{AT_c}\tr{BT_c}
    =\sum_{c}\frac{a^c}{2}\frac{b^c}{2}
    =\frac{1}{2}\tr{AB}\,.
\end{equation}
This completeness relation can be used to derive the following identities
\begin{alignat}{2}
    d\indices{_{ab}^e}d_{cde} =&\, 2\tr{\{T_a,T_b\}\{T_c,T_d\}}\,, \qquad
    &&\, d\indices{_{ab}^e}f_{cde} = \frac{2}{i}\tr{\{T_a,T_b\}[T_c,T_d]}\,,\\
    f\indices{_{ab}^e}d_{cde} =&\, \frac{2}{i}\tr{[T_a,T_b]\{T_c,T_d\}}\,, \qquad
    &&\,f\indices{_{ab}^e}f_{cde} = -2\tr{[T_a,T_b][T_c,T_d]}    \nonumber\,.
\end{alignat}}

The strategy will be to expand the imaginary part of the amplitudes in flavor partial waves and compare them with the parametrizations of the previous section. Let us work this out for the case of four pions. In this case we have
\begin{equation}\label{eq:4piFL}
    \im\,\T_{abcd} = d\indices{_{ab}^e}d_{cde}\,\im\,M^{(C=+)}(s,u) + 
    f\indices{_{ab}^e}f_{cde}\,\im\,M^{(C=-)}(s,u)\,.
\end{equation}
It is important to stress that this is not the most general decomposition of a tensor with indices $a,b,c,d$ that we can write down; it only accounts for exchanges of physical mesons in the $\R=\text{adj}$ representation, while in general all the representations in $\text{adj}\times \text{adj}$ would appear. All these additional representations have vanishing spectral density at large $N$. For this reason, when we compare \eqref{eq:4piFL} with the parametrization \eqref{eq:T4pi} we find
\begin{subequations}
\begin{align}
    \im\, M_{4\pi}(s,t) = &\, \frac{1}{2}\left(\im\,M^{(C=+)}(s,u) - \im\,M^{(C=-)}(s,u)\right)\,,\\
    \im\, M_{4\pi}(s,u) = &\, \frac{1}{2}\left(\im\,M^{(C=+)}(s,u) + \im\,M^{(C=-)}(s,u)\right)\,,\\
    \im\, M_{4\pi}(t,u) = &\, 0\,.
\end{align}
\end{subequations}
That is, there are \textit{no $t$-channel poles in $M_{4\pi}$}. This was already appreciated in \cite{Albert:2022oes}, and it is a consequence of the \textit{large-N selection rules}. Note that this is compatible with the OZI rule~\cite{OKUBO1963165,Zweig:1964jf,Iizuka}, which suppresses glueball (or ``closed strings'') exchanges, but it also encodes the suppression of exotic mesons ---both consequences of the large~$N$ counting.

We can proceed similarly with the remaining amplitudes to obtain the flavor expansion for every case. This is worked out in detail in the ancillary file. The upshot is that keeping only the contributions by $\R=\text{adj}$ from the outset directly implements the large~$N$ selection rules. In this way we find that other disk amplitudes from the mixed system, like $M_{\pi\gamma\gamma\pi}^{(1)}(s,u)$, $M_{\pi\gamma\gamma\pi}^{(2)}(s,u)$ and $M_{3\pi\gamma}(s,u)$, have no $t$-channel poles.

\subsection{Little group}
For the second part of the partial waves, we need to deal with the little group representations. A general prescription for doing so in any space-time dimensions was introduced in \cite{Caron-Huot:GravPW} --- here we review their method for $D=4$ and extend it to the case of mixed correlators. In $D=4$ one could also use Wigner $d$-matrices for the little-group partial waves (see e.g.\ \cite{Hebbar:2020ukp,Caron-Huot:2022ugt} for a recent review), but the approach of \cite{Caron-Huot:GravPW} has the advantage of being explicitly covariant and it extends naturally to the mixed-correlator case (with flavor dependence).

Integer-spin representations of $SO(3)$ correspond to symmetric-traceless tensors of $J$ $SO(3)$-vector indices. This means that the little group vertices we should construct are invariant tensors of the form $v^{i,((m_1,...,m_J))}$, where the $m_i$ are $SO(3)$-vector indices and $((...))$ denotes symmetric-traceless. As with the flavor partial waves, we will construct three types of vertices; for two pions, a pion and a photon and for two photons. The local data that we have at our disposal in a vertex involving the particles $1,2$ are at least the momenta $p_1^\mu,p_2^\mu$ and the polarization tensors $\epsilon_1^\mu,\epsilon_2^\mu$ when the corresponding photon is involved. These vectors satisfy
\begin{equation}
	p_i^2=0\,, \qquad \epsilon_i\cdot p_i=0\,.
\end{equation}

We can construct $SO(3)$ indices $m$ from $SO(3,1)$ indices $\mu$ by ``removing the time component''. To do this covariantly, we define $P^\mu\equiv p_1^\mu+p_2^\mu$ and make any index $\mu$ orthogonal to it. Although the index still takes $4$ values, this makes it effectively an $SO(3)$ index $m$. One can check that in the center of mass frame, this literally corresponds to removing the time component from $\mu$. It is therefore convenient to work directly with objects that are orthogonal to $P$ \cite{Caron-Huot:GravPW},
\begin{equation}\label{eq:defs}
	n^\mu\equiv p_2^\mu-p_1^\mu\,,
	\qquad e_i^\mu\equiv \epsilon_i^\mu - p_i^\mu \frac{\epsilon_i\cdot P}{p_i\cdot P}\,,
\end{equation}
which further satisfy $n\cdot e_i=0$. Note that the polarization tensors $e_i$ are gauge invariant $e_i\simeq e_i + \alpha p_i$, so they comprise only the \textit{physical} degrees of freedom of the photons.

Similarly, we can obtain the $SO(3)$-invariant metric that we will use to contract $SO(3)$ indices and the $SO(3)$ Levi-Civita symbol from their $SO(3,1)$ counterparts,
\begin{equation}\label{eq:etahat}
	\hat \eta_{\mu\nu} \equiv \eta_{\mu\nu} + \frac{{P}_\mu{P}_\nu}{s}\,, \qquad \hat \varepsilon_{\nu\rho\sigma}\equiv \frac{1}{\sqrt{s}} P^\mu\varepsilon_{\mu\nu\rho\sigma}\,.
\end{equation}
In both cases this projects out the components parallel to $P^\mu$. We will only abbreviate with a dot the contractions with the full Minkowski metric $\eta_{\mu\nu}$, not $\hat \eta_{\mu\nu}$. We will also use the $SO(3)$ Gram determinant $\varepsilon_{mnr}\varepsilon^{abc}=\delta_{mnr}^{abc}$, which follows from $P^2=-s$ and $\varepsilon_{\mu\nu\rho\sigma}\varepsilon^{\alpha\beta\gamma\delta}=-\delta_{\mu\nu\rho\sigma}^{\alpha\beta\gamma\delta}$.

The prescription now is to write down all the possible tensors of $J$ indices that can be built out of $n^\mu,e_1^\mu,e_2^\mu$ and satisfy all of the following:
\begin{itemize}
	\item are linear in the polarization vector $e_i$ of every external photon,
	\item are symmetric in all indices and
	\item are traceless under contractions with $\hat\eta_{\mu\nu}$.
\end{itemize}
For ease of notation, it is convenient to write these representations in terms of Young tableaux, e.g.\ for an $\hat \eta$-traceless symmetric tensor of three indices, $a^{((\mu}b^\nu c^{\rho))}\equiv \tyoung{abc}$.

Let us start with the vertex involving two pions. The only data available at the vertex is $n^\mu$, so the only tensors that we can construct are
\begin{equation}
    n^{((\mu_1}n^{\mu_2}\cdot\cdot\cdot n^{\mu_J))}= \underbrace{\myoung{\n\cdots\n}}_{J}\equiv {\myoung{\bullet}}_J\,.
\end{equation}
For the pion-photon vertex, we add $e_2$ as an ingredient and we find two structures,
\begin{equation}
    {\myoung{\etwo\bullet}}_{J-1}\,, \qquad
    \myoung{\etwo\bullet,n}{\scriptstyle J-1}\,,
\end{equation}
where the dot stands for the factors of $n$ needed to fill out the $J$ columns. The former is the symmetric-traceless product of $e_2$ with $(J-1)$ factors of $n$ while the latter involves the vector $\tyoung{\etwo,n}=\hat \varepsilon_{\mu\nu\rho}e_2^\nu n^\rho$.

The case of two external photons is more involved. In this case, a priori we have six structures,
\begin{equation}\label{eq:2gNaive}
	e_1\cdot e_2{\myoung{\bullet}}_J\,, \quad{\myoung{\eone\etwo\bullet}}_{J-2}\,,\quad
 \myoung{\eone\bullet,\etwo,n}{}^J\,, \quad
	\myoung{\eone\bullet,\etwo}{\scriptstyle J-1}\,, \quad
	\myoung{\eone\etwo\bullet,n}{\scriptstyle J-2}\,, \quad
	\myoung{\eone\etwo\bullet,nn}{\scriptstyle J-2}\,,
\end{equation}
where $\tyoung{\eone,\etwo,n}=\hat \varepsilon_{\mu\nu\rho}e_1^\mu e_2^\nu n^\rho$ is a singlet. But not all of these structures are independent. For example, $\tyoung{\eone,\etwo}$ has a component that can only lie in the $n$ direction, so the third and fourth vertices in \eqref{eq:2gNaive} are equivalent. Also, the product of epsilon symbols in $\tyoung{\eone\etwo,nn}$ can be evaluated via a Gram determinant to show that
$$\myoung{\eone\etwo\bullet,nn}{\scriptstyle J-2} = - {\myoung{\eone\etwo\bullet}}_{J-2} - \eone\cdot\etwo {\myoung{\bullet}}_J\,.$$
Finally, when we discuss the crossing properties of these vertices, instead of the fifth vertex alone we will need the combinations
$$\myoung{\eone\etwo\bullet,n}{\scriptstyle J-2}
\pm\myoung{\etwo\eone\bullet,n}{\scriptstyle J-2}\,.$$
Of these two, only the upper sign gives a genuine new vertex; the lower one does not in virtue of
\begin{equation}
	\myoung{\eone\etwo,n} + \myoung{n\eone,\etwo} + \myoung{\etwo\n,\eone}=0\,,
\end{equation}
which follows from antisymmetry.

To summarize, we have the following little group vertices:
\begin{subequations}\label{eq:lgPW}
\begin{itemize}
    \item \textbf{Two pions:}
    \begin{equation}
        v_{\pi\pi}^1 = {\myoung{\bullet}}_{J} \qquad P=(-1)^{J}\,,
    \end{equation}
    
    \item \textbf{Pion-photon:}
    \begin{alignat}{2}
        v_{\pi\gamma}^1 =&\, {\myoung{\etwo\bullet}}_{J-1} \qquad &&P=-(-1)^J\,,\\
	v_{\pi\gamma}^2 =&\, i\, \myoung{\etwo\bullet,n}{\scriptstyle J-1} \qquad &&P=-(-1)^{J+1}\,,
    \end{alignat}
    
    \item \textbf{Two photons:}
    \begin{alignat}{2}
        v_{\gamma\gamma}^1 =&\,e_1\cdot e_2{\myoung{\bullet}}_J  && P=(-1)^J\,,\\
	v_{\gamma\gamma}^2 =&\,{\myoung{\eone\etwo\bullet}}_{J-2}  &&  P=(-1)^J\,,\\
	v_{\gamma\gamma}^3 =&\, i \myoung{\eone\bullet,\etwo,n}{}^J  && P=(-1)^{J+1}\,,\\
	v_{\gamma\gamma}^4 =&\, i\left(\myoung{\eone\etwo\bullet,n}{\scriptstyle J-2} + (e_1 \leftrightarrow e_2)\right)  \qquad && P=(-1)^{J+1}\,.
    \end{alignat}
\end{itemize}
\end{subequations}
The parity of the internal state in each vertex is easy to evaluate. We consider a sign flip for every component transverse to $P_s^\mu$, and thus we have that the parity of a Young tableau is given by $P=P_0^{(1)}P_0^{(2)}(-1)^{\#\,\text{of boxes}}$, with $P_0^{(i)}$ the intrinsic parity transformation of the $i$th external leg (we have $P_0^{(\pi)}=-1$ and $P_0^{(\gamma)}=1$). The factors of $i$ in the above vertices accompanying every $\hat\varepsilon$ symbol are introduced to satisfy $CPT$ symmetry. Since our vertices are already $C$ and $P$ invariant, we just need to further impose invariance under $T$, which can be taken to send $i\mapsto -i$ and $P^\mu\mapsto -P^\mu$ while keeping all the transverse vectors unchanged.

\subsection{More large \texorpdfstring{$N$}{N} selection rules}\label{sec:MoreSelectionRules}
The next step is to combine the flavor and little group parts into the full vertices and impose Bose symmetry, i.e.\ invariance under the exchange of identical particles. This puts further constraints on the allowed $\pi\pi$ and $\gamma\gamma$ vertices. In particular, we require invariance under the simultaneous exchange of $a\leftrightarrow b$, $n\leftrightarrow - n$ and $e_1 \leftrightarrow e_2$. When the dust settles, we get the list of full three-point couplings of table \ref{tab:vertices}, organized in $J^{PC}$ quantum numbers.

\begin{table*}[h!]\centering
\begin{tabular}{c||c|c|c|c||c|c|c|c|}
& \multicolumn{4}{|c||}{$ C = + $} & \multicolumn{4}{|c|}{$ C = - $}\\
\hline
& $J=0$ & $J=\text{even}$ & $J=1$ & $J=\text{odd}$ & $J=0$ & $J=\text{even}$ & \multicolumn{2}{|c|}{$J=1,\text{odd}$}\\
\cline{2-9} $P=+$
&$\begin{array}{c} d\cdot v_{\pi\pi}^1 \\ d\cdot v_{\gamma\gamma}^1 \end{array}$
& $\begin{array}{c} d\cdot v_{\pi\pi}^1 \\ f\cdot v_{\pi\gamma}^2 \\ d\cdot v_{\gamma\gamma}^1 \\ d\cdot v_{\gamma\gamma}^2 \end{array}$
& $\begin{array}{c} f\cdot v_{\pi\gamma}^1 \end{array}$
& $\begin{array}{c} f\cdot v_{\pi\gamma}^1 \\ d\cdot v_{\gamma\gamma}^{4+} \end{array}$
& \cellcolor{red!10}
& \cellcolor{red!10}$\begin{array}{c} d\cdot v_{\pi\gamma}^2 \end{array}$
& \multicolumn{2}{|c|}{$\begin{array}{c} d\cdot v_{\pi\gamma}^1 \end{array}$} \\
\hline
& $J=0$ & $J=\text{even}$ & \multicolumn{2}{|c||}{$J=1,\text{odd}$} & $J=0$ & $J=\text{even}$ & \multicolumn{2}{|c|}{$J=1,\text{odd}$}\\
\cline{2-9} $P=-$
& $d\cdot v_{\gamma\gamma}^3$
& $\begin{array}{c} f\cdot v_{\pi\gamma}^1 \\ d\cdot v_{\gamma\gamma}^3 \end{array}$
& \multicolumn{2}{|c||}{\cellcolor{red!10} $\begin{array}{c} f\cdot v_{\pi\gamma}^2 \end{array}$}
& \cellcolor{red!10}
& $\begin{array}{c} d\cdot v_{\pi\gamma}^1 \end{array}$
& \multicolumn{2}{|c|}{$\begin{array}{c} f\cdot v_{\pi\pi}^1 \\ d\cdot v_{\pi\gamma}^2 \end{array}$}\\
\hline
\end{tabular}
\caption{Classification of three-point vertices for the mixed system of pion and photon scattering in terms of the $J,P,C$ quantum numbers of the intermediate state. $J=\text{even}$ and $J=\text{odd}$ omit 0 and 1 respectively. Each vertex is composed by a little group part from \eqref{eq:lgPW} and a flavor part from \eqref{eq:flPW}, schematically denoted by $f$ or $d$. Each non-empty cell defines a \textit{sector} of intermediate states in pion-photon scattering. The quantum numbers of the cells shaded in red are not compatible with the quark model and so the corresponding vertices are suppressed at large $N$.
\label{tab:vertices}}
\end{table*}

Each cell in table \ref{tab:vertices} defines an independent sector of the partial wave expansion \eqref{eq:PWexp} corresponding to the exchange of a state with quantum numbers $J^{PC}$. However, not every combination of $J^{PC}$ can be realized by single-meson states at large $N$. In this limit, mesons are strictly $q\bar q$ bound states and their spin $J$ should be the total angular momentum $J=L\times S$ of the bound state. Since $C$ and $P$ are related to the spin $S$ and orbital angular momentum $L$ of the bound state by
\begin{equation}
    P=(-1)^{L+1}\,, \qquad C=(-1)^{L+S}\,,
\end{equation}
it turns out that not every combination of $J^{PC}$ eigenvalues can be realized by single-meson states. The allowed eigenvalues for the meson spectrum are given in table \ref{tab:spectrum}, together with their standard family names. Comparing it to table \ref{tab:vertices} we realize that two of the vertices, while allowed by group theory, cannot be realized by single-meson states. These have been shaded in red in table \ref{tab:vertices}. This implies that the corresponding exchanges are suppressed at large $N$ and so they can be neglected in the partial wave expansion \eqref{eq:PWexp}. These two new sets of large $N$ selection rules (together with the ones for the flavor part that we derived above) encode the essence of the large $N$ limit, and they result in more constraining bounds than one would otherwise get for a generic scattering process.\footnote{It is worth noting that these selection rules are not new ---it is just the observation that at large $N$ exotic mesons (and glueballs) are absent. In the old literature (see e.g.\ \cite{Close:1988bw}), mesons with $\R\neq \text{adj}$ are known as exotic mesons \textit{of the first kind}, whereas those with $J^{PC}$ outside of table \ref{tab:spectrum} are known as exotic mesons \textit{of the second kind}. The fact that the exchanges of either kind of exotics are very suppressed in experiment traditionally serves as an indication that the large $N$ expansion might not be such a bad approximation of the real world after all.}

\begin{table}
$$\begin{array}{c|c |c}
 PC & \text{Spin} & \text{Family names}  \\
 \hline
 ++ & J = 0,1,2,3... & a_J,\quad f_J,f_J'\\
  \hline
 +- & J = 1,3,5... & b_J,\quad h_J,h_J'\\
  \hline
 -+ & J = 0,2,4... & \pi_J,\quad \eta_J,\eta_J'\\
  \hline
 -- & J = 1,2,3,4... & \rho_J,\quad \omega_J,\phi_J
\end{array}$$
\caption{\label{tab:spectrum}Meson spectrum along with standard naming conventions. The names given to real-world $N_f=3$ mesons distinguish the different isospin projections. At large $N$, the whole $U(N_f)$ multiplet becomes degenerate and so we make no distinction between the different names of each family.}
\end{table}

\subsection{Gluing vertices}
We are ready to discuss the gluing of these vertices and construct the partial waves. Using the shorthand notation $v^i_{ab}(n,e_1,e_2)$ for a generic $1_a,2_b\to X$ vertex (where $i$ labels the different entries of table \ref{tab:vertices}), we have
\begin{equation}
	\left(\pi^\R_J\right)^{ij}=\Big(\overline{v_{cd}^{i}}(n',e_3,e_4), v_{ab}^j(n,e_1,e_2)\Big)\,.
\end{equation}
Here the parenthesis denotes the $SO(3)\times U(N_f)$ inner product of \eqref{eq:gluing}. The prescription for constructing $\overline{v_{cd}^{i}}(n',e_3,e_4)$ from the vertices above is as follows. We take $v_{ab}^j(n,e_1,e_2)$ and replace $a\to c$, $b\to d$, $e_1\to e_3$, $e_2\to e_4$ and $n\to n'\equiv p_3-p_4$.\footnote{This replaces all the quantum numbers of the particles $1\to 3$ and $2\to 4$, but with a crucial minus sign in $p_1\to -p_3$, $p_2\to -p_4$ due to our \textit{all-incoming} conventions.} Then we take its complex conjugate.

We have already discussed the gluing of the flavor part of these vertices in section \ref{sec:flavor}, let us now address the little group part. The gluing of any two structures can be obtained from the partial waves of scalar scattering \cite{Caron-Huot:GravPW},
\begin{equation}\label{eq:scalarPW}
    \Big(\underbrace{\myoung{\np\cdots\np}}_{J}, \underbrace{\myoung{\n\cdots\n}}_{J}\Big) = \left(n'_{\mu_1}\cdots\n'_{\mu_J}-\text{traces}\right)n^{\mu_1} \cdots n^{\mu_J} = \frac{(D-3)_J}{2^J(\tfrac{D-3}{2})_J}|n'|^J|n|^J\legP_J(x)\,,
\end{equation}
where $x=\frac{n\cdot n'}{|n||n'|}=1+\frac{2u}{s}$ ($=\cos\theta$ in the center of mass frame), and $\legP_J(x)$ is the Gegenbauer polynomial
\begin{equation}
	\legP_J(x)={}_2F_1(-J,J+D-3,\tfrac{D-2}{2},\tfrac{1-x}{2})\,,
\end{equation}
which reduces to the usual Legendre polynomial for $D=4$. To do so, we just act on \eqref{eq:scalarPW} with derivatives in $n$ and $n'$ and use the identities
\begin{subequations}
\begin{align}
	\frac{v^\mu}{J} \frac{\partial}{\partial n^\mu}{\myoung{\bullet}}_J=& \,{\myoung{v\bullet}}_{J-1}\\
	\frac{v^\mu w^\nu}{J(J-1)} \frac{\partial}{\partial n^\mu}\frac{\partial}{\partial n^\nu}{\myoung{\bullet}}_J=& \,{\myoung{vw\bullet}}_{J-2}\,,
\end{align}
\end{subequations}
where $v^\mu$ and $w^\mu$ are two generic vectors.

As a non-trivial example, we show here how to glue two vertices $v_{\pi\gamma}^1$, needed for the partial wave expansion of $\pi\gamma\to\pi\gamma$.
\begin{align}
    \Big(\underbrace{\myoung{\efour\np\cdots\np}}_{J},& \underbrace{\myoung{\etwo\n\cdots\n}}_{J}\Big) = \frac{e_4^\mu}{J} \frac{\partial}{\partial n'^\mu} \frac{e_2^\nu}{J} \frac{\partial}{\partial n^\nu} \Big(\underbrace{\myoung{\np\cdots\np}}_{J}, \underbrace{\myoung{\n\cdots\n}}_{J}\Big)\\
    =&\, \frac{J!}{2^J(\tfrac{1}{2})_J}\frac{1}{J^2}e_4^\mu e_2^\nu \frac{\partial}{\partial n'^\mu} \frac{\partial}{\partial n^\nu}|n'|^J|n|^J\legP_J(x)\,,\nonumber\\
    =&\, \frac{J!}{2^J(\tfrac{1}{2})_J} \frac{1}{J^2} |n'|^{J}|n|^{J} \left( \frac{e_2\cdot e_4}{|n'||n|} \legP_J'(x) + \frac{(n'\cdot e_2)(n\cdot e_4)}{|n'|^2|n|^2} \legP_J''(x) \right)\,,\nonumber
\end{align}
where we used that $e_2\cdot n = e_4\cdot n'=0$. The overall coefficient can be simplified by a suitable normalization of the three-point vertices. See the ancillary numerical file for details and further examples.

With these definitions, we proceed to glue the different vertices of table \ref{tab:vertices} to construct the partial waves $\pi_J^{\R}$. Since we want to preserve both $P$ and $C$, we only allow for gluings of vertices with the same quantum numbers. This splits the partial waves into nine different matrices $\pi_J^{PC}$ ---one for each cell of table \ref{tab:vertices}--- so that
\begin{equation}\label{eq:JPCPW}
	\mathcal T = \sum_{J,P,C}n_J^{PC}\, \tr{a_J^{PC}(s)\pi_J^{PC}}\,.
\end{equation}
Each of the corresponding spectral densities $\rho_J^{PC}(s)=\im\, a_J^{PC}(s)$ is independently positive semidefinite. The strategy is then to expand each entry of the matrices $\pi_J^{PC}$ in the relevant basis of polarization structures $E_i$ and flavor traces $\tr{T_a...}$ and compare to the parametrizations of section \ref{sec:Parametrizations}.\footnote{In general, only the full trace of \eqref{eq:JPCPW} can be expanded in the basis of polarization and flavor structures, since individual entries may not satisfy all the symmetries of the basis. However, our choice of factors of $i$ in \eqref{eq:flPW} and \eqref{eq:lgPW} to satisfy $CPT$ invariance is such that makes all coupling constants real. As a result, all spectral densities $\rho_J^{PC}$ are symmetric and we are allowed to symmetrize $\pi_J^{PC}$. This allows us to expand the partial wave matrices entry by entry.} In this way we obtain a partial wave expansion for every reduced amplitude $M_i(s,u)$ (and their inequivalent crossed versions) of the form
\begin{equation}\label{eq:ReducedPW}
    \im\, M_i(s,u) = \sum_{J,P,C}n_J^{PC}\, \tr{\rho_J^{PC}(s)\,\left[\pi_i\right]_J^{PC}(s,u)}\,.
\end{equation}

The explicit expressions for the $\left[\pi_i\right]_J^{PC}(s,u)$ can be obtained from the ancillary file accompanying this work, which provides a pedagogical construction of the partial waves following the steps we outlined above. One can check that the final expressions are indeed polynomial in $u$, as we would expect for a good choice of the basis of polarization structures which does not induce spurious poles on the reduced amplitudes (see footnote \ref{foot:Poly}).

As a final remark, we note that the normalization constant $n_J^{PC}$ of \eqref{eq:JPCPW} can be fixed from the completeness relation of the two-particle Hilbert space and the normalization of the three-point vertices (see appendix C of \cite{Caron-Huot:GravPW}). Since we are only interested in the positivity of the spectral density, though, $n_J^{PC}$ will be irrelevant to us. However, while not strictly necessary, it may still be useful to (ortho)normalize the vertices \eqref{eq:lgPW} as \cite{Caron-Huot:GravPW}
\begin{equation}
    \sum_{\text{Pols}} \Big(v^{i}(n,e_1,e_2)^*, v^j(n,e_1,e_2)\Big) = \delta^{ij}\,,
\end{equation}
to produce cleaner expressions. Here the sum runs over the polarizations of $e_1,e_2$; associated to the external photons in $v^{i}$ and $v^{j}$ (if any).

\section{Dispersion relations}
\label{sec:Dispersion}
Let us summarize what we have achieved so far. In section \ref{sec:Parametrizations} we parameterized the amplitudes for all the processes involved in mixed pion-photon scattering in terms of \textit{reduced amplitudes} $M_{i}(s,u)$ that are functions of the Mandelstam invariants alone. Then in section~\ref{sec:PartialWaves} we constructed the corresponding partial waves and expanded the imaginary parts of each of the reduced amplitudes $M_i(s,u)$ in the different channels, studying along the way the implications of unitarity and large $N$ selection rules. To bring all these elements at play, we need two more ingredients. On the one hand, we need the high-energy behavior of every reduced amplitude in the Regge limit
\begin{equation}
    |s|\to\infty \,,\qquad \text{at fixed } u\lesssim 0\,.
\end{equation}
This is usually inferred from requirements of causality and unitarity, but for large $N$ QCD this can be derived systematically using \textit{Regge theory}. On the other hand, we need the low-energy expansions of these functions, which are where the effective field theory (EFT) enters the game.

Once we have these two ingredients we can proceed as usual to write down dispersion relations for each of the $M_{i}(s,u)$, linking the low-energy data to high-energy averages via \textit{sum rules} and imposing crossing symmetry in terms of \textit{null constraints} \cite{Tolley:2020gtv,Caron-Huot:2020cmc}. In what follows we first explain how to assign a Regge behavior to each of the reduced amplitudes (subsection \ref{sec:ReggeBehavior}) and we then discuss their low-energy expansions (subsection \ref{sec:LowEnergies}). After that we move on to a brief discussion about sum rules in subsection \ref{sec:SumRules} and we give a systematic algorithm to derive null constraints in subsection \ref{sec:NullConstraints}. Appendix \ref{app:Regge} reviews some background for the discussion of section \ref{sec:ReggeBehavior}.

\subsection{High energies: Regge behavior}\label{sec:ReggeBehavior}
To write down dispersion relations for $M_i(s,u)$ and drop the contour at infinity, we need the amplitude to behave well-enough as we move around the complex plane $|s|\to \infty$ (at fixed $u$). In particular we need
\begin{equation}\label{eq:spink-Regge}
    \lim_{|s|\to\infty} \frac{M_i(s,u)}{s^{k_i}} =0\,, \qquad \text{at fixed }u\lesssim 0\,.
\end{equation}
We refer to this behavior as a ``spin $k_i$ Regge behavior''. This polynomial boundedness is expected on general grounds from arguments involving causality and unitarity \cite{Jin:1964zza,Martin:1965jj}, but in QCD-like theories ---in which the spectrum arranges in Regge trajectories--- the behavior \eqref{eq:spink-Regge} is very much under control. Predicting this behavior is one of the successes of Regge theory. For reference, we have included in Appendix \ref{app:Regge} a review of the old literature on it. We just quote the main result here. For an amplitude $M(s,u)$ describing the scattering of four scalars, the Regge behavior at fixed $u<0$ is
\begin{equation}\label{eq:scalar-Regge}
    \lim_{|s|\to\infty} M(s,u)\sim s^{\alpha_0(u)}\,,
\end{equation}
where $\alpha_0(u)$ is the \textit{leading Regge trajectory} of the $u$ channel exchanges. Its behavior in the forward limit $u\sim 0$ is therefore controlled by the intercept $\alpha_0(0)$ of the leading trajectory.

When dealing with external spinning particles, as opposed to scalars, a similar statement goes through for the reduced amplitudes $M_i(s,u)$, but the power may receive corrections from stripping off the polarization dependence of the amplitude \cite{Caron-Huot:GravPW, rhos}. Ultimately, these corrections depend on the normalization we chose for the polarization structures of section \ref{sec:Parametrizations}, but they can be systematically obtained as follows. The behavior \eqref{eq:scalar-Regge} for scalar amplitudes comes from the Regge theory result (see~\eqref{eq:ReggePole})
\begin{equation}
    M(s,u)\sim \legP_{\alpha_0(u)}\left(1+\tfrac{2s}{u}\right)\,,
\end{equation}
which evaluates the fixed-$u$ partial wave Legendre polynomial at the ``continuous spin'' $\alpha_0(u)$. This result readily generalizes to spinning amplitudes if we upgrade the Legendre $\legP_J(x)$ to the polynomials $\left[\pi_i\right]_J^{PC}$ of the partial wave expansion for the reduced amplitudes $M_i(s,u)$, which we constructed in the previous section. Studying their large-$|s|$, fixed $u$ behavior for a spin $J=\alpha_0(0)$ determines the Regge behavior of the reduced amplitude.

\begin{figure}[ht]
\centering
\includegraphics[scale=0.85]{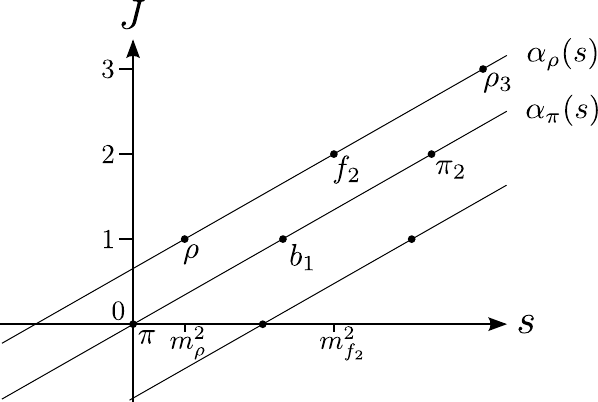}
\caption{Sketch of the first Regge trajectories expected in large $N$ QCD. In real-world QCD the trajectories are approximately linear for low spin but begin to curve higher up. It is unclear whether Regge trajectories curve at large $N$ \cite{Mandelstam:1974fq,Caron-Huot:2016icg,Veneziano:2017cks}.}
\label{fig:regge-traj}
\end{figure}

In QCD, the leading meson Regge trajectory is that of the rho meson; going through the $\rho$, the $f_2$, the $\rho_3$, etc, as sketched in figure \ref{fig:regge-traj}. Experimentally, its intercept is measured at $\alpha_{\rho}(0) \simeq 0.52$ \cite{Pelaez:2003ky}. The important point for us is that it is well below 1, which of course has to be the case for a monotonous function $\alpha_\rho(m^2)$ going through a \textit{massive} spin-one particle, i.e.\ $\alpha_\rho(m_\rho^2)=1$. So even at large $N$, we expect the intercept of the leading Regge trajectory to be strictly smaller than one,
\begin{equation}
    \alpha_\rho(0) < 1\,.
\end{equation}
This implies that scalar amplitudes exchanging the rho Regge trajectory, such as $M_{4\pi}(s,u)$, will satisfy a spin-one Regge behavior \eqref{eq:spink-Regge}. Since all other Regge trajectories are subleading compared to the rho, we can safely assume that all Regge intercepts satisfy $\alpha_0(0)<1$, in any process. Taking care of the polarization dependence in each case, this then implies for the reduced amplitudes (and all their crossed versions) the following Regge behaviors:\footnote{In the cases where the imaginary part along the $t$ channel vanishes, we are still allowed to assume the Regge behavior of the other channels, see the discussion at the end of section \eqref{app:exchange-degeneracy}.}
\begin{gather}\label{eq:MRegge}
    \left\{ M_{4\pi}(s,u), M_{4\gamma}^{(1,2,3)}(s,u), M_{\pi\gamma\pi\gamma}^{(1,2)}(s,u), M_{\pi\gamma\gamma\pi}^{(1,2)}(s,u)\right\}: \quad k_i=1\,,\\
    \left\{ M_{3\pi\gamma}(s,u)\right\}: \quad k_i=0\,.\nonumber
\end{gather}

\subsubsection*{Improved Regge channels}
While true, the assumption that all Regge intercepts satisfy $\alpha_0(0)<1$ is not optimal. The reason is that the rho Regge trajectory is not always exchanged in the crossed channel. If we find some combination of amplitudes such that the quantum numbers of the rho trajectory are forbidden in the $u$ channel, we are guaranteed that the Regge behavior for that combination will be controlled by the next subleading trajectory. In QCD this is the trajectory of the pion, which by construction has zero intercept in the chiral limit, $\alpha_\pi(0)=0$ (see the sketch of figure \ref{fig:regge-traj}). Since we will write down dispersion relations at fixed $u\lesssim 0$ (approaching zero from the left), this intercept will grant us one step better Regge behavior, which in turn will give us an extra set of constraints.

To find these \textit{improved Regge channels}, we look for linear combinations of the different amplitudes in each process whose partial waves in the $u$-channel vanish exactly for the sectors $J^{PC}=(1,\text{odd})^{--},\text{even}^{++}$. These are the quantum numbers of the mesons rho, $f_2$, $\rho_3$, etc that compose the rho trajectory. In the four-photons process, we find the combination
\begin{subequations}\label{eq:Mimp}
\begin{equation}
    M_{4\gamma}^{(\text{imp})}(s,u) \equiv M_{4\gamma}^{(1)}(s,u) + M_{4\gamma}^{(1)}(t,u) + 4 M_{4\gamma}^{(2)}(s,u) - 2 M_{4\gamma}^{(3)}(s,u)\,,
\end{equation}
while among the $2\pi2\gamma$ amplitudes we find
\begin{align}
    M_{2\pi 2\gamma}^{(\text{imp-1})}(s,u) &\,\equiv M_{\pi\gamma\gamma\pi}^{(1)}(s,u) + M_{\pi\gamma\pi\gamma}^{(2)}(t,u)\,, \\
    M_{2\pi 2\gamma}^{(\text{imp-2})}(s,u) &\,\equiv M_{\pi\gamma\pi\gamma}^{(1)}(s,u) + M_{\pi\gamma\gamma\pi}^{(2)}(t,u)\,,
\end{align}
\end{subequations}
as well as $M_{2\pi 2\gamma}^{(\text{imp-1})}(t,u)$ and $M_{2\pi 2\gamma}^{(\text{imp-2})}(t,u)$. There are no such improved Regge channels in the remaining processes.

In the $u$-channel, the combinations above exchange only states in the $(0,\text{even})^{-+}$ and $(1,\text{odd})^{+-}$ sectors, which are precisely the ones that compose the Regge trajectory of the pion. We have thus $\alpha_0(0)=0$ in this case and the Regge behavior for the improved amplitudes at $u\lesssim 0$ is
\begin{gather}\label{eq:MimpRegge}
    \left\{ M_{4\gamma}^{(\text{imp})}(s,u), M_{2\pi 2\gamma}^{(\text{imp-1})}(s,u), M_{2\pi 2\gamma}^{(\text{imp-2})}(s,u) \right\}: \quad k_i=0\,.
\end{gather}

\subsection{Low energies: The chiral Lagrangian}\label{sec:LowEnergies}
As usual, we parameterize the physics at low energies with an effective field theory (EFT). In this case, the relevant EFT is the chiral Lagrangian with $U(1)_Q$ sources \eqref{eq:gaugeLch}, which describes the effective interactions of a system of pions and photons. Since all the higher mesons have been integrated out, it only holds up to the mass $M$ of the first meson in the spectrum; the $\rho$ meson. The mass $M$ thus serves as a cutoff for the validity of the EFT description. At the level of scattering amplitudes, the unknown couplings of \eqref{eq:gaugeLch} translate into the coefficients in the low-energy expansion $s,u\sim 0$ of the reduced amplitudes $M_i^{\text{low}}(s,u)$. Here we work this out explicitly for the first terms in the expansion of all the reduced amplitudes, process by process.

\paragraph{Four pions.}
For four-pion scattering, the only relevant interactions are four-point contact terms, at different orders in the derivative expansion. We recall that the photon is treated as a probe and therefore there is no photon pole. As a result, the four--pion amplitude $M_{4\pi}^{\text{low}} (s, u)$ is analytic around $s,u\sim 0$ and it admits an expansion
\begin{equation}\label{eq:LE-M4pi}
    M_{4\pi}^{\text{low}} (s, u)  =  \, \sum_{n=1}^\infty \sum_{\ell = 0}^{[n/2]} g_{n, \ell} \, (s^{n-\ell} u^\ell + u^{n-\ell} s^\ell) \\
= \, g_{1, 0} (s+u) + g_{2, 0} (s^2 + u^2) + 2 g_{2, 1} \, su + \dots.
\end{equation}
where in terms of the couplings in \eqref{eq:gaugeLch},
\begin{equation}\label{eq:gpiLow}
    g_{1,0}=\frac{1}{2f_\pi^2}\,,\qquad g_{2,0}=\frac{2\kappa_1+4\kappa_2}{2f_\pi^4}\,,\qquad g_{2,1}=\frac{4\kappa_2}{f_\pi^4}\,.
\end{equation}

\paragraph{Four photons.}
At low energies, the first contribution to the four-photon amplitude is the pion exchange, induced by the Wess-Zumino-Witten term when coupled to the background photon (recall the first term in \eqref{eq:gaugeWZWpion}). Although a pion exchange in the chiral limit comes with a massless pole, the derivatives in the couplings conspire to save the amplitude from diverging at small energies. The contribution to the amplitude of the pion pole is
\begin{align}
    {\mathcal T}^{\lambda_1\lambda_2\lambda_3\lambda_4}  = &- \left(\frac{e^2k}{2\pi^2 f_\pi}\right)^2 
    \frac{1}{2}\tr{Q^4}
    \left(\frac{1}{s}\left(F_1\wedge F_2\right)\left(F_3\wedge F_4\right)\right.\nonumber\\
    &\qquad\left. +\frac{1}{u}\left(F_1\wedge F_3\right)\left(F_2\wedge F_4\right) + \frac{1}{t}\left(F_1\wedge F_4\right)\left(F_2\wedge F_3\right)\right)\,,
\end{align}
where we defined $F_i\wedge F_j\equiv \frac{1}{4}{F_i}_{\mu\nu}\varepsilon^{\mu\nu\rho\sigma}{F_j}_{\rho\sigma}$ and we used the expressions in footnote \ref{foot:completeness} to evaluate the flavor structure.

We can rewrite this result in terms of the basis \eqref{eq:4gPolStr} using the identity
\begin{equation}
    \left(F_1\wedge F_2\right)\left(F_3\wedge F_4\right) = H_{1324} - \frac{1}{4}H_{13}H_{24} - \frac{1}{4}H_{14}H_{23} =
    s^2 (E_3^{4\gamma} + E_{1,s}^{4\gamma})\,,
\end{equation}
which gives
\begin{equation}
    \MM_{4\gamma}(1^{\lambda_1},2^{\lambda_2},3^{\lambda_3},4^{\lambda_4}) = - \Ach^2\left(s\,E_{1,s}^{4\gamma} + u\,E_{1,u}^{4\gamma} + t\,E_{1,t}^{4\gamma}\right)\,, \qquad \Ach^2\equiv \left(\frac{e^2k}{8\pi^2 f_\pi}\right)^2
\end{equation}
where we used conservation of momentum to set $s+u+t=0$. We see that the anomaly contributes at linear order in the Mandelstams and so it only shows up in the amplitude that is not fully crossing symmetric, $M_{4\gamma}^{(1)}(s,u)$. What is especially interesting about this coupling is that it can be fixed by anomaly matching, as we reviewed in section \eqref{sec:anomalyMatching}. For QCD we have $k=N$, which makes the $N$ dependence explicitly enter the game.

After this contribution, there is an infinite series of four-photon interactions with higher powers of derivatives, coming with unfixed coefficients. For example, at order $O(\partial^4)$ we have two operators\footnote{\label{foot:4gComparison}See \cite{Henriksson:2021ymi,Henriksson:2022oeu,Haring:2022sdp} for a more detailed discussion about photon EFTs. To ease the comparison, we note that in terms of our reduced amplitudes, the amplitudes of \cite{Henriksson:2022oeu} and \cite{Haring:2022sdp} read
\begin{gather}
    f(s,t,u) = \Phi_2(s,t,u) = M_{4\gamma}^{(1)}(s,t) + M_{4\gamma}^{(1)}(s,u) + M_{4\gamma}^{(1)}(t,u) + 4 M_{4\gamma}^{(2)}(s,u) - 2 M_{4\gamma}^{(3)}(s,u)\,,\nonumber\\
    g(s|t,u) = \Phi_1(s,t,u) = M_{4\gamma}^{(1)}(t,u)\,, \qquad    h(s,t,u) = \Phi_5(s,t,u) = -M_{4\gamma}^{(2)}(s,u)\,,
\end{gather}
and so the low-energy couplings are related by
\begin{equation}
    f_2 = a_2^{(1)} - 2 a_2^{(3)}\,, \quad f_3 = -3 a_3^{(1)} + 4 a_3^{(2)} - 2 a_3^{(3)}\,, \quad g_2 = a_2^{(1)}\,, \quad g_3 = -a_3^{(1)}\,, \quad h_3 = -a_3^{(2)}\,.
\end{equation}}
\begin{equation}\label{eq:F4coupling}
    (F_{\mu\nu} F^{\mu\nu})(F_{\alpha\beta} F^{\alpha\beta})\,, \quad \text{and} \quad F_{\mu\nu} F^{\nu\rho} F_{\rho\sigma} F^{\sigma\mu}\,.
\end{equation}
At the level of the amplitude, these (and higher) operators map to the different terms in the low-energy expansions. The first terms for the four-photon reduced amplitudes at low energies are thus
\begin{subequations}\label{eq:4glow}
\begin{align}
    M_{4\gamma}^{(1)\text{low}}(s,u) &\,= \Ach^2(s+u) + a^{(1)}_{2}(s+u)^2 + a^{(1)}_{3}(s+u)^3 + \dots\,,\\
    M_{4\gamma}^{(2)\text{low}}(s,u) &\,= a^{(2)}_3 stu + \dots\,, \\
    M_{4\gamma}^{(3)\text{low}}(s,u) &\,= a^{(3)}_2(s^2+t^2+u^2) + a^{(3)}_3 stu + \dots\,.
\end{align}
\end{subequations}
The dots here indicate terms of order $O(p^8)$ and higher. Note that these are not the most general expansions consistent with their symmetries that one can write down. The reason is that at low order in derivatives not every term can be generated independently with local interactions. For instance, at order $O(p^4)$ we in principle have four independent $a_2$ couplings but there exist only the two local interactions of \eqref{eq:F4coupling}. One can find the correct low-energy expansion by working out explicitly the map from the Lagrangian to the amplitude. In practice, though, this is not necessary because the partial waves already know about the locality of interactions. We can start with the most general expansion for the low energy amplitudes and write down sum rules for all of their couplings (as we will review below). Couplings that would correspond to non-local interactions will have exactly vanishing sum rules and can then be set to zero.

\paragraph{Two pions and two photons.}
The first contribution to the scattering of two pions and two photons at low energies comes from the pion exchange and a contact term induced by the kinetic term of the Chiral Lagrangian \eqref{eq:gaugeLch}. There is no photon exchange because we are treating the photon as a probe, a non-dynamical background $U(1)_Q$ gauge field. Again, the derivatives in the couplings conspire to balance the massless pion propagator and make the amplitude well-behaved at low energies. With the basis of polarization structures defined above the pion exchange reads
\begin{equation}
    {\cal T}_{ad}^{\lambda_2\lambda_3} = 4e^2 \tr{[Q,T_a][Q,T_c]}E_{1,t}^{2\pi2\gamma}\,.
\end{equation}
The next order contributions to this process come from the contact interactions hidden in the $\kappa_3$ and $\kappa_4$ terms of \eqref{eq:gaugeLch}, which have the same form,
\begin{equation}
     {\cal T}_{ad}^{\lambda_2\lambda_3} = -2 \gOne \tr{[Q,T_a][Q,T_c]} t E_{2,t}^{2\pi2\gamma}\,, \qquad \gOne\equiv 4e^2 \frac{\kappa_3+\kappa_4}{f_\pi^2}\,.
\end{equation}

After that come contributions from higher-order contact interactions that we omitted in \eqref{eq:gaugeLch}. They start at order $O(\partial^4)$ with the operators
\begin{alignat}{2}\label{eq:2F2piCoupling}
&F_{\mu\nu}F^{\nu\rho}\,\tr{Q^2\{D_\rho U,D^\mu U^\dagger\}}\,, \qquad &&F_{\mu\nu}F^{\nu\rho}\,\tr{QD_\rho U Q D^\mu U^\dagger}\,, \\
&(F_{\mu\nu}F^{\mu\nu})\tr{Q^2\{D_\alpha U,D^\alpha U^\dagger\}}\,, \qquad &&(F_{\mu\nu}F^{\mu\nu})\tr{QD_\alpha U Q D^\alpha U^\dagger}\,,\nonumber
\end{alignat}
and continue to higher order in derivatives. They all come with unfixed couplings, which are mapped one to one to the coefficients in the low-energy expansion of the reduced amplitudes (defined in \eqref{eq:covExp}),
\begin{subequations}\label{eq:2p2gLow}
\begin{align}
    M_{\pi\gamma\pi\gamma}^{(1)\text{low}}(s,u) &\,= -2e^2 + c^{(1)}_2su + \cdots \,, \\
    M_{\pi\gamma\gamma\pi}^{(1)\text{low}}(s,u) &\,= 2e^2 + d^{(1)}_2su + \cdots \,, \\
    M_{\pi\gamma\pi\gamma}^{(2)\text{low}}(s,u) &\,= -\gOne(s+u) + c^{(2)}_2(s+u)^2 + \cdots \,, \\
    M_{\pi\gamma\gamma\pi}^{(2)\text{low}}(s,u) &\,= \gOne(s+u) + d^{(2)}_2(s+u)^2 + \cdots \,.
\end{align}
\end{subequations}
Again, we have low-energy expansions that are not quite the most general ones allowed by the symmetries. It is important to distinguish between two types of constraints. On the one hand, we have coefficients (or linear combinations thereof) that vanish exactly. These are terms that cannot be realized by a local term in the Lagrangian (see~\eqref{eq:2F2piCoupling} for the $O(p^4)$ terms), and the constraints apply independently to the different reduced amplitudes. The partial waves already know about these constraints and thence set the corresponding sum rules to zero.

On the other hand, the relation between the $O(p^2)$ terms (with coupling $\gOne$) across the reduced amplitudes $M_{\pi\gamma\pi\gamma}^{(2)\text{low}}(s,u)$ and $M_{\pi\gamma\gamma\pi}^{(2)\text{low}}(s,u)$ is of a different flavor. In terms of pion fields we can write two independent local operators,
\begin{equation}
    (F_{\mu\nu}F^{\mu\nu})\pi^a \pi^b\tr{Q^2\{T_a,T_b\}}\,, \qquad (F_{\mu\nu}F^{\mu\nu})\pi^a \pi^b\,\tr{Q T_a Q T_b}\,.
\end{equation}
So it would seem that the $O(p^2)$ terms in the two amplitudes should be independent. However, when writing operators in terms of the sigma model matrix $U\in U(N_f)$, we realize that only
\begin{equation}\label{eq:GBnature}
    (F_{\mu\nu}F^{\mu\nu})\tr{Q U Q U^\dagger}
\end{equation}
is possible due to the unitarity condition $UU^\dagger=1$. This relation is intrinsically due to the Goldstone boson nature of the pions, similar in spirit to the Adler zero for $M_{4\pi}(s,u)$. The partial waves do not know about it and equating the two sum rules for $\gOne$ gives a non-trivial new constraint, encoding some of the rich features of QCD. We call this a \textit{Goldstone constraint}, which explains our notation for the coupling $\gOne$.

\paragraph{Three pions and one photon.} The first contribution to this process is induced by the WZW term, recall \eqref{eq:gaugeWZWpion}. The associated contribution to the amplitude is
\begin{align}
    \mathcal T_{abc}^{\lambda_4}  = -\frac{ek}{3\pi^2 f_\pi^3}
    \Big[&\tr{T_bT_aQT_c} - \tr{T_aT_bQT_c}     +\tr{T_cT_bQT_a}  \\
    - &\tr{T_bT_cQT_a} + \tr{T_aT_cQT_b} - \tr{T_cT_aQT_b}\Big]
    \varepsilon^{\mu\nu\rho\sigma}{p_1}_\mu{p_2}_{\nu} {p_3}_\rho {\epsilon_4}_\sigma\,. \nonumber
\end{align}
The following contributions come from higher-derivative odd terms. Comparing to \eqref{eq:T3pi1g}, we conclude that at low energies we have the expansion
\begin{equation}\label{eq:BchLow}
    M_{3\pi\gamma}^{\text{low}}(s,u) = -\Bch + b_1 (s+u) +\cdots\,, \qquad \Bch\equiv \left(\frac{ek}{3\pi^2 f_\pi^3}\right)\,.
\end{equation}
For QCD the first coefficient is fixed by the chiral anomaly to be $k=N$ while the remaining low-energy couplings are unknown. This is the second instance where the anomaly appears in the scattering amplitudes. We denote it with a different name from $\Ach^2$ to highlight that from the EFT point of view these are two unrelated couplings.

\subsection{Sum rules}\label{sec:SumRules}
It is now straightforward to derive sum rules for the low-energy couplings introduced above. For each of the reduced amplitudes we write
\begin{equation}
    \frac{1}{2 \pi i } \oint_\infty ds'\,\frac{M_i(s', u)}{s'^{k+1}} = 0\,, \qquad k\geq k_i\,,
\end{equation}
and we shrink the contour towards the singularities on the real axis,
\begin{equation}
    \frac{1}{2 \pi i } \oint_0 ds'\,\frac{M_i^\text{low}(s', u)}{s'^{k+1}} = 
    \frac{1}{\pi} \int_{M^2}^\infty dm^2 \left(\frac{\text{Im}_s\,M_i (m^2, u) }{ m^{2(k+1)} }
    + (-1)^k\frac{\text{Im}_t\,M_i (m^2, u) }{ (m^2+u)^{k+1} }\right)\,.
\end{equation}
Around the origin we plug the low-energy expansions of the previous subsection, and for the cuts above the cutoff we use the partial wave expansions discussed in section \ref{sec:PartialWaves}. Sum rules are then obtained by expanding the result in the forward limit $u\sim 0$ and matching coefficients. They have the following schematic form:
\begin{equation}
    g_j^{(i)} = \frac{1}{\pi} \sum_{J }  n_J
 \int_{M^2}^\infty \frac{dm^2}{m^2}  \tr{\rho_J(m^2) f(J,m^2)}\,,
\end{equation}
for some (matrix-valued) function $f(J,m^2)$ coming from expanding the $s$- and $t$-channel partial waves in $u$.

We recall at this point that, due to the symmetries of the problem, the spectral density $\rho_J(m^2)$ is block diagonal with a block per sector $J^{PC}$ in table \ref{tab:vertices}. Since each of these blocks is a positive-semidefinite matrix on its own, it is convenient to introduce the following high-energy averages sector by sector,
\begin{equation}\label{eq:HEavg}
    \avg{\Big(\cdots\Big)}_{J^{PC}} \equiv \frac{1}{\pi} \sum_{J\in\, \text{sector}}  n_J^{PC}
 \int_{M^2}^\infty \frac{dm^2}{m^2}\, \tr{\rho_{J}^{PC}(m^2) \Big(\cdots\Big)}\,.
\end{equation}
Note that here we could choose a different cutoff for each sector, corresponding to the mass of the first resonance with those quantum numbers. For simplicity, though, we keep in all sectors the same cutoff $M$; the mass of the lowest meson ---the rho. With these definitions, sum rules are given by a sum of high-energy averages in different sectors; each of which weighted by a positive-semidefinite measure. For example, the first sum rule from $M_{4\pi}(s,u)$ reads\footnote{This is to be compared with the sum rules in \cite{Albert:2022oes}, which used only the $\pi\pi\to\pi\pi$ spectral density.}
\begin{gather}
    g_{1,0} = \avg{\begin{psmallmatrix}\frac{1}{m^2} & 0\\ 0 & 0\end{psmallmatrix}}_{0^{++}}
        + \quad\avg{\begin{psmallmatrix}\frac{1}{m^2} & 0 & 0 & 0\\ 
                0 & 0 & 0 & 0\\0 & 0 & 0 & 0\\0 & 0 & 0 & 0\end{psmallmatrix}}_{\text{even}^{++}}
        + \quad\avg{\begin{psmallmatrix}\frac{1}{m^2} & 0\\ 0 & 0\end{psmallmatrix}}_{1,\text{odd}^{--}}\,.
\end{gather}
The remaining sum rules for the couplings we will be interested in are listed in table \ref{tab:sumrules}.

\begin{table}[t]\centering
\caption{Sum rules for the lowest couplings in the EFT expansions of section \ref{sec:LowEnergies}. The sum rules are given by the sum of the corresponding high-energy averages. We use the shorthand notation $\J^2\equiv J(J+1)$ for the Casimir of $SO(3)$.
\label{tab:sumrules}}
\makebox[\textwidth][c]{
\begin{tabular}{|c||c|c|c|c|c|c|c|c|c|}
\hline
$PC$ & $++$ & $++$ & $++$ & $++$ & $-+$ & $-+$ & $+-$ & $--$ & $--$\\
$J$ & $0$ & even & $1$ & odd & $0$ & even & $1,$odd & even & $1,$odd\\
\hline
\hline
$g_{1,0}$& $\begin{psmallmatrix}\frac{1}{m^2} & 0\\ 0 & 0\end{psmallmatrix}$
        & $\begin{psmallmatrix}\frac{1}{m^2} & 0 & 0 & 0\\ 
                0 & 0 & 0 & 0\\0 & 0 & 0 & 0\\0 & 0 & 0 & 0\end{psmallmatrix}$
        & ---
        & ---
        & ---
        & ---
        & ---
        & ---
        & $\begin{psmallmatrix}\frac{1}{m^2} & 0\\ 0 & 0\end{psmallmatrix}$\\
\hline
$g_{2,0}$& $\begin{psmallmatrix}\frac{1}{m^4} & 0\\ 0 & 0\end{psmallmatrix}$
        & $\begin{psmallmatrix}\frac{1}{m^4} & 0 & 0 & 0\\ 
                0 & 0 & 0 & 0\\0 & 0 & 0 & 0\\0 & 0 & 0 & 0\end{psmallmatrix}$
        & ---
        & ---
        & ---
        & ---
        & ---
        & ---
        & $\begin{psmallmatrix}\frac{1}{m^4} & 0\\ 0 & 0\end{psmallmatrix}$\\
\hline
$2g_{2,1}$& ---
        & $\begin{psmallmatrix}\frac{\J^2}{m^4} & 0 & 0 & 0\\ 
                0 & 0 & 0 & 0\\0 & 0 & 0 & 0\\0 & 0 & 0 & 0\end{psmallmatrix}$
        & ---
        & ---
        & ---
        & ---
        & ---
        & ---
        & $\begin{psmallmatrix}\frac{\J^2}{m^4} & 0\\ 0 & 0\end{psmallmatrix}$\\
\hline
\end{tabular}}

\vspace{11pt}

\makebox[\textwidth][c]{
\begin{tabular}{|c||c|c|c|c|c|c|c|c|c|}
\hline
$PC$ & $++$ & $++$ & $++$ & $++$ & $-+$ & $-+$ & $+-$ & $--$ & $--$\\
$J$ & $0$ & even & $1$ & odd & $0$ & even & $1,$odd & even & $1,$odd\\
\hline
\hline
$\Ach^2$& $\begin{psmallmatrix}0 & 0\\ 0 & \frac{-1}{m^2}\end{psmallmatrix}$
        & $\begin{psmallmatrix}0 & 0 & 0 & 0\\ 
                0 & 0 & 0 & 0\\0 & 0 & \frac{-1}{m^2} & 0\\0 & 0 & 0 & \frac{1}{m^2}\end{psmallmatrix}$
        & ---
        & $\begin{psmallmatrix}0 & 0\\ 0 & \frac{1}{m^2}\end{psmallmatrix}$
        & $\frac{-1}{m^2}$
        & $\begin{psmallmatrix}0 & 0\\ 0 & \frac{-1}{m^2}\end{psmallmatrix}$
        & ---
        & ---
        & ---\\
\hline
$a_2^{(1)}$& $\begin{psmallmatrix}0 & 0\\ 0 & \frac{1}{m^4}\end{psmallmatrix}$
        & $\begin{psmallmatrix}0 & 0 & 0 & 0\\ 
                0 & 0 & 0 & 0\\0 & 0 & \frac{1}{m^4} & 0\\0 & 0 & 0 & \frac{1}{m^4}\end{psmallmatrix}$
        & ---
        & $\begin{psmallmatrix}0 & 0\\ 0 & \frac{1}{m^4}\end{psmallmatrix}$
        & $\frac{1}{m^4}$
        & $\begin{psmallmatrix}0 & 0\\ 0 & \frac{1}{m^4}\end{psmallmatrix}$
        & ---
        & ---
        & ---\\
\hline
$a_3^{(1)}$& $\begin{psmallmatrix}0 & 0\\ 0 & \frac{-1}{m^6}\end{psmallmatrix}$
        & $\begin{psmallmatrix}0 & 0 & 0 & 0\\ 
                0 & 0 & 0 & 0\\0 & 0 & \frac{-1}{m^6} & 0\\0 & 0 & 0 & \frac{1}{m^6}\end{psmallmatrix}$
        & ---
        & $\begin{psmallmatrix}0 & 0\\ 0 & \frac{1}{m^6}\end{psmallmatrix}$
        & $\frac{-1}{m^6}$
        & $\begin{psmallmatrix}0 & 0\\ 0 & \frac{-1}{m^6}\end{psmallmatrix}$
        & ---
        & ---
        & ---\\
\hline
$a_3^{(2)}$& ---
        & $\begin{psmallmatrix}0 & 0 & 0 & 0\\ 
                0 & 0 & 0 & 0\\0 & 0 & 0 & \frac{\sqrt{\J^2(\J^2-2)}}{2m^6}\\0 & 0 & \frac{\sqrt{\J^2(\J^2-2)}}{2m^6} & 0\end{psmallmatrix}$
        & ---
        & ---
        & ---
        & ---
        & ---
        & ---
        & ---\\
\hline
$a_2^{(3)}$& ---
        & $\begin{psmallmatrix}0 & 0 & 0 & 0\\ 
                0 & 0 & 0 & 0\\0 & 0 & 0 & 0\\0 & 0 & 0 & \frac{1}{2m^4}\end{psmallmatrix}$
        & ---
        & $\begin{psmallmatrix}0 & 0\\ 0 & \frac{1}{2m^4}\end{psmallmatrix}$
        & $\frac{1}{m^4}$
        & $\begin{psmallmatrix}0 & 0\\ 0 & \frac{1}{m^4}\end{psmallmatrix}$
        & ---
        & ---
        & ---\\
\hline
$a_3^{(3)}$& ---
        & $\begin{psmallmatrix}0 & 0 & 0 & 0\\ 
                0 & 0 & 0 & 0\\0 & 0 & 0 & \frac{\sqrt{\J^2(\J^2-2)}}{2m^6}\\0 & 0 & \frac{\sqrt{\J^2(\J^2-2)}}{2m^6} & \frac{11-2\J^2}{2m^6}\end{psmallmatrix}$
        & ---
        & $\begin{psmallmatrix}0 & 0\\ 0 & \frac{11-2\J^2}{2m^6}\end{psmallmatrix}$
        & $\frac{3}{m^6}$
        & $\begin{psmallmatrix}0 & 0\\ 0 & \frac{3-2\J^2}{m^6}\end{psmallmatrix}$
        & ---
        & ---
        & ---\\
\hline
\end{tabular}}

\vspace{11pt}

\makebox[\textwidth][c]{
\begin{tabular}{|c||c|c|c|c|c|c|c|c|c|}
\hline
$PC$ & $++$ & $++$ & $++$ & $++$ & $-+$ & $-+$ & $+-$ & $--$ & $--$\\
$J$ & $0$ & even & $1$ & odd & $0$ & even & $1,$odd & even & $1,$odd\\
\hline
\hline
$\Bch$& ---
        & $\begin{psmallmatrix}0 & 4\sqrt{\frac{\J^2}{m^6}} & 0 & 0\\ 
                4\sqrt{\frac{\J^2}{m^6}} & 0 & 0 & 0\\0 & 0 & 0 & 0\\0 & 0 & 0 & 0\end{psmallmatrix}$
        & ---
        & ---
        & ---
        & ---
        & ---
        & ---
        & $\begin{psmallmatrix}0 & -4\sqrt{\frac{\J^2}{m^6}} \\ 
                -4\sqrt{\frac{\J^2}{m^6}} & 0 \end{psmallmatrix}$\\
\hline
$b_1$& ---
        & $\begin{psmallmatrix}0 & -4\sqrt{\frac{\J^2}{m^{10}}} & 0 & 0\\ 
                -4\sqrt{\frac{\J^2}{m^{10}}} & 0 & 0 & 0\\0 & 0 & 0 & 0\\0 & 0 & 0 & 0\end{psmallmatrix}$
        & ---
        & ---
        & ---
        & ---
        & ---
        & ---
        & $\begin{psmallmatrix}0 & 4\sqrt{\frac{\J^2}{m^{10}}} \\ 
                4\sqrt{\frac{\J^2}{m^{10}}} & 0 \end{psmallmatrix}$\\
\hline
\end{tabular}}
\end{table}
\flushbottom

\begin{table*}[!t]\centering
\makebox[\textwidth][c]{
\begin{tabular}{|c||c|c|c|c|c|c|c|c|c|}
\hline
$PC$ & $++$ & $++$ & $++$ & $++$ & $-+$ & $-+$ & $+-$ & $--$ & $--$\\
$J$ & $0$ & even & $1$ & odd & $0$ & even & $1,$odd & even & $1,$odd\\
\hline
\hline
$c_2^{(1)}$& ---
        & $\begin{psmallmatrix}0 & 0 & 0 & 0\\ 
                0 & \frac{-2}{m^4} & 0 & 0\\0 & 0 & 0 & 0\\0 & 0 & 0 & 0\end{psmallmatrix}$
        & $\frac{-2}{m^4}$
        & $\begin{psmallmatrix}\frac{-2}{m^4} & 0\\ 0 & 0\end{psmallmatrix}$
        & ---
        & $\begin{psmallmatrix}\frac{-2}{m^4} & 0\\ 0 & 0\end{psmallmatrix}$
        & $\frac{-2}{m^4}$
        & $\frac{-2}{m^4}$
        & $\begin{psmallmatrix}0 & 0\\ 0 & \frac{-2}{m^4}\end{psmallmatrix}$\\
\hline
$d_2^{(1)}$& ---
        & $\begin{psmallmatrix}0 & 0 & 0 & 0\\ 
                0 & \frac{2}{m^4} & 0 & 0\\0 & 0 & 0 & 0\\0 & 0 & 0 & 0\end{psmallmatrix}$
        & $\frac{2}{m^4}$
        & $\begin{psmallmatrix}\frac{2}{m^4} & 0\\ 0 & 0\end{psmallmatrix}$
        & ---
        & $\begin{psmallmatrix}\frac{2}{m^4} & 0\\ 0 & 0\end{psmallmatrix}$
        & $\frac{-2}{m^4}$
        & $\frac{-2}{m^4}$
        & $\begin{psmallmatrix}0 & 0\\ 0 & \frac{-2}{m^4}\end{psmallmatrix}$\\
\hline
$c_{1}^{(2)}$& ---
        & $\begin{psmallmatrix}0 & 0 & 0 & 0\\ 
                0 & \frac{-1}{m^2} & 0 & 0\\0 & 0 & 0 & 0\\0 & 0 & 0 & 0\end{psmallmatrix}$
        & $\frac{-1}{m^2}$
        & $\begin{psmallmatrix}\frac{-1}{m^2} & 0\\ 0 & 0\end{psmallmatrix}$
        & ---
        & $\begin{psmallmatrix}\frac{1}{m^2} & 0\\ 0 & 0\end{psmallmatrix}$
        & $\frac{1}{m^2}$
        & $\frac{-1}{m^2}$
        & $\begin{psmallmatrix}0 & 0\\ 0 & \frac{-1}{m^2}\end{psmallmatrix}$\\
\hline
$c_{1}^{(2)}$
        & $\begin{psmallmatrix}0 & \frac{1}{\sqrt{2}m^2}\\ 
                \frac{1}{\sqrt{2}m^2} & 0\end{psmallmatrix}$
        & $\begin{psmallmatrix}0 & 0 & \frac{1}{\sqrt{2}m^2} & 0\\ 
                0 & \frac{-1}{m^2} & 0 & 0\\\frac{1}{\sqrt{2}m^2} & 0 & 0 & 0\\0 & 0 & 0 & 0\end{psmallmatrix}$
        & $\frac{-1}{m^2}$
        & $\begin{psmallmatrix}\frac{-1}{m^2} & 0\\ 0 & 0\end{psmallmatrix}$
        & ---
        & $\begin{psmallmatrix}\frac{1}{m^2} & 0\\ 0 & 0\end{psmallmatrix}$
        & $\frac{-1}{m^2}$
        & $\frac{1}{m^2}$
        & $\begin{psmallmatrix}0 & 0\\ 0 & \frac{1}{m^2}\end{psmallmatrix}$\\
\hline
$c_2^{(2)}$& $\begin{psmallmatrix}0 & \frac{1}{\sqrt{2}m^4}\\ \frac{1}{\sqrt{2}m^4} & 0\end{psmallmatrix}$
        & $\begin{psmallmatrix}0 & 0 & \frac{1}{\sqrt{2}m^4} & 0\\ 
                0 & \frac{1}{m^4} & 0 & 0\\\frac{1}{\sqrt{2}m^4} & 0 & 0 & 0\\0 & 0 & 0 & 0\end{psmallmatrix}$
        & $\frac{1}{m^4}$
        & $\begin{psmallmatrix}\frac{1}{m^4} & 0\\ 0 & 0\end{psmallmatrix}$
        & ---
        & $\begin{psmallmatrix}\frac{-1}{m^4} & 0\\ 0 & 0\end{psmallmatrix}$
        & $\frac{1}{m^4}$
        & $\frac{-1}{m^4}$
        & $\begin{psmallmatrix}0 & 0\\ 0 & \frac{-1}{m^4}\end{psmallmatrix}$\\
\hline
$d_2^{(2)}$& ---
        & $\begin{psmallmatrix}0 & 0 & 0 & 0\\ 
                0 & \frac{-1}{m^4} & 0 & 0\\0 & 0 & 0 & 0\\0 & 0 & 0 & 0\end{psmallmatrix}$
        & $\frac{-1}{m^4}$
        & $\begin{psmallmatrix}\frac{-1}{m^4} & 0\\ 0 & 0\end{psmallmatrix}$
        & ---
        & $\begin{psmallmatrix}\frac{1}{m^4} & 0\\ 0 & 0\end{psmallmatrix}$
        & $\frac{1}{m^4}$
        & $\frac{-1}{m^4}$
        & $\begin{psmallmatrix}0 & 0\\ 0 & \frac{-1}{m^4}\end{psmallmatrix}$\\
\hline
\end{tabular}}
\end{table*}

\subsection{Null constraints}\label{sec:NullConstraints}
These sum rules link EFT couplings to the high-energy data. Unitarity (or rather positivity) is encoded in the positive measure of the high-energy averages. This results in some inequalities for the low-energy couplings, which we will explore below. Further constraints are obtained by imposing \textit{crossing symmetry}. In this language crossing symmetry is encoded in \textit{null constraints}~\cite{Tolley:2020gtv,Caron-Huot:2020cmc}. Null constraints are usually derived by equating different sum rules for low-energy couplings that are required to be the same by crossing symmetry. The downside of this approach is that it requires having the low-energy expansions to arbitrary order and, as we discussed in section \ref{sec:LowEnergies}, these are not always easy to write down. Instead, we can take a more low-energy-agnostic approach by engineering a contour integral that directly kills low-energy contributions.

Consider a function $f(s,u)$ (a placeholder for the reduced amplitudes $M_i(s,u)$) which accepts a polynomial expansion
\begin{equation}
    f_{\text{low}}(s,u) = \sum_{i,j}a_{i,j} s^iu^j
\end{equation}
around the origin and whose only singularities in the complex $s$-plane (at fixed $u<0$) are along the real axis with $s>M^2$ and $s<-M^2-u$. At this point, we make no crossing-symmetry assumptions for this function, but we assume that it satisfies a ``spin-$k$ Regge behavior'', i.e.\
\begin{equation}
    \lim_{|s|\to\infty} \frac{f(s,u)}{s^{k}} =0\,, \qquad \text{at fixed }u< 0\,.
\end{equation}

Consider now the following double contour integral,
\begin{equation}\label{eq:NCint}
    \oint_0 \frac{du}{2\pi i}\oint_\infty \frac{ds}{2\pi i}\frac{1}{s u}\left(\frac{f(s,u)}{s^{n-\ell}u^\ell} - \frac{f(u,s)}{u^{n-\ell}s^\ell}\right)=0\,, \quad \begin{cases}\text{for }n-\ell\geq k\,,\\ \text{and }\ell\geq k\,.\end{cases}
\end{equation}
The integral in $s$ implements a fixed-$u$ dispersion relation whereas the integral in $u$ picks a particular coefficient in the corresponding $u\sim 0$ (forward limit) expansion. As long as we have enough subtractions of $s$ in both terms, this integral vanishes like any other dispersion relation. The beauty of this particular integrand is that it is such that all low-energy contributions vanish,
\begin{equation}
    \mathop{\mathrm{Res}}_{u = 0}\mathop{\mathrm{Res}}_{s = 0} \left(\frac{f_{\text{low}}(s,u)}{s^{n-\ell+1}u^{\ell+1}}-\frac{f_{\text{low}}(u,s)}{u^{n-\ell+1}s^{\ell+1}}\right) = 0\,.
\end{equation}

The only contributions to this integral come from the high-energy cuts. Collecting all of them we obtain
\begin{align}\label{eq:NullConstr}
    0 =  
    \mathop{\mathrm{Res}}_{u = 0}\left[\frac{1}{u}\right.&\int_{M^2}^\infty dm^2 \left(\frac{\text{Im}\, f(m^2, u) }{ m^{2(n-\ell+1)} u^\ell} -\frac{\text{Im}\, f(u, m^2) }{ u^{n-\ell} m^{2(\ell+1) } }\right)\\
    -&\left.\frac{1}{u}\int_{M^2}^\infty dm^2 \left(\frac{\text{Im}\, f(-m^2-u, u) }{ (-m^2-u)^{(n-\ell+1)} u^\ell} -\frac{\text{Im}\, f(u,-m^2-u) }{ u^{n-\ell} (-m^2-u)^{(\ell+1) } }\right)\right]\,.\nonumber
\end{align}
Now we can simply expand the imaginary part of the different crossed versions of $f(s,u)$ in partial waves and rewrite this expression in terms of the high-energy averages of \eqref{eq:HEavg} to get null constraints of the form
\begin{equation}
    0 = \sum_{\text{sectors}} \avg{\mathcal N_{n,\ell}(m^2,J)}_{J^{PC}} \,, \qquad 
    \begin{aligned}
        n & = 2k, 2k+1, \ldots \\
        \ell & = k, \ldots, n-k \,.
    \end{aligned}
\end{equation}

With this algorithmic approach it is straightforward to derive null constraints for all of the reduced amplitudes $M_i(s,u)$. We just have to identify $f(s,u)$ with (a crossed version of) $M_i(s,u)$ and use the partial wave expansions \eqref{eq:ReducedPW} for the imaginary parts of $f(s,u)$. But not all these null constraints are independent. The number of independent null constraints depends on the symmetries of $M_i(s,u)$. We distinguish two cases:
\begin{itemize}
    \item If $M_i(s,u)$ is fully $s\leftrightarrow t \leftrightarrow u$ symmetric, we set $f(s,u)=M_i(s,u)$ in \eqref{eq:NullConstr} and all the numerators are the same. In our setup, this is the case only for the reduced amplitudes $M_{4\gamma}^{(2)}(s,u)$ and $M_{4\gamma}^{(3)}(s,u)$, which both obey spin $k=1$ Regge behavior. One can check that the non-trivial independent null constraints in both cases are given by the range of parameters
    \begin{equation*}
        f(s,u) = M_i(s,u)\,, \qquad \left\{n = 4,5\ldots\,, \: \ell = 1,\ldots \left[\tfrac{n-1}{3}\right]\right\}\,.
    \end{equation*}

    \item If $M_i(s,u)$ instead is only $s\leftrightarrow u$ symmetric, we have two inequivalent choices; $f(s,u)=M_i(s,u)$ and $f(s,u)=M_i(-s-u,u)$. These two choices effectively implement dispersion relations at fixed $u$ and at fixed $t$ respectively. Many of our reduced amplitudes fall into this category. First, with a spin $k=1$ Regge behavior we have $M_{4\pi}(s,u)$, $M_{4\gamma}^{(1)}(s,u)$, $M_{\pi\gamma\pi\gamma}^{(1)}(s,u)$, $M_{\pi\gamma\pi\gamma}^{(2)}(s,u)$, $M_{\pi\gamma\gamma\pi}^{(1)}(s,u)$ and $M_{\pi\gamma\gamma\pi}^{(2)}(s,u)$. In each of these cases, the full set of independent null constraints is spanned by\footnote{For $M_{4\pi}(s,u)$ one can check that the null constraints obtained in this way reproduce the SU and ST null constraints of \cite{Albert:2022oes}.} 
    \begin{alignat*}{2}
        f(s,u) \,&= M_i(s,u)\,, \qquad &&\left\{n = 3,4\ldots\,, \: \ell = 1,\ldots \left[\tfrac{n-1}{2}\right]\right\}\,,\\
        f(s,u) \,&= M_i(-s-u,u)\,, \qquad &&\left\{n = 2,3\ldots\,, \: \ell = 1,\ldots \left[\tfrac{n}{2}\right]\right\}\,.
    \end{alignat*}
    Second, with a spin $k=0$ Regge behavior we have $M_{3\pi\gamma}(s,u)$, the null constraints for which correspond to
    \begin{alignat*}{2}
        f(s,u) \,&= M_{3\pi\gamma}(s,u)\,, \qquad &&\left\{n = 1,2\ldots\,, \: \ell = 0,\ldots \left[\tfrac{n-1}{2}\right]\right\}\,,\\
        f(s,u) \,&= M_{3\pi\gamma}(-s-u,u)\,, \qquad &&\left\{n = 0,1\ldots\,, \: \ell = 0,\ldots \left[\tfrac{n}{2}\right]\right\}\,.
    \end{alignat*}
\end{itemize}

\section{Positivity bounds}
\label{sec:Positivity}
We are finally ready to put bounds on low-energy coefficients. As usual, the strategy is to write down a semidefinite problem combining sum rules and null constraints and exploit unitarity (in the form of positivity of the spectral density) to obtain bounds for the low-energy coefficients.
We start with a brief review of the general framework used in \cite{Albert:2022oes} to systematically derive bounds for four-pion scattering, generalizing it to the case where the spectral density is matrix-valued. We also point out some subtleties that one encounters in the choice of the normalizing denominator, which serve as guidelines for the bounds in the processes that we subsequently address.
We then describe our results, 
starting with the
simpler closed subsystems for the $\gamma \gamma \to \gamma \gamma $ and $\pi \gamma \to \pi \gamma$ processes and
then  graduating to the full mixed  system and to bounds on the chiral anomaly.

\subsection{Generalities}\label{sec:generalities}
We start by pointing out, from the sum rules in table \ref{tab:sumrules}, that the three couplings in four-pion scattering are obviously positive;
\begin{equation}
    g_{1,0}\geq 0\,, \qquad g_{2,0}\geq 0\,, \qquad g_{2,1}\geq 0\,.
\end{equation}
In the case of matrix-valued sum rules this is diagnosed by noticing that they involve only averages of positive-semidefinite matrices and $\avg{M}_{J^{PC}} \sim \tr{\rho_J^{PC}(m^2) M}\geq 0$ if $M$ is positive semidefinite by positivity of the spectral density. One can next look for linear combinations of sum rules which still have positive-semidefinite matrices in all sectors and thus derive bounds on different ratios of $g$'s.

All bounds will be \textit{homogeneous} with respect to rescalings of the couplings. This comes from the fact that we are only using \textit{positivity} of the spectral density $\rho_J(s)\succeq 0$, which is invariant under rescalings of $\rho_J(s)$. This means we can choose one of the couplings (say $g_{1,0}$) as normalization and place two-sided bounds on ratios of other couplings with it (together with the powers $M^2$ required by dimensional analysis).
When choosing the normalization coupling, there are two points to keep in mind. First, it should have a positive-semidefinite sum rule. Otherwise, the coupling could become 0 for nontrivial amplitudes and the normalized ratio would blow up. Second, it should be of lower (or equal) Mandelstam order\footnote{We recall that in the low-energy expansions of the reduced amplitudes, we organize the couplings in increasing powers of momenta (see e.g.\ \eqref{eq:LE-M4pi}), defining the so-called ``Mandelstam order''. Since the Mandelstam order of a low-energy coupling is related to the number of subtractions/derivatives that we need to capture it in dispersion relations, we in general have that their sum rules scale as $g_{n,k}\sim \frac{1}{m^{2n}}$.} than the other couplings involved in the bound. A coupling with a positive-semidefinite sum rule can only vanish when all the states in the theory are infinitely heavy, $m^2\to \infty$. If we use as normalization a coupling of higher Mandelstam order (which decays faster with $m^2\to \infty$) the bound will again diverge.

Once a suitable normalization coupling $g_{\mathbb{1}}$ has been chosen, we can systematize the search for bounds on the low-energy couplings by rewriting it as a semidefinite problem \cite{Caron-Huot:2020cmc} amenable to a solver such as SDPB \cite{Simmons-Duffin:2015qma}. This was already discussed in detail in \cite{Albert:2022oes}, here we briefly discuss the minor modifications needed when the spectral density is matrix-valued. At the level of four-pion scattering this is of course unnecessary, but it becomes essential when dealing with spinning particles and/or mixed correlators.

The strategy to bound a ratio $g_{n}M^{\#}/g_{\mathbb{1}}$ is the following. First, we define the vectors
\begin{equation}
	\vec{v}_\mathbb{1}=\begin{pmatrix}
	1\\ 0 \\0\\ \vdots\\ 0
	\end{pmatrix}\,,\qquad
	\vec{v}_n=\begin{pmatrix}
	0\\ 1 \\0\\ \vdots\\ 0
	\end{pmatrix}\,,\qquad
	\vec{v}_J^{PC}\left(m^2,J\right)=\begin{pmatrix}
	-g_{\mathbb{1}}\left(m^2,J\right)_{J^{PC}}\\ -g_{n}\left(m^2,J\right)_{J^{PC}} \\ \mathcal{N}_{1}\left(m^2,J\right)_{J^{PC}}\\ \mathcal{N}_{2}\left(m^2,J\right)_{J^{PC}}\\ \vdots
	\end{pmatrix}\,,
\end{equation}
where the vectors $\vec{v}_J^{PC}\left(m^2,J\right)$ (defined sector by sector) have matrix-valued components corresponding to the contributions of the relevant sum rules and null constraints to the average in the sector $J^{PC}$. Their length is controlled by the number of null constraints we choose to include.\footnote{We also organize null constraints by the Mandelstam order $n$ of the dispersion relation \eqref{eq:NCint} that they come from. In numerical implementations we include all null constraints up to a certain order $n_\text{max}$ in $\vec{v}_J^{PC}\left(m^2,J\right)$. Taking higher $n_\text{max}$ can only make the bounds stronger.} Note that we have set $M^2=1$, it can be restored at any point by dimensional analysis. These vectors satisfy the following ``bootstrap equation'':
\begin{equation} \label{eq:bootstrpeq}
	g_{\mathbb{1}} \,\vec v_\mathbb{1} + g_{n}\, \vec v_n+ \sum_{\text{sectors}}\avg{\vec v_J^{PC}\left(m^2,J\right)}_{J^{PC}}=0\,.
\end{equation}

Now we just have to look for a ``functional'' $\vec\alpha$ such that
\begin{enumerate}
	\item it is normalized as $\vec \alpha \cdot \vec v_n=\left\{\begin{matrix}
	+1\quad \text{for upper bound}\\
	-1\quad \text{for lower bound,}
	\end{matrix}\right.$
	\item its action on every $J^{PC}$ sector separately gives a positive-semidefinite \textit{matrix},
 $$\vec \alpha \cdot \vec v_J^{PC}\left(m^2,J\right)\succeq 0\qquad 
 \forall J\in \text{sector}, \; m^2\geq M^2\,,$$
	\item and it maximizes $\vec \alpha \cdot \vec v_\mathbb{1}$.
\end{enumerate}
Once that functional is found, it is immediate to find a bound by applying it on equation \eqref{eq:bootstrpeq} and using the positivity of high-energy averages. The two-sided bound reads
\begin{equation}
	\vec\alpha_{(-)}\cdot\vec v_{\mathbb{1}}\leq \frac{g_nM^{\#}}{g_{\mathbb{1}}} \leq -\left(\vec \alpha_{(+)} \cdot \vec v_\mathbb{1}\right)\,,
\end{equation}
where the subscript $(\pm)$ indicates the choice of normalization for $\vec\alpha$.
It is straightforward to generalize this algorithm to include more than one coupling and carve out a multidimensional exclusion plot as in \cite{Caron-Huot:2020cmc,Albert:2022oes}.

\subsection{The \texorpdfstring{$4 \gamma$}{four-photon} subsystem}\label{sec:4photonsPlots}
The four-photon system is closed on its own under crossing and unitarity. We can therefore derive bounds for ratios of the couplings in \eqref{eq:4glow} independently from the other amplitudes in the mixed system. This system has already been studied with positivity methods in \cite{Henriksson:2021ymi,Henriksson:2022oeu} and with full non-perturbative unitarity in \cite{Haring:2022sdp}, although in a somewhat different context.

\begin{figure}[ht]
\centering
\includegraphics[scale=1]{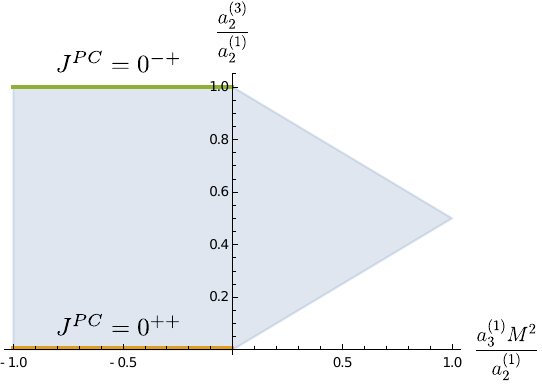}
\caption{Exclusion plot in the space of couplings $(a_3^{(1)}, a_2^{(3)})$ normalized by $a_2^{(1)}$. The bottom (orange) and top (green) lines are ruled in respectively by a single exchange of the scalar $\phi$ and pseudo-scalar $\chi$ of \eqref{eq:2gammaInt}.}
\label{fig:a31a23}
\end{figure}

\begin{figure}[ht]
\centering
\includegraphics[scale=1]{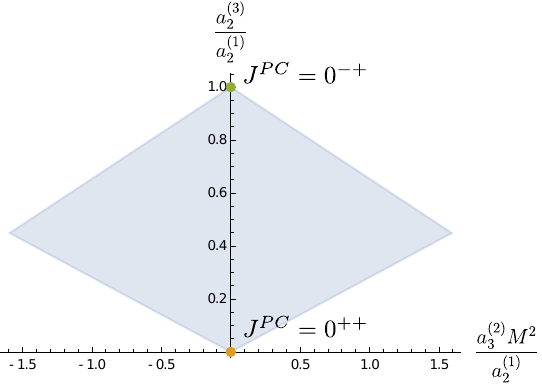}
\caption{Exclusion plot in the space of couplings $(a_3^{(2)}, a_2^{(3)})$ normalized by $a_2^{(1)}$. The bottom (orange) and top (green) corners are ruled in respectively by a single exchange of the scalar $\phi$ and pseudo-scalar $\chi$ of \eqref{eq:2gammaInt}.}
\label{fig:a32a23}
\end{figure}

\begin{figure}[ht]
\centering
\includegraphics[scale=1]{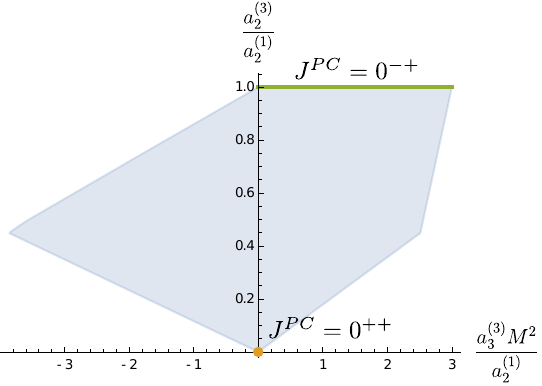}
\caption{Exclusion plot in the space of couplings $(a_3^{(3)}, a_2^{(3)})$ normalized by $a_2^{(1)}$. The bottom (orange) corner and the top (green) edge are ruled in respectively by a single exchange of the scalar $\phi$ and pseudo-scalar $\chi$ of \eqref{eq:2gammaInt}.}
\label{fig:a33a23}
\end{figure}

From table \ref{tab:sumrules}, we see that the couplings with lowest Mandelstam order \textit{and} a positive-semidefinite sum rule are
\begin{equation}
    a_2^{(1)}\geq 0\,, \qquad a_2^{(3)}\geq 0\,.
\end{equation}
We can therefore use either of them as the normalization coupling $g_{\mathbb{1}}$ to derive bounds on ratios of higher couplings with them, following the systematic method discussed above. In figures \ref{fig:a31a23}, \ref{fig:a32a23} and \ref{fig:a33a23} we show three different 2-dimensional sections of the space of large $N$ 4-photon effective amplitudes consistent with crossing and unitarity. We have chosen the coupling $a_2^{(1)}$ as normalization and considered couplings up to order six in derivatives, whose sum rules are listed in table \ref{tab:sumrules}. These plots were obtained with $n_\text{max}=6$, but they seem to have converged.

Although we are assuming a softer Regge behavior  compared to \cite{Henriksson:2021ymi,Henriksson:2022oeu} our results are compatible with these previous works. In fact, our methods reproduce the exclusion plots (without gravity) of \cite{Henriksson:2022oeu} exactly. It thus seems that the theories saturating their bounds already satisfy a softer Regge behavior.\footnote{This is indeed the case for some bounds that we rule in below, but another logical possibility is that due to some numerical obstruction, SDPB is ignoring the additional null constraints accessed with the softer Regge behavior. We elaborate on this point in section \ref{sec:sharper}.} A key difference is that with a softer Regge behavior, we have access to the lowest coefficient in \eqref{eq:4glow}, the chiral anomaly $\Ach^2$. As we have emphasized, this coupling is an interesting bootstrap target because it can be fixed by anomaly matching. Unfortunately, while one can derive a sum rule for it (see~table \ref{tab:sumrules}), it does not couple to any positivity bounds. The reason is that it comes at lowest Mandelstam order but its sum rule is not explicitly positive. As a result, any positivity bounds involving $\Ach^2$ are either trivial or diverge. It is conceivable  that this coupling may be probed by the non-perturbative S-matrix bootstrap, exploiting the unitarity upper bound on the spectral density. This would be a finite $N$ effect.

\subsubsection*{Ruling in} 
Some of the corners in figures \ref{fig:a31a23}, \ref{fig:a32a23} and \ref{fig:a33a23} can be explained by simple analytic solutions to crossing, as already appreciated in \cite{Henriksson:2021ymi}. The first obvious candidates are theories with a single scalar. A single scalar is allowed on its own because we are assuming a spin-one Regge behavior, and it will be crossing symmetric as long as we include all $s$, $t$ and $u$ exchanges. Looking at table \ref{tab:vertices} we see that there are two independent $J=0$ particles that are kinematically allowed to interact with two photons. There is a scalar $\phi^a$ with $J^{PC}=0^{++}$ and a pseudo-scalar (or axion) $\chi^a$ with $J^{PC}=0^{-+}$.\footnote{In the real world these would correspond to an $f_0$ and a $\pi_0$ meson respectively. See table \ref{tab:spectrum} for more details.} Their interactions are given by
\begin{equation}\label{eq:2gammaInt}
    \mathcal L_{\text{int}} = \frac{g_{\gamma\gamma\phi}}{m_\phi}\, \phi^a F_{\mu\nu}F^{\mu\nu} \tr{T_a Q^2} + \frac{g_{\gamma\gamma\chi}}{2m_\chi}\, \chi^a\, \varepsilon^{\mu\nu\rho\sigma}F_{\mu\nu}F_{\rho\sigma} \tr{T_a Q^2}\,,
\end{equation}
and the corresponding reduced amplitudes are
\begin{subequations}\label{eq:4gRulingIn}
\begin{align}
    M_{4\gamma}^{(1)}(s,u) &\,= g_\phi^2 \frac{t}{m_\phi^2-t} + g_\chi^2 \frac{t}{m_\chi^2-t}\,,\\
    M_{4\gamma}^{(2)}(s,u) &\,= 0\,, \\
    M_{4\gamma}^{(3)}(s,u) &\,= g_\chi^2 \left(\frac{s}{m_\chi^2-s} + \frac{t}{m_\chi^2-t} + \frac{u}{m_\chi^2-u}\right)\,,
\end{align}
\end{subequations}
which clearly have the right crossing properties and satisfy a spin-one Regge behavior. They are also unitary as long as the coupling constants are real. Expanding at low energies and comparing with \eqref{eq:4glow} we find
\begin{gather}
    \Ach^2 = -\frac{g_\phi^2}{m_\phi^2} - \frac{g_\chi^2}{m_\chi^2}\,, \quad 
    a_2^{(1)} = \frac{g_\phi^2}{m_\phi^4} + \frac{g_\chi^2}{m_\chi^4}\,, \quad
    a_3^{(1)} = -\frac{g_\phi^2}{m_\phi^6} - \frac{g_\chi^2}{m_\chi^6}\,,\nonumber\\
    a_3^{(2)} = 0\,, \quad
    a_2^{(3)} = \frac{g_\chi^2}{m_\chi^4}\,, \quad
    a_3^{(3)} = 3\frac{g_\chi^2}{m_\chi^6}\,,
\end{gather}
in agreement with the sum rules of table \ref{tab:sumrules}. Now by exploring different combinations of values for the couplings $g_\phi,g_\chi$ and the masses $m_\phi,m_\chi$ we see that these two amplitudes saturate some of the bounds, as indicated by the color lines/dots in figures \ref{fig:a31a23}, \ref{fig:a32a23} and \ref{fig:a33a23}.

The remaining corners are not as easy to rule in. In analogy with \cite{Albert:2022oes} one could try to UV complete a spin-one exchange by adding infinitely heavy states, but the symmetries prevent four photons from exchanging any spin-one state (cf.\ table \ref{tab:vertices}). Another simple candidate would be an ``$stu$-pole'' theory like the ones identified in \cite{Caron-Huot:2020cmc,Albert:2022oes}, but it is non-trivial to write down one such amplitude that is compatible with the different polarization structures. It would be interesting to nail down the theories that saturate these bounds, in particular the bounds with nonzero $a_3^{(2)}$, since its sum rule indicates that it only receives contributions from higher spin states.

\subsection{The \texorpdfstring{$\pi \gamma \to \pi \gamma$}{two-pion two-photon} subsystem}\label{sec:2pi2gamma}
In the processes involving two pions and two photons, there are two positive channels; $\pi\gamma \to \pi\gamma$ and $\pi\gamma\to \gamma \pi$. The third channel, $\pi\pi \to \gamma \gamma$, does not have a positive-semidefinite spectral density ---it only enters as an off-diagonal component of the full spectral density. This means that if we find a subset of couplings that only receive contributions from the $\pi\gamma \to \pi\gamma$ spectral density, we may derive bounds independently from the other processes. This is the case for all the couplings in the low-energy expansion of $\MM_{\pi\gamma\gamma\pi}$, since the large~$N$ selection rules suppress its $\pi\pi \to \gamma \gamma$ channel (as is clear from figure \ref{fig:2p2g}), but it is also the case for some of the couplings in $\MM_{\pi\gamma\pi\gamma}$. They can be read off directly from the sum rules in table \ref{tab:sumrules} by looking for couplings with no off-diagonal entries.

\begin{figure}[ht]
\centering
\includegraphics[scale=1]{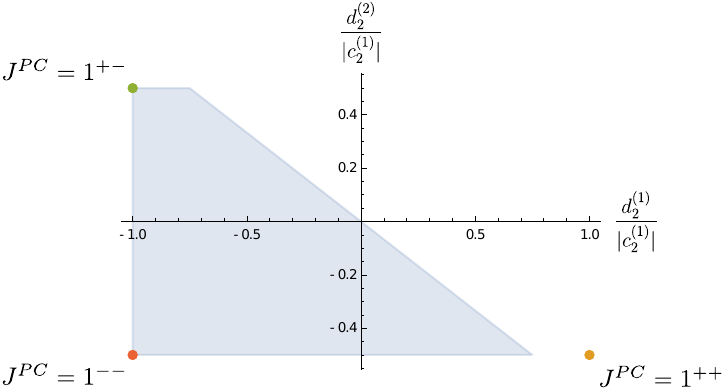}
\caption{Exclusion plot in the space of couplings $(d_2^{(1)}, d_2^{(2)})$ normalized by $-c_2^{(1)}$ at order $n_\text{max}=6$ (although it seems to have converged). The upper (green) and lower (red) left corners are saturated by single spin-one exchanges with $PC=+-,--$ respectively. The $1^{++}$ exchange is well outside the bounds.}
\label{fig:d21d22}
\end{figure}

Up to Mandelstam order 2, there are four couplings in this positive subsystem; $\gOne$, $c_2^{(1)}$, $d_2^{(1)}$, $d_2^{(2)}$. Out of these, only one has a sign-definite sum rule:
\begin{equation}
    c_2^{(1)}\leq 0\,.
\end{equation}
We can therefore use this as normalization and, using the usual strategy, carve out the space of allowed EFTs in the directions $d_2^{(1)}/|c_2^{(1)}|,d_2^{(2)}/|c_2^{(1)}|$. The corresponding exclusion plot is shown in figure \ref{fig:d21d22}. Like $\Ach^2$ before, $\gOne$ decouples from all positivity bounds because it comes at lowest order but its sum rule is not sign definite. This is unfortunate, because $\gOne$ has been measured experimentally for real-world QCD \cite{Gasser:1984gg,Bijnens:1987dc}. As far as we are aware, the corresponding data for the higher-derivative couplings, which we \textit{can} bound, are not available in the literature.

\subsubsection*{Ruling in}
The exclusion plot in figure \ref{fig:d21d22} has some sharp corners. Can any of them be ``ruled in'' by simple solutions to crossing? Spin-zero exchanges cannot do the job in this case, since we see from table \ref{tab:sumrules} that none of these couplings receives $J=0$ contributions. The next candidates are \textit{spin-one} exchanges. On their own, spin-one exchanges are not good enough, for we are demanding the Regge behavior to be \textit{strictly better than one}. However, it was shown in \cite{Albert:2022oes} that one can cure the Regge behavior of a single rho meson exchange (in four-pion scattering) by simply adding higher states with infinite mass. In the current system there are three possible spin-one exchanges: $J^{PC} = 1^{++}, 1^{+-}, 1^{--}$, corresponding to the physical $a_1$, $b_1$ and $\rho$ mesons. One may hope that a similar cure goes through for each of these mesons.

The interactions of the three spin-one mesons are given by
\begin{equation}\label{eq:spin1ints}
    \mathcal L_{\text{int}} = \partial_\mu \pi^a F^{\mu\nu}\left(\frac{2g_{\pi\gamma a}}{m_a}f_{aQb}\, a^b_\nu
    + \frac{2g_{\pi\gamma b}}{m_b}d_{aQb}\, b^b_\nu\right)
    +\frac{g_{\pi\gamma \rho}}{m_\rho} \,\epsilon^{\mu\nu\kappa\sigma}\partial_\mu \pi^a F_{\nu\kappa} \, d_{aQb}\, \rho_\sigma^b\,,
\end{equation}
and the corresponding mixed amplitudes are\footnote{Here we have corrected the exchanges by an analytic piece so that they are just above the assumed Regge behavior but the low-energy form \eqref{eq:2p2gLow} remains valid.}
\begin{align}\label{eq:J1RuleIn}
    M_{\pi\gamma\pi\gamma}^{(1)}(s,u) &\, = - \frac{g_{\pi\gamma a}^2}{m_a^2} \frac{su}{m_a^2 - s} 
    - \frac{g_{\pi\gamma b}^2}{m_b^2} \frac{su}{m_b^2 - s} 
    - \frac{g_{\pi\gamma \rho}^2}{m_\rho^2} \frac{su}{m_\rho^2 - s} + (s\leftrightarrow u)\,,\\
    M_{\pi\gamma\gamma\pi}^{(1)}(s,u) &\, = + \frac{g_{\pi\gamma a}^2}{m_a^2} \frac{su}{m_a^2 - s} 
    - \frac{g_{\pi\gamma b}^2}{m_b^2} \frac{su}{m_b^2 - s} 
    - \frac{g_{\pi\gamma \rho}^2}{m_\rho^2} \frac{su}{m_\rho^2 - s} + (s\leftrightarrow u)\,,\nonumber\\
    M_{\pi\gamma\pi\gamma}^{(2)}(s,u) &\, = - g_{\pi\gamma a}^2 \frac{t}{m_a^2 - s} 
    - g_{\pi\gamma b}^2 \frac{t}{m_b^2 - s} 
    + g_{\pi\gamma \rho}^2 \frac{t}{m_\rho^2 - s} + (s\leftrightarrow u)\,,\nonumber\\
    M_{\pi\gamma\gamma\pi}^{(2)}(s,u) &\, = + g_{\pi\gamma a}^2 \frac{t}{m_a^2 - s} 
    - g_{\pi\gamma b}^2 \frac{t}{m_b^2 - s} 
    + g_{\pi\gamma \rho}^2 \frac{t}{m_\rho^2 - s} + (s\leftrightarrow u)\,,\nonumber
\end{align}
which, upon expanding at small energies give
\begingroup
\allowdisplaybreaks
\begin{align}\label{eq:J1LE}
    c_2^{(1)} &\,= -2 \frac{g_{\pi\gamma a}^2}{m_a^4} -2 \frac{g_{\pi\gamma b}^2}{m_b^4} -2 \frac{g_{\pi\gamma \rho}^2}{m_\rho^4} \,, \\
    d_2^{(1)} &\,= +2 \frac{g_{\pi\gamma a}^2}{m_a^4} -2 \frac{g_{\pi\gamma b}^2}{m_b^4} -2 \frac{g_{\pi\gamma \rho}^2}{m_\rho^4} \,, \nonumber \\
    c_2^{(2)} &\,= + \frac{g_{\pi\gamma a}^2}{m_a^4} + \frac{g_{\pi\gamma b}^2}{m_b^4} - \frac{g_{\pi\gamma \rho}^2}{m_\rho^4} \,, \nonumber\\
    d_2^{(2)} &\,= - \frac{g_{\pi\gamma a}^2}{m_a^4} + \frac{g_{\pi\gamma b}^2}{m_b^4} - \frac{g_{\pi\gamma \rho}^2}{m_\rho^4} \,. \nonumber
\end{align}
\endgroup

From here we see that the exchange of the $b_1$ ($J^{PC}=1^{+-}$) and the rho meson ($J^{PC}~=~1^{--}$) saturate respectively the top and bottom left corners of figure \ref{fig:d21d22}. Their exchanges can be UV-completed by adding masses at infinity in a manner analogous to \cite{Albert:2022oes}. On the other hand, the $a_1$ ($J^{PC}=1^{++}$) is outside the bounds ---it cannot be UV-completed in the same way. The reason is that the would-be UV completion of the $a_1$ exchange necessitates higher-spin resonances from the sectors suppressed by the \textit{large $N$ selection rules} discussed in section \ref{sec:MoreSelectionRules}. Indeed, if one re-runs the plot allowing for the sectors shaded in red in table \ref{tab:vertices}, figure \ref{fig:d21d22} becomes a perfect rectangle stretching between the $1^{+-},1^{--}$ and $1^{-+},1^{++}$ exchanges. The slanted line of figure \ref{fig:d21d22} is thus a consequence of the large $N$ selection rules, very intrinsic to the large $N$ kinematics. It would be interesting to understand what theory saturates this line, perhaps by constructing explicitly the UV completions of the spin-one exchanges and studying what contributions they get from each sector.

While admitting healthy UV completions, the $b_1$ and rho meson exchanges are not quite the QCD-like amplitudes that we are after. Indeed, expanding the last two reduced amplitudes at low energies we find
\begin{align}
    M_{\pi\gamma\pi\gamma}^{(2)}(s,u) &\, = \left(2g_{\pi\gamma b}^2
    - 2g_{\pi\gamma \rho}^2\right) \left(s+u\right) + \cdots\,,\\
    M_{\pi\gamma\gamma\pi}^{(2)}(s,u) &\, = \left(2g_{\pi\gamma b}^2
    - 2g_{\pi\gamma \rho}^2\right) \left(s+u\right) + \cdots\,.
\end{align}
Compared to \eqref{eq:2p2gLow}, we see that there is a missing relative sign in the $\gOne$ coefficient. This means that the $b_1$ and rho meson single exchanges are incompatible with the chiral Lagrangian, since the identification of these two couplings (with a relative sign) was intrinsic to the sigma model nature of pion interactions (recall the discussion around \eqref{eq:GBnature}).

The two left corners of figure \ref{fig:d21d22} should thus be ruled out by including this ``Goldstone constraint'' in the semidefinite problem. But this is not the case in practice. When including this Goldstone constraint, which corresponds to the difference of the two $\gOne$ sum rules of table \ref{tab:sumrules}, SDPB sets to zero its contribution and yields the same figure \ref{fig:d21d22}. We discuss further this (and related) issue(s) in section \ref{sec:sharper}, where we also speculate on some possible workarounds, but it remains an open question how to incorporate numerically this Goldstone constraint.

\subsection{The full mixed system}\label{sec:MixedSys}
In the previous subsection, we derived bounds for the couplings in the positive subsystem of $2\pi 2\gamma$ scattering. The story changes when we consider the coupling $c_2^{(2)}$, whose sum rule has off-diagonal entries. These correspond to contributions from the $\pi\pi \to \gamma\gamma$ spectral density, which has no positivity properties on its own. As a result, it is not possible to bound a ratio of $c_2^{(2)}$ with any other $2\pi2\gamma$ coupling. Indeed, there is no positive-semidefinite matrix one can construct involving only the $c_2^{(2)}$ sum rule and other sum rules and/or null constraints from the $2\pi2\gamma$ amplitudes.

To obtain bounds involving $c_2^{(2)}$ we need to graduate to the full mixed-correlator system and bring in couplings from the four-pion and four-photon amplitudes. In particular, we need to consider couplings whose sum rules fill in the diagonals of $c_2^{(2)}$. Since we do not want to spoil positivity in other sectors by doing this, the positive four-pion and four-photon couplings $g_{1,0}$, $a_2^{(1)}$ are good candidates. By combining these couplings into a positive-semidefinite sum rule (in every sector) we find the following bound(s),
\begin{equation}\label{eq:g24Abound}
    \frac{1}{\sqrt{2}}g_{1,0} + \frac{1}{\sqrt{2}}a_2^{(1)}M^2 - \frac{1}{2}c_2^{(1)}M^2 \pm c_2^{(2)}M^2\geq 0\,.
\end{equation}
This bound requires some interpretation. We start by noting that it transforms in a nontrivial way under rescalings of the spectral density. Since the spectral density is now in general matrix-valued, we can consider rescaling its components by
\begin{equation}
    \rho_{4\pi} \to \rho_{4\pi}\,, \quad \rho_{4\gamma} \to e^4\rho_{4\gamma}\,,\quad  \rho_{2\pi 2\gamma} \to e^2\rho_{2\pi 2\gamma}\,, \quad \rho_{3\pi\gamma} \to e\rho_{3\pi \gamma}\,,
\end{equation}
which does not spoil its positive-semidefiniteness. Under such a rescaling, \eqref{eq:g24Abound} becomes
\begin{equation}
    \frac{e^4}{\sqrt{2}}g_{1,0} + \frac{1}{\sqrt{2}}a_2^{(1)}M^2 - \frac{e^2}{2}c_2^{(1)}M^2 \pm e^2 c_2^{(2)}M^2\geq 0\,.
\end{equation}
This makes the dependence on the gauge coupling explicit and it becomes manifest that the bound is invariant under rescalings of the photon field. The same goes through for rescalings of the pion, and any bound involving couplings of the mixed system is in fact \textit{doubly homogeneous}.

By appropriate rescalings of the spectral density we can thus bring \eqref{eq:g24Abound} to
\begin{equation}
    \frac{|c_2^{(2)}|M}{\sqrt{g_{1,0}\,a_2^{(1)}}} \leq \sqrt{2} - \frac{1}{2}\frac{c_2^{(1)}M}{\sqrt{g_{1,0}\,a_2^{(1)}}} \,.
\end{equation}
The takeaway is that to put bounds on couplings of the mixed system we need to choose \textit{two normalization conditions}. Indeed the bound above can be reproduced numerically with an algorithm analogous to that of section \ref{sec:generalities} but where we set both $g_{1,0}=1$ and $a_2^{(1)}=1$.

\begin{figure}[ht]
\centering
\includegraphics[scale=1]{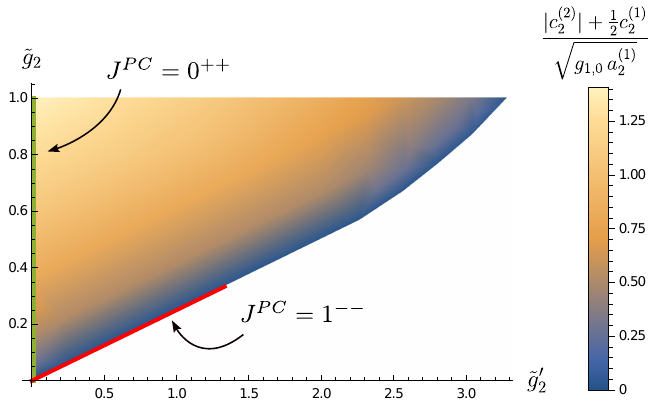}
\caption{Extension of the $(\tilde g_2', \tilde g_2)$ exclusion plot from \cite{Albert:2022oes} in a (color-coded) third direction $|c_2^{(2)}|+\frac{1}{2}c_2^{(1)}$ normalized by $g_{1,0}, a_2^{(1)}$, at order $n_\text{max}=5$. The bound is saturated at the top left corner and decreases monotonically to zero towards the right-side boundary. The bound along the vertical line $\tilde g_2=0$ (green) is still saturated by a single scalar exchange with $PC=++$. A single rho (i.e.\ $J^{PC}=1^{--}$) exchange saturates (part of) the ``Skyrme line'' (red).}
\label{fig:c22-c21}
\end{figure}

What is most interesting of the bounds in the mixed system, however, is that they can be combined with the four-pion bounds of \cite{Albert:2022oes}. By choosing as normalization $g_{1,0}=a_2^{(1)}=1$, and spanning the allowed values of
\begin{equation}
    \tilde g_2 \equiv g_{2,0}\frac{M^2}{g_{1,0}}\,, \qquad \tilde g_2' \equiv 2g_{2,1}\frac{M^2}{g_{1,0}}\,,
\end{equation}
we can extend the $(\tilde g_2',\tilde g_2)$ exclusion plot of \cite{Albert:2022oes} in a third dimension corresponding to some mixed coupling $\sim g_{2\pi2\gamma}/(g_{1,0}\, a_2^{(1)})^{\frac{1}{2}}$. As an example, we show in \ref{fig:c22-c21} the three-dimensional extension of the $(\tilde g_2',\tilde g_2)$ plot by the bound on the combination of couplings $|c_2^{(2)}|+\frac{1}{2}c_2^{(1)}$.

As a word of caution, we mention that not every (combination of) coupling(s) $g_{2\pi2\gamma}$ may be bounded with the choice of normalization $g_{1,0}=a_2^{(1)}=1$. From table \ref{tab:sumrules} we see that the $2\pi2\gamma$ couplings receive contributions from sectors that completely decouple from the four-pion and four-photon amplitudes. There can exist UV-complete theories which receive contributions only from these sectors and therefore have $g_{2\pi2\gamma}\neq 0$ but vanishing $g_{1,0}$, $a_2^{(1)}$; ``ruling in'' $g_{2\pi2\gamma}/(g_{1,0}\, a_2^{(1)})^{\frac{1}{2}}\to \infty$ and trivializing the bound. What can be bounded with this choice of normalization are combinations of couplings that cancel the contributions from these decoupled sectors, like $|c_2^{(2)}|+\frac{1}{2}c_2^{(1)}$ above. It is not obvious what combinations are allowed, but one can get some intuition from the ``ruled-in'' amplitudes we discuss below.

\subsubsection*{Ruling in}
We end the discussion of the mixed system by ``ruling in'' some bounds of the exclusion plot in figure \ref{fig:c22-c21}. Like in the previous subsections, and following \cite{Caron-Huot:2020cmc,Albert:2022oes}, we check whether some simple solutions to crossing and unitarity saturate any of the bounds. The first candidate is the exchange of a $J^{PC}=0^{++}$ scalar $\phi$, coupling to two photons as in \eqref{eq:2gammaInt} and to two pions by\footnote{This normalization of the on-shell coupling is chosen to simplify the final expressions.}
\begin{equation}\label{eq:pipiphiInt}
    \mathcal L_{\text{int}} = \sqrt{2}\,\frac{g_{\pi\pi\phi}}{m_\phi}\, \partial_\mu \pi^a \partial^\mu \pi^b \phi^c d_{abc}\,.
\end{equation}
It cannot couple to a pion and a photon. Such an exchange contributes only to one of the mixed amplitudes,
\begin{equation}
    M_{\pi\gamma\pi\gamma}^{(2)}(s,u) = \sqrt{2}\,g_{\pi\pi\phi}g_{\gamma\gamma\phi}\,\frac{t}{m_\phi^2 - t}\,,
\end{equation}
and it results in a four-pion amplitude
\begin{equation}
    M_{4\pi}(s,u) = g_{\pi\pi\phi}^2 \left(\frac{s}{m_\phi^2 - s} + \frac{u}{m_\phi^2 - u}\right)\,.
\end{equation}
Expanding around zero we find the low-energy couplings
\begin{equation}
    g_{1,0} = \frac{g_{\pi\pi\phi}^2}{m_\phi^2}\,, \qquad
    g_{2,0} = \frac{g_{\pi\pi\phi}^2}{m_\phi^4}\,, \qquad
    c_2^{(2)} = \sqrt{2}\,\frac{g_{\pi\pi\phi} g_{\gamma\gamma\phi}}{ m_\phi^4}\,, \qquad
    a_2^{(1)} = \frac{g_{\gamma\gamma\phi}^2}{m_\phi^4}\,.
\end{equation}
This exchange was shown to rule in the vertical line (and in particular the top-left corner) of the $(\tilde g_2', \tilde g_2)$ exclusion plot in \cite{Albert:2022oes}. One can check now that it also saturates the upper bound on $(|c_2^{(2)}|+\frac{1}{2}c_2^{(1)})/(g_{1,0}\, a_2^{(1)})^{\frac{1}{2}}$ (shown in figure \ref{fig:c22-c21}) along this line. This suggests that this is the only theory living in the top-left corner of the $(\tilde g_2', \tilde g_2)$ plot, with no degeneracies.

We turn now to the ``Skyrme line'' $\tilde g_2 = \frac{1}{4}\tilde g_2'$ of figure \ref{fig:c22-c21}. In \cite{Albert:2022oes} it was shown that this is the line where the Skyrme model lies. It was further shown that part of this line can be ruled in by the exchange of a single rho meson UV completed at infinity. Later, \cite{Fernandez:2022kzi} argued that as one takes $n_\text{max}\to\infty$ the numerical kink will go down to meet the ruled in amplitude, so that the whole straight line can be ``explained'' by a single rho. It is interesting to study this line in higher-dimensional sections of the space of pion-photon EFTs to understand whether this is the unique exchange that saturates the Skyrme line or if it gets extended by exchanges in sectors that decouple from pure pion scattering. In figure \ref{fig:c22-c21} the bound along this line reads $|c_2^{(2)}|+\frac{1}{2}c_2^{(1)}\leq 0$. We see from \eqref{eq:J1LE} that this is indeed saturated by a single rho meson ($J^{PC}=1^{--}$) exchange, indicating no such extension.

\subsection{Bounds on the anomaly}

\label{sec:anomalybounds}

Finally, we look for bounds on the chiral anomaly. As discussed in sections \ref{sec:WZWterm} and \ref{sec:LowEnergies}, the chiral anomaly manifests itself in our EFT amplitudes as two different couplings; $\Ach$ and $\Bch$ ---the neutron pion decay and the $3\pi 1\gamma$ contact term. In contrast to $\Ach$, which decouples from all positivity bounds (see the discussion at the end of section \ref{sec:4photonsPlots}), $\Bch$ does actually participate in some bounds. To see how, we first note that the sum rule for $\Bch$ (listed in table \ref{tab:sumrules}), involves only off-diagonal entries coming from the $\pi\pi \to \pi\gamma$ spectral density. Since bounds require constructing positive-semidefinite matrices, we need to provide sum rules that fill in the $\pi\pi\to \pi\pi$ and $\pi\gamma\to\pi\gamma$ diagonals. These should provide the \textit{two normalization conditions} that are needed for the mixed system.

As usual, we choose $g_{1,0}=1$ as the normalization of the pion field. For the photon field we can choose any sign-definite coupling of the $\pi\gamma\to\pi\gamma$ processes. One option is to use $c_2^{(1)}\leq 0$, like in section \ref{sec:2pi2gamma}. Solving the corresponding semidefinite problem at order $n_{\text{max}}=5$ yields the bound
\begin{equation}\label{eq:Bchbound}
    |\widetilde B_{\text{ch}}|\equiv \frac{\left|\Bch\right|}{\sqrt{-g_{1,0}c_2^{(1)}}} \leq 10.5830052443\,.
\end{equation}
The interest of this bound lies in the fact that $\Bch$ can be determined explicitly by matching the chiral anomaly in the UV, as reviewed in section \ref{sec:anomalyMatching}. Plugging in the values of \eqref{eq:gpiLow} and \eqref{eq:BchLow}, we can reinterpret this result as a bound on the $2\pi2\gamma$ coupling $c_2^{(1)}$,
\begin{equation} \label{eq:nonhomogeneousbound}
    \sqrt{-c_2^{(1)}/e^2}\gtrsim \frac{1}{10.583}\frac{2 N}{3\pi^2 f_\pi^2}\,.
\end{equation}
We see that the anomaly brings the $N$ dependence explicitly into the game, producing \textit{inhomogeneous} bounds! Out of amusement, we may plug in the experimentally measured value $f_\pi\simeq 92\,\text{MeV}$ and set $N=3$ to find the real-world bound $\sqrt{-c_2^{(1)}/e^2}\gtrsim 2.262\, \text{GeV}^{-2}$. This should hold up to finite $N$ corrections. It would be interesting to compare this to experimental results and/or lattice computations but, as far as we are aware, these cannot be found in the literature.

It is worth pointing out that $c_2^{(1)}$ is not the only coupling that can be used to normalize the bound on $\Bch$, we may use any other sign-definite combination. For example, using $c_2^{(1)}+d_2^{(1)}\leq 0$ we obtain
\begin{equation}\label{eq:Bchbound2}
    \frac{\left|\Bch\right|}{\sqrt{-g_{1,0}\big(c_2^{(1)}+d_2^{(1)}\big)}} \leq 7.4833147735\,.
\end{equation}
Another possibility is to use $d_2^{(2)}+\frac{2}{3}d_2^{(1)}\leq 0$, which is not obvious from the sum rules in table \ref{tab:sumrules} but corresponds to the slanted line of figure \ref{fig:d21d22} ---the bound due to the large $N$ selection rules. This yields
\begin{equation}\label{eq:Bchbound3}
    \frac{\left|\Bch\right|}{\sqrt{-g_{1,0}\big(d_2^{(2)}+\frac{2}{3}d_2^{(1)}\big)}} \leq 9.7979589711\,.
\end{equation}
Similar arguments as before can be used to translate these results into inhomogeneous bounds for the different pion-photon couplings involving $N$ explicitly.

\begin{figure}[ht]
\centering
\includegraphics[scale=1]{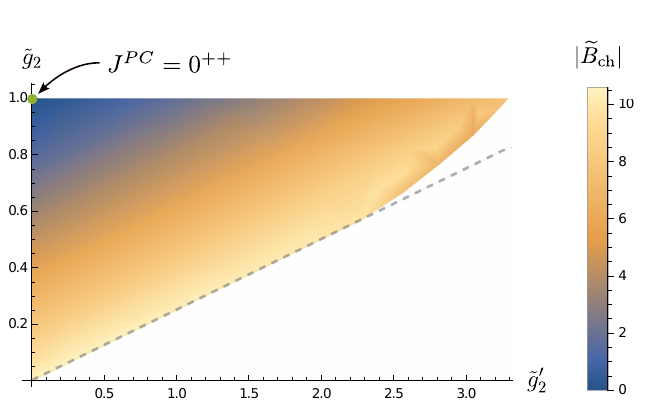}
\caption{Bound on the normalized anomaly $|\widetilde B_{\text{ch}}|$ (defined in \eqref{eq:Bchbound}) across the $(\tilde g_2', \tilde g_2)$ plane. This plot was made with $n_\text{max}=5$. The bound is saturated along the Skyrme line $\tilde g_2 = \frac{1}{4}\tilde g_2'$ up to the kink and then decreases away from it. At the top-left corner (green) the bound drops to zero, which agrees with a single $J^{PC}=0^{++}$ exchange with mass at the cutoff $M^2$.}
\label{fig:Bch3D}
\end{figure}

Aside from absolute bounds, it is also interesting to study how these bounds compare to other directions in the space of EFT couplings. In particular, we may ask \textit{what is the (normalized) bound on $\Bch$ at every point of the $(\tilde g_2', \tilde g_2)$ plot of \cite{Albert:2022oes}?} This carves out the three-dimensional convex space of figure \ref{fig:Bch3D} (where we used $c_2^{(1)}$ to normalize $\Bch$). We see that the bound \eqref{eq:Bchbound} is saturated along the Skyrme line, where it remains constant, and it vanishes at the top left corner. Using the normalizations of \eqref{eq:Bchbound2} and \eqref{eq:Bchbound3} instead produces plots qualitatively identical to figure \ref{fig:Bch3D} but saturating the corresponding bounds on the Skyrme line.

\subsubsection*{Ruling in}
We conclude this section by discussing some features of the theories that might saturate the above bounds on the anomaly. The $\tilde g_2'=0$ vertical line of figure \ref{fig:Bch3D} is known to correspond to amplitudes involving only $J^{PC}=0^{++}$ scalar exchanges \cite{Albert:2022oes}. It is interesting to study this line in higher-dimensional sections to find out whether there are other theories sitting on top of it. Scalars with $J^{PC}=0^{++}$ couple to two pions by \eqref{eq:pipiphiInt}, but cannot couple to a pion and a photon (as can be read off from table \ref{tab:vertices}). This means that they do not contribute to $\pi\pi\to\pi\gamma$ and so they have $\Bch=0$. This only saturates the bound of figure \ref{fig:Bch3D} strictly at the top left corner, suggesting no degeneracy for the amplitudes on this corner, but showing that along the vertical line there must exist extended theories with more that a single $J^{PC}=0^{++}$ exchange.

We can perform a similar analysis for the Skyrme line $\tilde g_2 = \frac{1}{4}\tilde g_2'$. The bounds are constant along this line and saturate the absolute bounds (\ref{eq:Bchbound},\ref{eq:Bchbound2},\ref{eq:Bchbound3}). Amusingly, these bounds match (up to at least 10 decimal points) with the following simple expressions,
\begin{gather}\label{eq:AnalyticalBounds}
    \frac{\left|\Bch\right|}{\sqrt{-g_{1,0}c_2^{(1)}}} \leq 4\sqrt{7}\,, \qquad \frac{\left|\Bch\right|}{\sqrt{-g_{1,0}\big(c_2^{(1)}+d_2^{(1)}\big)}} \leq 2\sqrt{14}\,, \nonumber\\
    \frac{\left|\Bch\right|}{\sqrt{-g_{1,0}\big(d_2^{(2)}+\frac{2}{3}d_2^{(1)}\big)}} \leq 4 \sqrt{6}\,.
\end{gather}
This calls for a simple analytic explanation for these bounds. The obvious candidate is a single rho meson ($J^{PC}=1^{--}$) exchange, suitably UV-completed by infinitely-heavy masses, since this is known to saturate the Skyrme line $\tilde g_2 = \frac{1}{4}\tilde g_2'$ \cite{Albert:2022oes,Fernandez:2022kzi}.

The rho couples to two pions by
\begin{equation}
    \mathcal L_{\text{int}} = \sqrt{2}\, g_{\pi\pi\rho} \pi^a \partial^\mu \pi^b \rho_\mu^c f_{abc}\,,
\end{equation}
and to a pion and a photon via the third term of \eqref{eq:spin1ints}. The corresponding $\pi\pi \to \pi\pi$ and $\pi\pi \to\pi\gamma$ amplitudes for a single rho exchange are
\begin{align}
    M_{4\pi}(s,u) &\, = g_{\pi\pi\rho}^2 \left(\frac{s + 2u}{m_\rho^2 - s} + \frac{u + 2s}{m_\rho^2 - u}\right)\,,  \\
    M_{3\pi\gamma}(s,u) &\, = 8\sqrt{2}\, \frac{g_{\pi\pi\rho}g_{\pi\gamma\rho}}{m_\rho} \left(\frac{1}{m_\rho^2 - s} + \frac{1}{m_\rho^2 - u}\right)\,.
\end{align}
Its contributions to the $\pi\gamma \to \pi\gamma$ amplitudes were given in \eqref{eq:J1RuleIn}.

These amplitudes are not valid as they stand, since in the Regge limit they grow exactly like $M_{4\pi}\sim s$, $M_{3\pi\gamma}\sim \text{ct}$ and we assumed it was strictly better than this. But it was pointed out in \cite{Albert:2022oes} that $M_{4\pi}(s,u)$ can be UV-completed by higher poles as
\begin{equation}\label{eq:1rhoUV}
    M_{4\pi}(s,u) = g_{\pi\pi\rho}^2 \left[\frac{s + 2u}{m_\rho^2 - s}\left(\frac{m_\infty^2}{m_\infty^2-u}\right) + \frac{u + 2s}{m_\rho^2 - u}\left(\frac{m_\infty^2}{m_\infty^2-s}\right)\right]\,,
\end{equation}
where we send $m_\infty \to \infty$. This amplitude is now consistent with the Regge assumption and it is unitary at all residues. It stands to reason that similar UV completions exist for $M_{3\pi\gamma}(s,u)$ and the various $2\pi2\gamma$ amplitudes. If that is indeed the case, their corresponding low-energy couplings should be
\begin{gather}
    \frac{\left|\Bch\right|}{\sqrt{-g_{1,0}c_2^{(1)}}} = \frac{16}{\sqrt{3}}\,, \qquad \frac{\left|\Bch\right|}{\sqrt{-g_{1,0}\big(c_2^{(1)}+d_2^{(1)}\big)}} = \frac{8\sqrt{2}}{\sqrt{3}}\,, \nonumber\\
    \frac{\left|\Bch\right|}{\sqrt{-g_{1,0}\big(d_2^{(2)}+\frac{2}{3}d_2^{(1)}\big)}} = \frac{16\sqrt{2}}{\sqrt{7}}\,.
\end{gather}
While these values do not quite saturate the bounds in \eqref{eq:AnalyticalBounds}, they do have the same relative ratios. It is unclear what modifications would be needed to push these results up to the bound. Perhaps one needs to include other resonances apart from the rho, or perhaps there exists some clever UV completion of a single rho which modifies some of the low-energy couplings. It would be interesting to investigate this further.

\section{Sharper bounds from the new physical constraints?}
\label{sec:Discussion}

We have seen that, 
compared to the four-pion system, the mixed pion/photon system 
 enjoys more stringent constraints, of various conceptual origin. They capture some  very interesting physics
 and might be expected to get us much closer to large $N$ QCD.
 Unfortunately, we have encountered 
several obstacles
in implementing some of them in the semidefinite program.
In this section,  after a systematic
summary of the new constraints, we discuss our best understanding of the obstructions and speculate on  possible solutions.

\subsection{Summary of the new constraints}
\label{sec:newconstraints}

 As our discussion of the new constraints is scattered throughout the text, we provide here an executive summary.

\paragraph{Large $N$ selection rules.} Requiring that intermediate states in any large $N$ scattering process correspond only to single-meson exchanges with physical quantum numbers results in a set of ``large $N$ selection rules''. These come in two types, both derived from the condition that mesons are pure $q\bar q$ pairs:
\begin{itemize}
    \item \textbf{Flavor:} Only the adjoint representation of $U(N_f)$ has non-vanishing spectral density. Other representations are suppressed at large $N$. This is the suppression of what are known as exotic mesons \textit{of the first kind} in the old literature. This is discussed in section~\ref{sec:flavor}.

    \item \textbf{Spin:} Not every combination of spin $J$, parity $P$ and charge conjugation $C$ quantum numbers can be realized in a $q \bar q$ bound state of quarks. Exchanges with such quantum numbers (the so-called exotic mesons \textit{of the second kind}) are suppressed at large $N$. See the discussion in section \ref{sec:MoreSelectionRules}.
\end{itemize}

In four-pion scattering, only the first type of these sum rules makes an appearance in what can be seen as a consequence of the OZI rule. As a result, $M_{4\pi}(s,u)$ has no $t$-channel poles, which imposes stringent constraints on the low-energy couplings. In the full system, these sum rules also imply the vanishing of the $t$-channel poles in some reduced amplitudes like $M_{\pi\gamma\gamma\pi}^{(1)}(s,u)$, $M_{\pi\gamma\gamma\pi}^{(2)}(s,u)$ and $M_{3\pi\gamma}(s,u)$. The second type only becomes relevant in the full mixed system. While their consequences are not as clear-cut in general, these sum rules are responsible for the slanted line of figure \ref{fig:d21d22}, which seems to prevent the UV completion of a single $J^{PC}=1^{++}$ exchange (see the discussion below \eqref{eq:J1LE}).

\paragraph{Goldstone constraints.} The fact that pions are Goldstone bosons of spontaneous chiral symmetry breaking has some consequences for their low-energy scattering. Famously, $M_{4\pi}(s,u)$ vanishes when $s=u=0$ ---the so-called Adler zero. This is clear from the chiral Lagrangian \eqref{eq:Lch}, since one cannot write down a non-trivial term without derivatives out of the unitary matrix $U(x)$ of the $U(N_f)$ non-linear sigma model. Goldstone bosons are thus always derivatively coupled. Similar constraints occur when coupling the chiral Lagrangian to background gauge fields. For example, there is only one non-trivial term proportional to $F_{\mu\nu}F^{\mu\nu}$ that one can write down with no additional derivatives (see \eqref{eq:GBnature}). This implies the equality of different low energy coefficients that were a priori independent.

While the Adler zero in $M_{4\pi}(s,u)$ is too low an effect to be reached with the dispersion relations allowed by the Regge behavior, we \textit{can} reach one such Goldstone constraint in the full mixed system. With the default Regge behavior granted by the intercept of the rho trajectory\footnote{We will shortly see that one can reach further Goldstone constraints in the improved Regge channels.} (see section \ref{sec:ReggeBehavior} and appendix \ref{app:Regge} for details) we obtain two independent sum rules for the coupling $\gOne$ defined in \eqref{eq:2p2gLow} (see table \ref{tab:sumrules}). Their difference generates a novel null constraint. We stress that this null constraint is not a consequence of crossing symmetry, but rather of the Goldstone boson nature of pions. One would therefore expect that it imposes stringent bounds that go beyond the simple solutions to crossing that tend to saturate the bounds (and indeed it can be used to rule out some candidate theories for the ruling in of figure \ref{fig:d21d22}). But unfortunately this constraint is not taken into account numerically by the semidefinite solver due to a potential numerical obstruction that we discuss below.

\paragraph{Improved Regge channels.} 
So far in the discussion and in the numerical bounds above we only used the constraints coming from the default Regge behavior \eqref{eq:MRegge}, which follows from the intercept of the rho trajectory. However, there exist some combinations of reduced amplitudes $M_i^{(\text{imp)}}(s,u)$ with no contributions from the rho Regge trajectory in the $u$ channel. We call these combinations \textit{improved Regge channels}, since Regge theory tells us that they enjoy a softer Regge behavior \eqref{eq:MimpRegge}. We identified three such channels in section \ref{sec:ReggeBehavior}. What does the improved Regge behavior buy us?

First, the improved Regge behavior allows us to write down a new dispersion relation with fewer subtractions for each case,
\begin{equation}
    \oint_\infty \frac{ds}{2\pi i}\frac{1}{s}M^{(\text{imp})}_*(s,u) =0\,.
\end{equation}
Expanding around $u\sim 0$ we find new sum rules for the low-energy couplings defined in section \ref{sec:LowEnergies}. Equating them with the previous sum rules then gives three new towers of null constraints, which can be obtained systematically like in section \ref{sec:NullConstraints} from the following contour integral,
\begin{equation}\label{eq:NCint-IMP}
    \oint_0 \frac{du}{2\pi i}\oint_\infty \frac{ds}{2\pi i}\frac{1}{s u}\left(\frac{M^{(\text{imp})}_*(s,u)}{u^\ell} - \frac{M^{(\text{imp})}_*(u,s)}{s^\ell}\right)=0\,, \quad \ell=1,2,...\,.
\end{equation}

Second, with fewer subtractions we reach lower couplings that were not accessible before. In particular, we now reach the gauge coupling $e^2$ with dispersion relations (see table \ref{tab:e2sumrules} for the corresponding sum rules). This is very exciting because we could now in principle use this coupling as the normalization for the photon field. Then, the bound on $\Bch$ \eqref{eq:Bchbound} would involve $e^2$ rather than the higher-derivative coupling $c_2^{(1)}$, turning \eqref{eq:nonhomogeneousbound} into an absolute lower bound for $f_\pi^2/ M_\rho^2 N$! Sadly, this turns out not to be possible because the sum rule for $e^2$ is not sign-definite and thus cannot be used to fix a normalization. We elaborate on this point and suggest some potential workarounds below.

\begin{table*}[!t]\centering
\caption{New sum rules for the low-energy couplings defined in section \ref{sec:LowEnergies} that can only be reached through the improved Regge channels. The sum rules are given by the sum of the corresponding high-energy averages.
\label{tab:e2sumrules}}
\makebox[\textwidth][c]{
\begin{tabular}{|c||c|c|c|c|c|c|c|c|c|}
\hline
$PC$ & $++$ & $++$ & $++$ & $++$ & $-+$ & $-+$ & $+-$ & $--$ & $--$\\
$J$ & $0$ & even & $1$ & odd & $0$ & even & $1,$odd & even & $1,$odd\\
\hline
\hline
$e^2$
        & $\begin{psmallmatrix}0 & \frac{1}{2\sqrt{2}}\\ 
                \frac{1}{2\sqrt{2}} & 0\end{psmallmatrix}$
        & $\begin{psmallmatrix}0 & 0 & \frac{1}{2\sqrt{2}} & 0\\ 
                0 & \frac{1}{2} & 0 & 0\\\frac{1}{2\sqrt{2}} & 0 & 0 & 0\\0 & 0 & 0 & 0\end{psmallmatrix}$
        & $\frac{1}{2}$
        & $\begin{psmallmatrix}\frac{1}{2} & 0\\ 0 & 0\end{psmallmatrix}$
        & ---
        & $\begin{psmallmatrix}\frac{-1}{2} & 0\\ 0 & 0\end{psmallmatrix}$
        & $\frac{1}{2}$
        & $\frac{-1}{2}$
        & $\begin{psmallmatrix}0 & 0\\ 0 & \frac{-1}{2}\end{psmallmatrix}$\\
\hline
$e^2$& ---
        & $\begin{psmallmatrix}0 & 0 & 0 & 0\\ 
                0 & \frac{1}{2} & 0 & 0\\0 & 0 & 0 & 0\\0 & 0 & 0 & 0\end{psmallmatrix}$
        & $\frac{1}{2}$
        & $\begin{psmallmatrix}\frac{1}{2} & 0\\ 0 & 0\end{psmallmatrix}$
        & ---
        & $\begin{psmallmatrix}\frac{-1}{2} & 0\\ 0 & 0\end{psmallmatrix}$
        & $\frac{-1}{2}$
        & $\frac{1}{2}$
        & $\begin{psmallmatrix}0 & 0\\ 0 & \frac{1}{2}\end{psmallmatrix}$\\
\hline
\end{tabular}}
\end{table*}

Finally, these improved Regge channels yield new \textit{Goldstone constraints}. For example, we do not just access the gauge coupling $e^2$ but we actually access it twice, meaning that we obtain two independent sum rules for it (see table \ref{tab:e2sumrules}). These come from the two appearances of $e^2$ in the low-energy expansions of \eqref{eq:2p2gLow}. These two couplings are equated for a similar reason than the two $\gOne$ were, namely one can only write a single relevant term in terms of the $U(x)$ matrix of the chiral Lagrangian. The difference of sum rules in table \ref{tab:e2sumrules} thus realizes a new Goldstone constraint.

\subsection{Numerical obstruction for null constraints}

\label{sec:sharper}
The new constraints that we have just reviewed capture many features intrinsic to large $N$ QCD, like its sigma model nature at low-energies or the intercepts of the first Regge trajectories. For this reason, one would expect that including these constraints would shrink the exclusion plots above, zooming in onto the low-energy EFT of large $N$ QCD. But this is not the case numerically. Incorporating the Goldstone constraints and the additional towers of null constraints derived from the improved Regge channels does not modify the exclusion plots above nor produce any new bounds.

This appears to be due to the following technical obstruction. The new null constraints derived in this way are \textit{oscillating in spin} at large $J$, meaning that as $J\to\infty$ there are sectors that contribute equally with opposite signs. For example, the new null constraint from \eqref{eq:NCint-IMP} with $\ell=2$ for $M_{4\gamma}^{(\text{imp})}(s,u)$ involves, in the $J\to \infty$ limit,
\begin{gather}
    \quad\avg{\begin{psmallmatrix}0 & 0 & 0 & 0\\ 
                0 & 0 & 0 & 0\\0 & 0 & \frac{J^4}{2m^4} & 0\\0 & 0 & 0 & \frac{-J^4}{12m^4}\end{psmallmatrix}}_{\text{even}^{++}}
        + \quad\avg{\begin{psmallmatrix}0 & 0\\ 0 & \frac{J^4}{12m^4}\end{psmallmatrix}}_{\text{odd}^{++}}
        + \quad\avg{\begin{psmallmatrix}0 & 0\\ 0 & \frac{-J^4}{2m^4}\end{psmallmatrix}}_{\text{even}^{-+}}\,.
\end{gather}
When running the semidefinite problem solver, we demand positivity of the functional on every sector independently (recall \ref{eq:bootstrpeq}). Being high-order polynomials in $J$, the new null constraints dominate as $J\to\infty$ (at any given order in $1/m^2$). But since they involve contributions with opposite signs, SDPB weights them by zero to satisfy the positivity requirement, thus ignoring them.

More generally, we note that oscillating null constraints imply nontrivial relations between spectral densities asymptotically in $J$. But this limit lies outside the scope of the standard formulation of the semidefinite problem, in which one truncates the spins as $J=0,1,2,...,J_{\text{max}}$.
This suggests that in order to incorporate oscillating null constraints numerically one would have to find a different truncation that probes the $J\to \infty$ limit systematically. It would be very interesting to find such a truncation in which oscillating constraints do not trivialize, for these new null constraints are bound to get us closer to large $N$ QCD.

\subsection{Accessing low-derivative couplings}
\label{sec:lowderivative}
One of the main obstacles of positivity bounds is reaching the coefficients of low-derivative EFT operators. The reason, of course, is that one can only assume boundedness of the amplitude up to a given Regge behavior. Then one can only write down dispersion relations with at least a fixed minimum of subtractions, which might fail to capture the lowest-derivative EFT couplings. Typically, one assumes only the conservative Regge behavior granted by Froissart's bound, which for scalar scattering reads $\lim M(s,u)/s^2\to 0$ as $|s|\to \infty$ with fixed $u<0$. This then leaves out any coupling with less than four derivatives.

We have seen that in the case of large $N$ QCD, Regge theory allows us to do better.
In the case of meson scattering, the Regge behavior is controlled by the intercept of the leading Regge trajectory ---the rho, which already goes beyond Froissart's bound. This grants us access to couplings like $g_{1,0}$ in four-pion scattering, $\Ach^2$ in four-photon scattering, $\gOne$ in $2\pi2\gamma$ scattering and $\Bch$ in $3\pi\gamma$ scattering. A more detailed examination of the Regge behavior then even allows us to reach $e^2$, as we just reviewed.

Clearly, these are the most interesting couplings in the chiral Lagrangian \eqref{eq:gaugeLch}. Not only because they control pion (and photon) scattering at lowest energies, and are therefore the first couplings to be measured experimentally, but also because they capture RG invariants like 't Hooft anomalies. Including all these couplings in positivity bounds would take us a long way and yield the strongest bounds to date.

Unfortunately, while we do have access to all these couplings via valid dispersion relations, we have found in practice that some of them decouple from all positivity bounds. This is the case for $\Ach^2$, $e^2$ and $\gOne$. The reason appears to be that they come at lowest-order but their sum rules are not sign definite. Recalling our discussion in section \ref{sec:generalities}, we know that the lowest-derivative couplings should be used as normalization for the external states in any positivity bound. Otherwise, theories with states only at $m^2\to \infty$ would trivialize the bounds. However, if the coupling is not sign-definite, it cannot be used as normalization, since it may vanish for non-trivial theories and, again, make the bounds diverge. If we could impose further selection rules that forbid some sectors and produce sign-definite sum rules, then they could be used to normalize other ratios. But we are not aware of any additional symmetry that would produce such a selection rule.

Perhaps a more promising direction is to look instead for new sum rules and null constraints. We recall at this point that we have analyzed only the simplest system of mixed correlators involving only pion states and \textit{on-shell} electromagnetic currents $J_{\text{em}}^{\mu}(x)$ ---which turned out to be already rather technically involved. There are two clear generalizations. First, one may take the photons (or the corresponding currents) \textit{off-shell}. And second, one may consider more general $A_L,A_R$ backgrounds. In either case, we would be able to write down new dispersion relations (either from new processes or new kinematic variables) that would result in new sum rules and null constraints. Importantly, these would not be constraints on new unfixed low-energy coefficients but \textit{the same} coefficients appearing in the chiral Lagrangian \eqref{eq:gaugeLch}. With this plethora of new constraints there is room for hope that one may be able to construct \textit{sign-definite} sum rules for the low-derivative coefficients, making them couple to positivity bounds. These generalizations remain exciting future directions, which should be now tractable with the machinery we have developed.

\section{Conclusions}
\label{sec:Conclusions}

In this paper we have continued the program \cite{Albert:2022oes} of carving out the space of large $N$ confining gauge theories by the systematic implementation 
of the general consistency conditions on $2 \to 2$ hadron scattering.  
The ultimate goal is to corner large $N$ QCD and  other theories of physical interest.  Given that the method is so general, a central question is how to input the physics of QCD into the bootstrap machinery.
This is our main motivation to start with the mesons --  their  large $N$ kinematics is more rigid,  and we can rely on real-world and lattice data
to inform our intuition.

We organize the problem using the language of effective field theory. In the first iteration \cite{Albert:2022oes}, we only included the massless pions among the light states. This simple setup was already powerful enough
to yield fairly stringent bounds on the higher-derivative coefficients of the chiral Lagrangian. The basic kinematic features of large $N$ pion scattering, namely the absence of $t$-channel poles in flavored-ordered amplitudes and
their milder Regge behavior (one subtraction suffices because the pomeron is absent), played an essential role. 

Here we have enlarged the set of observables, including probe on-shell photons as external states, which is really just a convenient way to study form factors of the electromagnetic current.
To conquer this much more involved kinematics we had to circumvent several technical hurdles. We dealt with them by working fully covariantly, both in the parametrization of the amplitudes
and in their partial wave expansion. 

The payout is that
this richer setup is  much more sensitive to the physics of large $N$ QCD. 
First,
it probes very thoroughly the
 peculiarities of large $N$ meson scattering. For starters, we could impose the ``large $N$ selection rules'' that encode planarity and the fact that  the intermediate states must be compatible
with the naive quark model (i.e.~only $q \bar q$ states are allowed). 
These are still rather generic features, as they are shared by several large $N$ theories with quarks in the fundamental representation. By a careful application of Regge theory, we were then able to identify ``improved Regge channels'' where no subtractions are needed in the dispersion relation,
because
 the rho Regge trajectory cancels out and we can rely on our phenomenological  knowledge that the next trajectory is that of the pion. This is precisely  the kind of more detailed physical input that should help us zoom in onto large~$N$ QCD.
 Second, the EFT of pions and photons encodes the Goldstone boson nature of the pion  in four-point couplings that are accessible to our dispersion relations. This allowed us to derive some  novel ``Goldstone constraints'' for the heavy spectral densities.
 Finally, the coefficient of the chiral anomaly also shows up at the four-point coupling level. Matching the anomaly 
   to the microscopic theory we could input into the bootstrap an exact quantitative feature of large $N$ QCD.

Following the blueprint of~\cite{Caron-Huot:2020cmc} (with the aid by semidefinite numerical methods \cite{Simmons-Duffin:2015qma}), we
 have obtained several exclusion plots for various slices of parameter space. One is able to constrain suitable ratios of Wilson 
coefficients, rendered dimensionless by powers of the EFT cut-off $M$ (which we identify with the mass of the rho meson). Perhaps our most interesting results are the bounds that 
involve the (appropriately normalized) chiral anomaly coefficient. If we treat it as any other EFT coupling, we can follow how its upper bound varies in the space of four-derivative $4 \pi$ couplings that we carved out in our first paper (figure \ref{fig:Bch3D}).
The anomaly reaches a maximum on  the straight  oblique segment of the exclusion boundary. We do not yet have an analytic understanding of its numerical value --  integrating out a
single $\rho$ in the $\pi \gamma \to \pi \gamma$ amplitude gives a value that is strictly smaller than our bound, suggesting the need for additional light states (which decouple in the $4 \pi$ system).
 A clear direction for future work is to play the analytic  ``ruling in'' game more thoroughly, as it seems likely that several features of our plots may be explained by relatively simple amplitudes. 
  If we remember the special role of the anomaly coefficient and match its value to the microscopic theory, we obtain very interesting ``inhomogeneous'' bounds with an explicit $N$ dependence. In essence, the anomaly allows us
  to fix the overall normalization of the Wilson coefficients. 

Our numerical results reflect many conceptually new features of the pion/photon system,  such as several large $N$ constraints and the 
ability to input the chiral anomaly, but unfortunately not all of them. As we  have discussed  at length in section \ref{sec:Discussion}, we have encountered some numerical obstructions in 
implementing the Goldstone and improved Regge channels constraints. We have also found that some interesting low-lying EFT coefficients, which would be outside the reach of dispersion relations by naive power counting, but that become accessible with our improved treatment of Regge theory, do not have positive definite sum rules
and thus decouple from the semidefinite program.
We have mentioned in section \ref{sec:Discussion}  two promising extensions of our work that may overcome some of these difficulties. One is 
to take the photons off-shell, i.e.~to study the full kinematics for the form factors of the electromagnetic current. The other is to enlarge the set of background gauge fields  from those for $U(1)_Q$  to those of the full $U(N_f)_L \times U(N_f)_R$
global symmetry group.

We hope to have convinced our patient readers that the large $N$ hadronic  bootstrap is a very powerful tool to study a classic problem in theoretical physics. 
This work represents a second step
in a comprehensive research program. A third step is taken in~\cite{rhos}, which  will include the rho meson among the light states and  study the full pion/rho mixed system.
While  significantly more involved than in~\cite{Albert:2022oes}, the calculations carried out in this and in the upcoming paper are still rather simpler than 
those routinely undertaken by the numerical conformal bootstrap community.  
We are beginning to understand how to input the physics of QCD into the
the large~$N$ hadronic bootstrap. 
We look forward to continuing this exploration.

\acknowledgments
We are grateful to Simon Caron-Huot, Yichul Choi, Matthew Forslund, Zohar Komargodski, Yue-Zhou Li, Julio Parra-Martinez, Jo\~ao Penedones and David Simmons-Duffin for  useful discussions, 
and especially  to Johan Henriksson and Alessandro Vichi 
for ongoing collaboration on related projects and  comments
on the draft.
This research was supported in part 
by the National Science Foundation under Grant No.~NSF PHY-2210533. LR is supported in
part  by Simons Foundation grants 397411 (Simons
Collaboration on the Nonperturbative Bootstrap) and 681267 (Simons Investigator
Award).
We thank the KITP, Santa Barbara, for hospitality during the workshop ``Bootstrapping quantum gravity'', when some of this work was carried out. KITP workshops are 
supported in part by the National Science Foundation under Grant No.~NSF PHY-1748958.

\appendix

\section{Review of Regge theory}\label{app:Regge}
In order to make the discussion in section \ref{sec:ReggeBehavior} self-contained, we summarize here 
some basic facts of Regge theory. None of the results of this appendix are new, and they are mostly drawn from Gribov's book \cite{Gribov:2003nw}. See also \cite{Penedones} for a nice recent pedagogical introduction, and \cite{Phillips:875552} for an excellent review of the ideas (and jargon) that were running around in the early seventies. 

Consider a scattering amplitude $M(s,u)$ of four arbitrary massless scalar mesons $a$,$b$,$c$,$d$. Since we have in mind a large $N$ amplitude, we will assume $M(s,u)$ to be a meromorphic function with poles corresponding only to the physical exchanges of mesons in the $s$, $t$ and $u$ channels (although most of the arguments below hold more generally). The question we will address is what is the behavior of $M(s,u)$ in the 
Regge limit
\begin{equation}
    |s|\to\infty \,,\qquad \text{at fixed } u\lesssim 0\,.
\end{equation}
As first suggested by Mandelstam, the high-energy behavior of this amplitude at fixed $u$ and large $s$ can be obtained from its $u$-channel $a+ c \to b + d$ partial wave expansion
\begin{equation}\label{eq:uPW}
    M(s,u) = \sum_{J=0}^\infty (2J+1) a_J^{u-\text{ch}}(u) \legP_J\left(z\right)\,, \qquad z = 1+\frac{2s}{u}\,.
\end{equation}
However, we cannot probe the Regge limit directly since this expansion is only valid for the physical $u$-channel region $u>0$, $s<0$. Indeed, taking $|s|\sim |z| \to \infty$ at fixed $u$ gives $\legP_J(z)\sim z^J$, which results in a divergent sum. We need a different representation of this partial wave expansion that can be safely continued to the region $ u <0$
and arbitrary $s$.

\subsection{The Sommerfeld-Watson transform}
\begin{figure}[ht]
\centering
\includegraphics[scale=0.70]{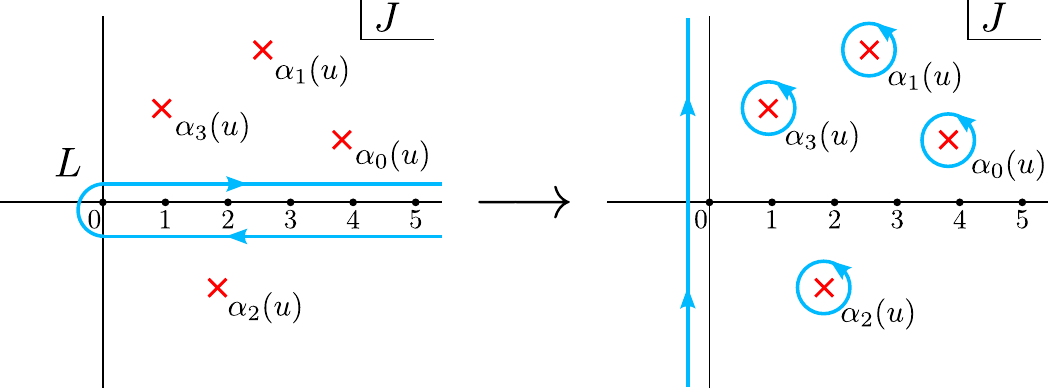}
\caption{Contour deformation for the Sommerfeld-Watson transform.}
\label{fig:SomWats}
\end{figure}

We can achieve such a representation if we find an analytic continuation of the coefficients $a_J^{u-\text{ch}}(u)$ into the complex $J$ plane that does not grow exponentially on the $\text{Re}\,J>0$ half plane. Indeed, if such a continuation is found, we can use the Sommerfeld-Watson transform to write
\begin{equation}
    M(s,u) = \frac{-1}{2\pi i}\int_L dJ\,\frac{\pi}{\sin \pi J}(2J+1)a_J^{u-\text{ch}}(u)\legP_J(-z)\,,
\end{equation}
where the contour $L$ is engineered to pick up all the poles due to $\sin \pi J=0$ but avoid any other singularities (see figure \ref{fig:SomWats}). Then we can deform the contour towards the imaginary axis, as sketched in figure \ref{fig:SomWats}, and drop the contour at infinity since
\begin{equation}\label{eq:PJasym}
    \frac{\legP_J(-z)}{\sin \pi J} \xrightarrow[|J|\to\infty]{} \frac{\cos(J (\pi - \theta))}{\sin \pi J}\,, \qquad z\equiv \cos\theta\,,
\end{equation}
decays exponentially away from the real $J$ axis. We thus obtain
\begin{equation}\label{eq:SomWats}
    M(s,u) = \text{residues} + \frac{i}{2}\int_{-i\infty}^{i\infty} dJ\,\frac{(2J+1)}{\sin \pi J}a_J^{u-\text{ch}}(u)\legP_J(-z)\,,
\end{equation}
where ``residues'' stands for all the residues and singularities of $a_J^{u-\text{ch}}(u)$ that we picked up when doing the contour deformation of figure \ref{fig:SomWats}. We will come back to these singularities momentarily.

In contrast with the initial partial wave expansion \eqref{eq:uPW}, which was defined for the physical $u$-channel angles $-1\leq z \leq 1$, the Sommerfeld-Watson representation \eqref{eq:SomWats} converges everywhere on the $z$ complex plane away from the $z>1$ real axis. This can be checked using the asymptotics of the Legendre polynomials \eqref{eq:PJasym} along the imaginary $J$ axis (see \cite{Gribov:2003nw} for details). Similar arguments can be used to continue $u$ to our region of interest, $u\lesssim 0$. We can therefore use this representation to probe the behavior in the Regge limit $|z|\to\infty$ (fixed $u\lesssim 0$) which, we will see, is controlled by the ``residues'' in \eqref{eq:SomWats}.

\subsection{Complex angular momentum}
\begin{figure}[ht]
\makebox[\textwidth][c]{
\centering
\includegraphics[scale=0.65]{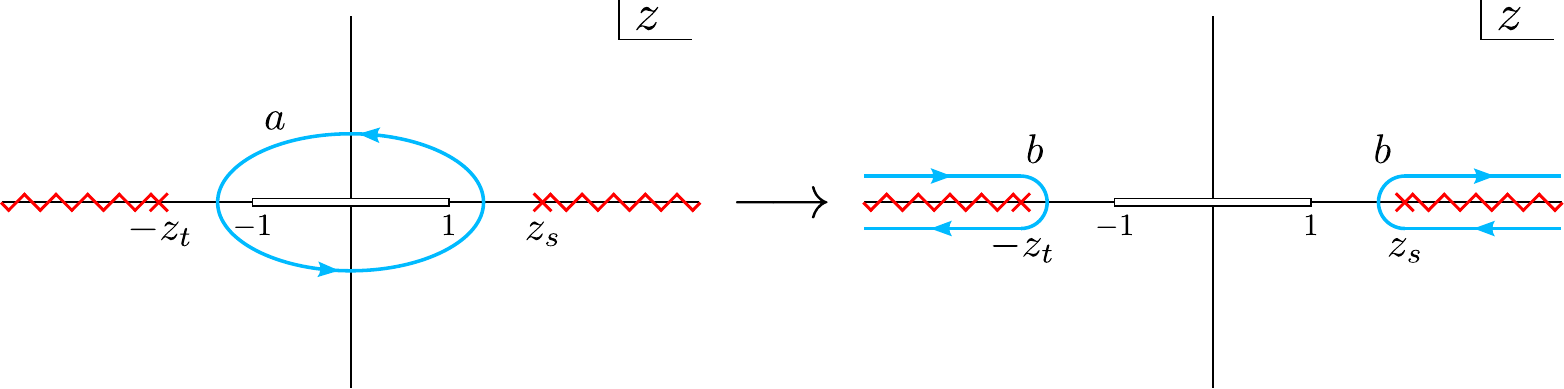}}
\caption{Contour deformation for the Froissart-Gribov representation.}
\label{fig:FroisGriv}
\end{figure}

The existence of the representation \eqref{eq:SomWats} hinges on the existence of a suitable analytic continuation of $a_J^{u-\text{ch}}(u)$ into the complex $J$ plane. This can be obtained via the Gribov-Froissart projection as follows. Using orthogonality of the Legendre polynomials on \eqref{eq:uPW}, we first write
\begin{equation}\label{eq:PJproj}
    a_J^{u-\text{ch}}(u) = \frac{1}{2}\int_{-1}^1 dz \,\legP_J(z) M(s(z),u)\,, \qquad s(z)\equiv \frac{u}{2}(z-1)\,.
\end{equation}
Then, recalling that the Legendre polynomials are related to the discontinuity of the Legendre functions of second kind $\legQ_J(z)$ (see e.g.\ \cite{Correia:2020xtr}), we have
\begin{equation}
    a_J^{u-\text{ch}}(u) = \frac{1}{2\pi i}\oint_a dz \,\legQ_J(z) M(s(z),u)\,,
\end{equation}
where the contour wraps the singularity at $z\in [-1,1]$, as shown in figure \ref{fig:FroisGriv}.

By blowing up the contour as in figure \ref{fig:FroisGriv}, we pick up the discontinuities on the positive and negative real $z$ axis. These start at $z_s\equiv 1+\frac{2M_s^2}{u}$ and $z_t=1+\frac{2M_t^2}{u}$ respectively and they correspond to the $s$-channel and $t$-channel physical exchanges. $M_s^2$ and $M_t^2$ denote the location of the first resonance (or ``cutoff'') in each channel. With a straightforward change of variables we may write
\begin{align}\label{eq:GribFrois}
    a_J^{u-\text{ch}}(u) =&\, \frac{1}{\pi}\frac{2}{u}\int_{M_s^2}^\infty ds \,\legQ_J(1+\tfrac{2s}{u}) \im_s M(s,u) \\
    +&\, \frac{(-1)^J}{\pi}\frac{2}{u}\int_{M_t^2}^\infty dt \,\legQ_J(1+\tfrac{2t}{u}) \im_t M(-t-u,u)\,, \nonumber
\end{align}
where we used $\legQ_J(-z)=(-1)^{J+1}\legQ_J(z)$. This expresses the partial wave coefficients in one channel in terms of the exchanges of the other two channels. To get to this expression we dropped the contour at infinity. This is justified for large-enough spin $J \geq J_0(u)$ if the amplitude is polynomially bounded by
\begin{equation}
    \lim_{|s|\to\infty} \frac{M(s,u)}{s^{J_0(u)}} = 0\,, \qquad \text{at fixed }u
\end{equation}
in virtue of the asymptotics $\legQ_J(z)\to 1/z^{J+1}$. This behavior is expected from general arguments involving causality and locality (see e.g.\ the discussion in appendix A.2 of \cite{Arkani-Hamed:2017jhn}), and in fact the Froissart-Martin bound \cite{Froissart:1961ux,Martin:1962rt} requires $J_0(u=0)\leq 2$ from unitarity (at least for theories with a mass gap) \cite{Jin:1964zza,Martin:1965jj}. 

The Gribov-Froissart projection \eqref{eq:GribFrois} is much more suitable for analytic continuation in $J$ than \eqref{eq:PJproj} because $\legQ_J(x)$ drops exponentially with $|J|\to \infty$ in the half plane $\text{Re}\, J>0$ (for $x>1$, where the integrals are supported). In contrast, $\legP_J(x)$ grows exponentially with $|J|$, making the continuation of \eqref{eq:PJproj} not valid for the Sommerfeld-Watson representation. The only issue of \eqref{eq:GribFrois} is the factor $(-1)^J$, which explodes due to the imaginary part of $J$ when continued into the complex plane. To avoid this issue, we define \textit{two} analytic continuations
\begin{equation}\label{eq:aJpm}
    a_J^{\pm}(u) \equiv \frac{1}{\pi}\frac{2}{u}\left(\int_{M_s^2}^\infty ds \,\legQ_J(1+\tfrac{2s}{u}) \im_s M(s,u)
    \pm \int_{M_t^2}^\infty dt \,\legQ_J(1+\tfrac{2t}{u}) \im_t M(-t-u,u)\right).
\end{equation}
From here on, we drop the superscript ``$u$-ch'' to avoid clutter; the channel should be clear from the argument. All in all, we have obtained two analytic continuations with the desired property that they do not grow too fast in the half plane $\text{Re}\, J>0$ which separately match the physical partial wave coefficients when $J$ takes even/odd integer values:
\begin{align}
    &a_J^{+}(u) = a_J(u)\;\; \text{for } J=\text{even}>J_0(u)\, \\
    &a_J^{-}(u) = a_J(u)\;\; \text{for } J=\text{odd}>J_0(u)\,. \nonumber
\end{align}

\subsection{Regge trajectories}
With $M(s,u)$ being meromorphic, $a_J(u)$ consists of a collection of poles corresponding to all the mesons of spin $J$ exchanged in the $u$-channel. The continuations $a_J^{\pm}(u)$ smoothly interpolate in spin between these physical resonances, realizing Regge trajectories $\alpha_n^{\pm}(u)$. Physical exchanges occur at the values of $u$ for which $\alpha_n^+(u)$ ($\alpha_n^-(u)$) take non-negative even (odd) integer values, corresponding to the spins of the resonances. We label trajectories with the subindex $n=0,1,2,\ldots$ decreasing with spin at a given mass. The leading Regge trajectories $\alpha_0^{\pm}(u)$, with the highest-spin mesons at every mass, will play a distinctive role.

The fact that the analytical continuations \eqref{eq:aJpm} are only valid for $J\geq J_0$ is related to the potential existence of elementary particles of spin $J<J_0$ falling outside of Regge trajectories \cite{Gribov:2003nw}. But it is a salient feature of QCD that all narrow resonances really do lie on Regge trajectories \cite{Phillips:875552}, even for lower spins. We will therefore assume that the continuations $a_J^{\pm}(u)$ can be extended to spins below $J_0$ and reproduce the physical low-spin resonances.

When exploring the complex $J$ plane, the poles of $a_J^{\pm}(u)$ become poles in $J$ of the form
\begin{equation}\label{eq:reggeon}
    a_J^{\pm}(u) \sim \frac{g_{ac}(J) g_{bd}(J)}{u - M_n^2(J)} \sim \frac{g'_{ac}(u) g'_{bd}(u)}{J-\alpha_n^{\pm}(u)}\,.
\end{equation}
The coefficients factorize for the whole Regge trajectory. This allows us to interpret it as the single exchange of some sort of state ---the ``Reggeon''. These poles are the singularities we pick up when blowing up the contour in the Sommerfeld-Watson transformation (figure \ref{fig:SomWats}). Filling in the ``residues'' of \eqref{eq:SomWats} we arrive at
\begingroup
\allowdisplaybreaks
\begin{align}\label{eq:ReggePole}
    M(s,u) =&\, \sum_n\beta^+_n(u)\frac{\legP_{\alpha_n^{+}(u)}(z) + \legP_{\alpha_n^{+}(u)}(-z)}{\sin \pi\alpha_n^+(u)} \nonumber \\
    &\, +\sum_n\beta^-_n(u)\frac{\legP_{\alpha_n^{-}(u)}(z) - \legP_{\alpha_n^{-}(u)}(-z)}{\sin \pi\alpha_n^-(u)} \nonumber \\
    &\,+ \frac{i}{2}\int_{-i\infty}^{i\infty} dJ\,\frac{(2J+1)}{\sin \pi J}a_J^{\pm}(u)\frac{\legP_J(-z)\pm\legP_J(z)}{2}\,,
\end{align}
\endgroup
where $\beta(u)$ contains the couplings of \eqref{eq:reggeon} and other factors. The combinations of Legendre polynomials ensure that only the physical zeroes of $\sin \pi\alpha(u)$ generate poles, i.e.\ the first term has poles only for $J=\text{even}$ while the second one for $J=\text{odd}$.

We are finally ready to answer what is the Regge behavior of $M(s,u)$. Keeping $u$ fixed and sending $|s|\sim |z|\to\infty$,
\begin{equation}\label{eq:AsymReggeon}
    M(s,u) \longrightarrow \beta_0^+(u)\frac{1 + e^{-i\pi \alpha_0^+(u)}}{\sin \pi\alpha_0^+(u)} z^{\alpha_0^+(u)} + \beta_0^-(u)\frac{1 - e^{-i\pi \alpha_0^-(u)}}{\sin \pi\alpha_0^-(u)} z^{\alpha_0^-(u)}\,.
\end{equation}
Clearly, the Regge growth is controlled by the \textit{leading Regge trajectory}, which has the highest spin. Out of $\alpha_0^+$ and $\alpha_0^-$, we should keep whichever is highest, giving
\begin{equation}
    \lim_{|s|\to\infty} M(s,u)\sim s^{\alpha_0(u)}\,, \qquad \text{at fixed }u\,.
\end{equation}
In particular, for $u\sim 0$, the Regge behavior is controlled by the Regge intercept $\alpha_0(0)$. An entirely analogous argument shows that the Regge behavior at fixed $t=-s-u$ is $M(s,u)\sim s^{\alpha_0(t)}$, where $\alpha_0(t)$ here is the leading Regge trajectory of the $t$-channel exchanges. The upshot is that the Regge behavior of the amplitude is controlled by (the intercept of) the leading Regge trajectory of the channel we keep fixed.

\subsection*{Exchange degeneracy}\label{app:exchange-degeneracy}
The previous account is what happens in a generic process. However, for large $N$ theories it is often the case that one of the channels is missing. This occurs when the quantum numbers in one of the channels are \textit{exotic} (i.e.\ not from $q\bar q$) and it is killed by the large $N$ selection rules (see the discussion at the end of section \ref{sec:flavor}). This is for example the case in the four-pion amplitude $M_{4\pi}(s,u)$ of section \ref{sec:param4Pi}. When all $t$-channel poles are missing, the second term in the Gribov-Froissart projection \eqref{eq:aJpm} vanishes and we obtain $a_J^+(u)=a_J^-(u)$. This implies that the plus and minus Regge trajectories are the same, $\alpha^+_n(u)=\alpha^-_n(u)$, a phenomenon known as \textit{exchange degeneracy} in the literature \cite{Phillips:875552}.

In the case of $M_{4\pi}(s,u)$, the leading two trajectories are that of the rho and $f_2$ mesons, with the following quantum numbers:
\begin{alignat*}{2}
    \alpha^-_0(u) =&\, \alpha_\rho(u)\,, \qquad &&\text{with } J^{PC}=\text{odd}^{--}\,, \\
    \alpha^+_0(u) =&\, \alpha_{f_2}(u)\,, \qquad &&\text{with } J^{PC}=\text{even}^{++}\,.
\end{alignat*}
We expect these two trajectories to coincide at large $N$, where Zweig's rule becomes sharp, but they actually agree very well already in the real world \cite{Phillips:875552} --- yet another instance of how well large $N$ approximates the real world.

The statement in the other direction also holds, and we have in fact
$$\text{exchange degeneracy} \Leftrightarrow \text{absence of $t$-channel poles.}$$
Indeed, if $u$-channel Regge poles are exchange-degenerate, meaning $\beta^+_n(u)=\beta^-_n(u)$ and $\alpha^+_n(u)=\alpha^-_n(u)$ in \eqref{eq:AsymReggeon}, we find
\begin{equation}
    M(s,u) \xrightarrow[|z|\to \infty]{} \beta_0(u)\frac{2}{\sin \pi\alpha_0(u)} z^{\alpha_0(u)}\,.
\end{equation}
This has a nonzero imaginary part $\im\, M(s,u)$ in the physical $s$-channel region ($z=1+\tfrac{2s}{u}\to -\infty$, $u<0$) but it is real in the physical $t$-channel region ($z=-1-\tfrac{2t}{u}\to +\infty$, $u<0$), indicating that $t$-channel exchanges are absent. We see that the absence of $t$-channel poles in $M_{4\pi}(s,u)$ is a consequence of a precise cancellation of the imaginary parts due to the plus and minus trajectories.

We end this appendix by discussing the Regge behavior at fixed $t$ of amplitudes, like $M_{4\pi}(s,u)$, whose $t$-channel poles are missing. When there are no physical exchanges in the channel that we keep fixed, the contour deformation of figure \ref{fig:SomWats}. does not pick up any poles until it approaches the imaginary axis. In particular, there is no $t$-channel leading Regge pole which would control the high-energy behavior. In this case, the arguments above cannot fix the precise scaling of $M(s,u)$, which will presumably be theory-dependent. It is nevertheless clear that it cannot be \textit{worse} than the Regge behavior in the other channels. To see this, we may trivially rewrite $M(s,-s-u)$ as the following linear combination,
\begin{equation}
    M(s,-s-u) = \frac{1}{2} \left(M^+(s,u) - M^-(s,u)\right)\,, \quad  M^\pm(s,u) \equiv M(s,u) \pm M(s,-s-u)\,.
\end{equation}
The fixed-$u$ Regge behavior of this combination probes the fixed-$t$ limit of $M(s,u)$. Applying the arguments above separately to the amplitudes $M^+(s,u)$ and $M^-(s,u)$ shows that each of them grows like $s^{\alpha_0(u)}$ because the $u$-channel poles are the first thing we encounter in the deformation of figure \ref{fig:SomWats}. Since $M(s,-s-u)$ is a linear combinations of polynomials $\sim s^{\alpha_0(u)}$ we conclude that the fixed-$t$ Regge behavior of $M(s,u)$ is at least \textit{as good as} its fixed-$u$ behavior. It is in fact bound to be better due to a cancellation of the leading term, but \textit{how much} better remains unclear.

\bibliographystyle{ytphys}
\bibliography{references}

\end{document}